\documentclass[prb,aps,amsmath,amssymb,showpacs,twocolumn]{revtex4-2}
\usepackage{CJK}
\usepackage{graphicx,amsmath,amssymb,bm,xcolor,upgreek}
\usepackage[utf8]{inputenc}
\usepackage{nicefrac}
\usepackage{multirow, makecell, ragged2e}
\usepackage{diagbox}
\usepackage[titletoc,title]{appendix}
\usepackage{chngcntr}
\usepackage{float}

\renewcommand{\vec}[1]{{\textbf{\textit{#1}}}}
\renewcommand{\vec}[1]{\mbox{\boldmath$#1$}}

\newcommand*{\rom}[1]{\expandafter\@slowromancap\romannumeral #1@}

\begin{document}
	
	\title{Origin of the $\nu=1/2$ fractional quantum Hall effect in wide quantum wells
	}
	\author{Tongzhou Zhao$^{1}$, William N. Faugno$^{1}$, Songyang Pu$^{1}$, Ajit C. Balram$^{2}$ and J. K. Jain$^{1}$}
	
	\affiliation{$^{1}$Department of Physics, 104 Davey Lab, Pennsylvania State University, University Park, Pennsylvania 16802, USA}
	\affiliation{$^{2}$Institute of Mathematical Sciences, HBNI, CIT Campus, Chennai 600113, India}

	\date{\today}
	
	\begin{abstract} 
		The nature of the fractional quantum Hall effect at $\nu=1/2$, observed in wide quantum wells almost three decades ago, is still under debate. Previous studies have investigated it by the variational Monte Carlo method, which assumes that the transverse wave function and the gap between the symmetric and antisymmetric subbands obtained in a local density approximation at zero magnetic field remain valid even at high perpendicular magnetic fields; this method also ignores the effect of Landau level mixing. We develop in this work a three-dimensional fixed phase diffusion Monte Carlo method, which gives, in a single framework, the total energies of various candidate states in a finite width quantum well, including Landau level mixing, directly in a large magnetic field. This method can be applied to one-component states, and also to two-component states in the limit where the symmetric and antisymmetric bands are nearly degenerate. Our three-dimensional fixed-phase diffusion Monte Carlo calculations find that the one-component composite-fermion Fermi sea and the one-component Pfaffian states are very close in energy for a range of quantum well widths and densities, suggesting that the observed 1/2 fractional quantum Hall state in wide quantum wells is likely to be the one-component Pfaffian state. We hope that this will motivate further experimental studies of this state.
	\end{abstract}
	\maketitle
		
	\tableofcontents
	
	\section{Introduction}
	
The field of fractional quantum Hall effect~\cite{Tsui82} (FQHE) has been the birthplace for a web of spectacular phenomena, exotic emergent particles, and nontrivial states, all arising as a result of the interaction between electrons. The FQHE is a rare example of a strongly correlated state for which we not only have a qualitative understanding of a large part of the prominent phenomenology but have achieved a detailed microscopic description that is quantitatively accurate~\cite{Jain07, Halperin20}. Nonetheless, the origin of a few experimentally observed states remains unsettled. This article aims to report on our theoretical investigations of one such state, namely the FQHE state at filling factor $\nu=1/2$ observed in wide quantum wells (WQWs)~\cite{Suen92, Suen92b, Suen94b, Luhman08, Shabani09a, Shabani09b, Shabani13, Liu14d, Hasdemir15, Mueed16, Drichko19}, the origin of which has been a topic of debate ever since its discovery. There are two motivations for our study. First, this observation is in stark contrast to the state at half-filling in narrow quantum wells, which is established to be a Fermi sea of composite fermions (CFs)~\cite{Halperin93, Jain07, Halperin20}.  The FQHE thus arises due to changes in the interaction arising from finite quantum well width, and thus constitutes an important challenge for our quantitative understanding of the FQHE. Second, the physical origin of the observed state can be potentially very interesting. 

A promising two-component state is the Halperin $(3,3,1)$ state~\cite{Halperin83}, which can be relevant because a very WQW behaves as a two-component system. [There is little doubt that the $1/2$ FQHE observed in {\it real} double-layer systems~\cite{Eisenstein92}, observed at the same time as the $1/2$ state in WQWs, is the two-component Halperin $(3,3,1)$ state~\cite{Faugno20}. Our focus in this article is on WQWs, not double layer systems.] However, another promising candidate is the one-component Pfaffian state, which is a paired state of composite fermions~\cite{Moore91, Read00}. This state is believed to be responsible for the FQHE at $\nu=5/2$~\cite{Willett87}, and one can ask if the changes in the inter-electron interaction due to finite width may stabilize this state at $\nu=1/2$ as well. The Halperin $(3,3,1)$ state supports Abelian quasiparticles, whereas the Pfaffian is believed to support non-Abelian quasiparticles. The latter has motivated many interesting theoretical
and experimental studies of the $5/2$ FQHE. If the 1/2 state in WQWs turns out to be the Pfaffian state, that would provide another venue where non-Abelian quasiparticles may be investigated.
	
	While the $1/2$ FQHE in WQWs has often been interpreted in terms of the $(3,3,1)$ state, arguments can also be given in favor of a one-component state. We provide here a summary of experimental results and their implications for the nature of the state:
	
	\begin{itemize}
		
		\item In a double layer system, which consists of two layers separated by a distance $d$, the situation is relatively clear~\cite{Park98,Papic10,Faugno20,Scarola01b, Scarola02b,Chakraborty87,Yoshioka89,He91,He93}. For zero layer separation, the two-component system of spin polarized electrons is formally equivalent to a single layer system of spinful electrons with zero Zeeman splitting. Here the state is a layer singlet Fermi sea of composite fermions~\cite{Park98,Balram15c,Balram17}. In variational calculations~\cite{Scarola01b, Scarola02b,Faugno20} this state survives in the range $d/l_B\lesssim1$. The $(3,3,1)$ state is predicted to occur for layer separations $1\lesssim d/l_B \lesssim 3$\cite{Scarola01b,Faugno20}, in general agreement with experiments. For layer separations $d/l_B  \gtrsim 3$ two uncoupled CF Fermi seas (CFFSs) are formed in each layer, with composite fermions now binding four vortices~\cite{Scarola01b,Faugno20}. In contrast, the 1/2 FQHE in WQWs is seen when the width is approximately $2.6 - 8$ $l_B$\cite{Suen92,Suen92b,Suen92c,Suen94b, Shayegan96,Manoharan96,Shabani09a,Shabani09b,Shabani13,Yang14b,Hasdemir15, Mueed16,Liu14d}. Although not conclusive, this points against the two-component $(3,3,1)$ state. 
		
		\item For quantum well widths and densities where the 1/2 FQHE is observed in WQWs, the behavior of FQHE states surrounding it is often consistent with single layer physics. In particular, the standard Jain sequences $n/(2n\pm 1)$~\cite{Jain89} are observed. Recently, Mueed {\it et al.}~\cite{Mueed15} have directly measured, from commensurability oscillations, the Fermi wave vector of composite fermions in the vicinity of filling factor $1/2$ and found that the Fermi sea is a one-component state. The fact that the states in the immediate vicinity of $\nu=1/2$ are one-component states makes it plausible that the $\nu=1/2$ FQHE also has a one-component origin. If not, it would be important to understand what is special about $\nu=1/2$ that makes a two-component state favorable.
		
		\item A phase diagram has been constructed as a function of the filling factor and $\Delta_{\rm SAS}$, the gap between the symmetric and antisymmetric subbands~\cite{Manoharan96}. The island of the 1/2 FQHE state straddles the boundary where many nearby FQHE states make a transition from a one-component state to an insulator, presumably a double layer crystal. However, the 1/2 FQHE island is contiguous, i.e. it is either all one component or two-component.
		
		\item The effect of asymmetry in the charge distribution is complex but worth mentioning here. An early work on 80 nm wide QW by Suen {\it et al.}~\cite{Suen92b, Suen94b} reported a monotonic decrease in the strength of the FQHE at $\nu=1/2$ as the charge distribution is made asymmetric, with the FQHE state disappearing at approximately 10\% imbalance. This may arise from either two-component nature or complicated changes in the effective interaction. Subsequently, Shabani {\it et al.}~\cite{Shabani09a, Shabani09b} found that in a 55 nm quantum well an asymmetry of the charge distribution favors FQHE at $1/2$. This suggests a one-component nature of the FQHE here. Numerical studies also show that in such asymmetric quantum wells around certain widths the one-component Pfaffian wave function has a large overlap with the ground state, although the $(3,3,1)$ state is also competitive~\cite{Thiebaut14, Liu14d, Peterson10}.

	\end{itemize}

	We next briefly review the theoretical studies of the 1/2 FQHE in WQWs and also provide a summary of the main results arising from the present study. In particular, we indicate how the theoretical phase diagram is sensitive to the various assumptions that go into the calculation.
	
	The problem has been addressed by exact diagonalization (ED)~\cite{Papic10,He93,Storni10,Peterson10,Balram20}. ED can often deal with only very small systems and is thus not likely to capture the thermodynamic behavior. This is especially the case for WQWs, for which the width may become comparable to the available lateral dimension of the system. The energy orderings of states are often seen to change as the system size increases. 
	
	\begin{figure}[H]
		\includegraphics[width=\columnwidth]{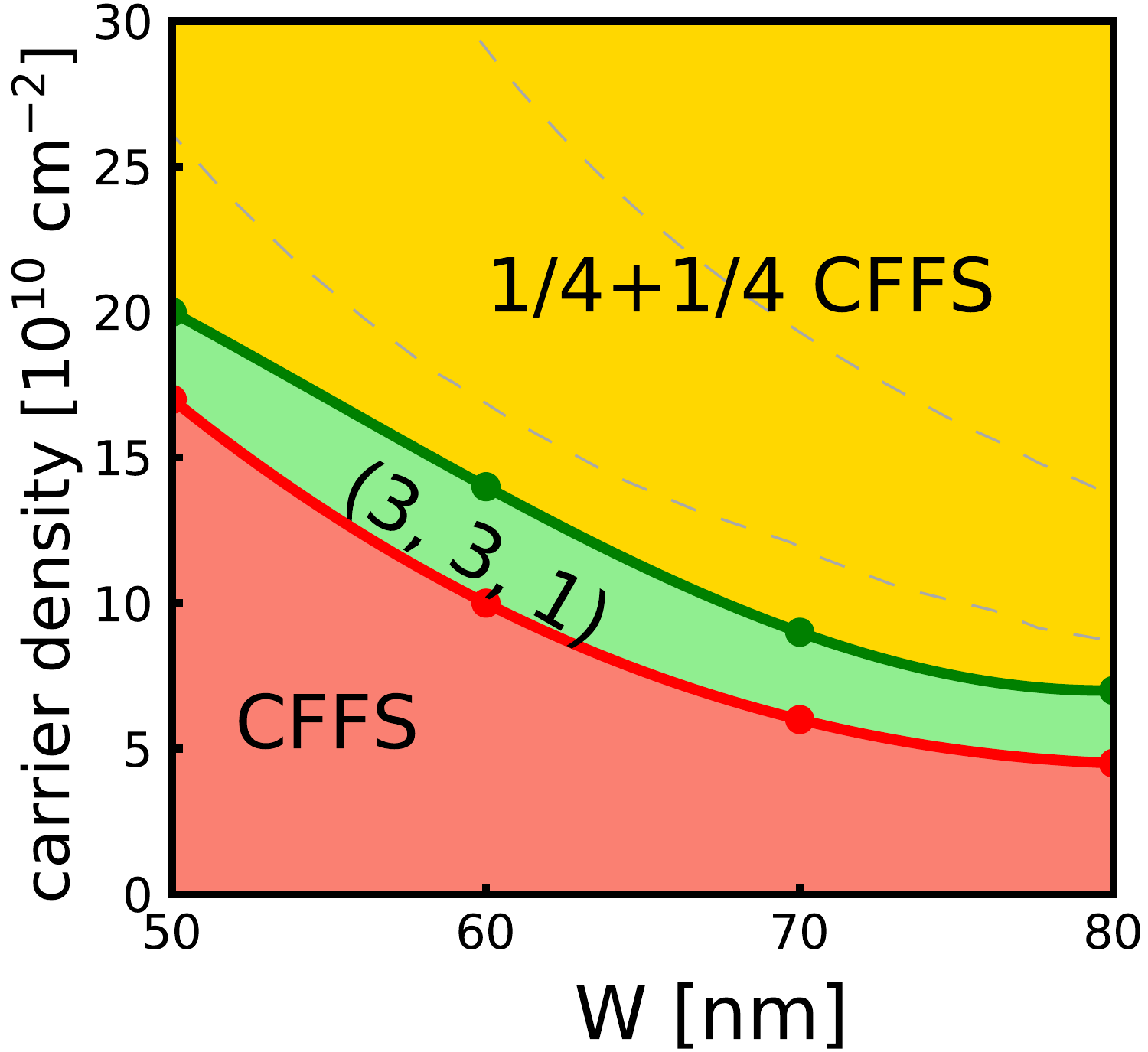}
		\caption{The phase diagram of states at $\nu=1/2$ obtained by the VMC method as a function of the quantum well width $W$ and the carrier density. The transverse wave function is assumed to have the form obtained from LDA at zero magnetic field. Both one-component and two-component states are included. The following states are seen to occur:  the one-component CFFS state (red), the $(3,3,1)$ state (green), and the state with two uncoupled $1/4$ CFFSs, labeled $1/4+1/4$ CFFS (yellow). The region where experiments find an incompressible state~\cite{Shabani09b} is indicated by light dashed lines.  For a given width, the uncertainty of the calculated transition densities is approximately $1\times 10^{10} \text{cm}^{-2}$. The overall phase boundary is obtained by smoothly joining the transition points at $W=50,60,70,80$ nm. The subband gap determined by LDA is used to to determine the total energies of the two-component states.
		}
		\label{VMC_PHASE_DIAGRAM_2}
	\end{figure}

	\begin{figure}[H]
		\includegraphics[width=\columnwidth]{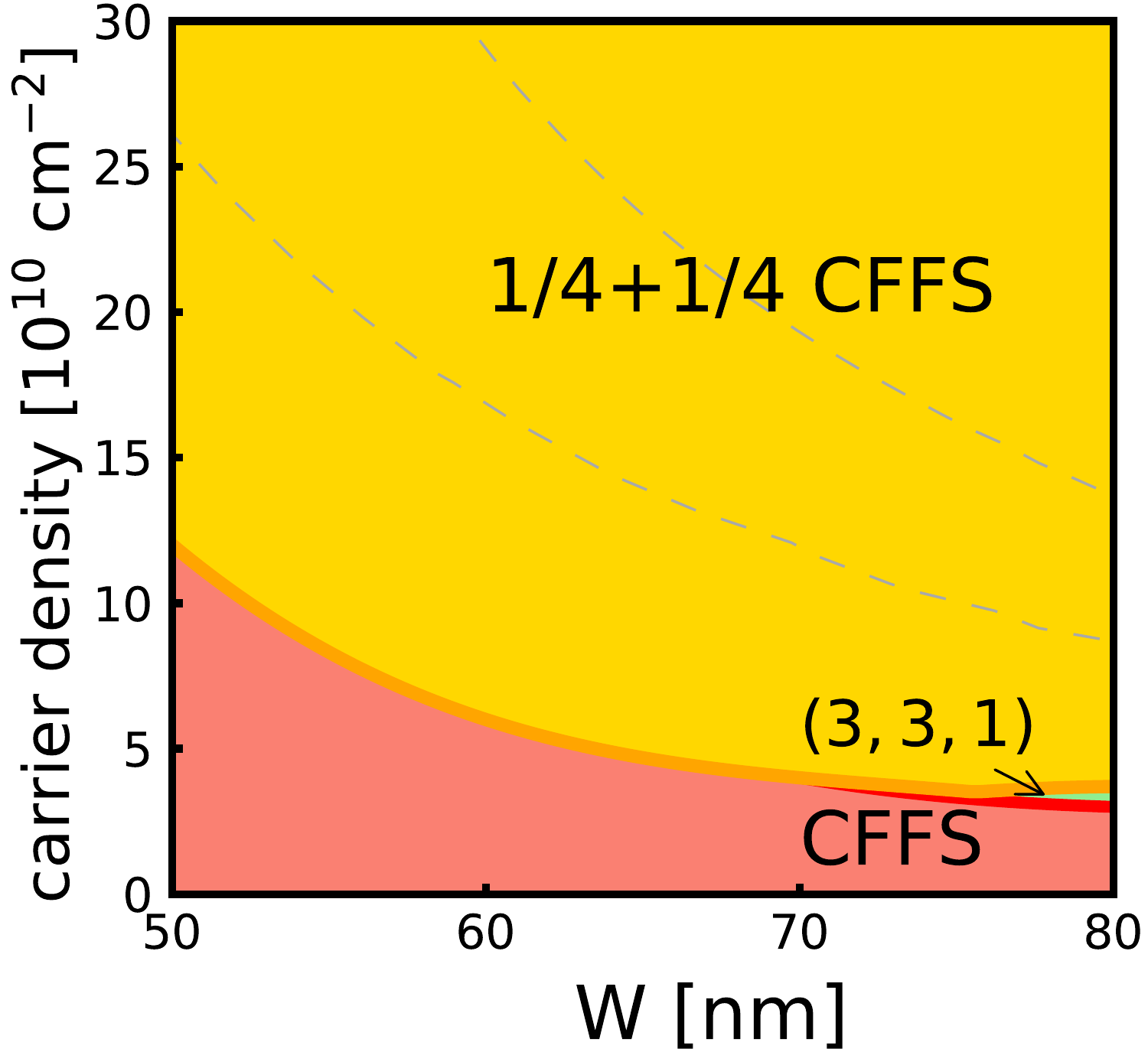}
		\caption{The phase diagram of states as a function of the quantum well width $W$ and the carrier density obtained from a 2D DMC calculation, which incorporates finite width corrections by using an LDA interaction derived at zero magnetic field.  This figure shows how the phase diagram in Fig.~\ref{VMC_PHASE_DIAGRAM_2} changes upon Landau level mixing. The region where experiments find an incompressible state~\cite{Shabani09b} is indicated by light dashed grey lines. For a given width, the uncertainty of the calculated transition densities is about $2\times 10^{10} \text{cm}^{-2}$.}
		\label{2D_DMC_BOUNDARY_ALL}
	\end{figure}
	
	This issue has also been investigated by variational Monte Carlo (VMC)\cite{Biddle13,Papic09, Scarola10,Thiebaut14,Thiebaut15,Faugno20}. During the course of this work, we have determined the phase diagram of $\nu=1/2$ in a WQW using the VMC method, shown in Fig.~\ref{VMC_PHASE_DIAGRAM_2}. The $(3,3,1)$ state is stabilized in a part of the phase diagram that qualitatively agrees with experiments. The phase boundary between one component CFFS and the $(3,3,1)$ state is consistent with earlier calculations~\cite{Thiebaut15}.

	However, the VMC calculations make the following assumptions. (i) The effect of finite width is incorporated through a transverse wave function for electrons, which modifies the interactions between them (see Eq.\,\ref{V_eff}). The transverse wave function is evaluated in local density approximation (LDA) at zero magnetic field~\cite{Park99b}, and it is assumed that it remains unaltered at a strong perpendicular magnetic field. Given that the nature of the transverse wave function depends on the state that the electrons form in two dimensions (for example, at zero magnetic field LDA assumes a Fermi sea state of electrons), one may wonder to what extent this assumption is valid.  (ii) The phase boundary between the one- and two-component states depends sensitively on $\Delta_{\rm SAS}$, i.e. the gap between the symmetric and antisymmetric subbands. One uncritically uses its value obtained at zero magnetic fields. However, this gap is typically very large compared to the Coulomb energy differences between the competing states, and even a few percent change in $\Delta_{\rm SAS}$ can substantially shift the phase boundaries.

The VMC calculation also does not incorporate the effect of Landau level mixing (LLM) directly. We have further investigated the role of LLM within the VMC method through a two-dimensional (2D) fixed-phase diffusion Monte Carlo (DMC) method developed by Ortiz, Ceperley and Martin ~\cite{Ortiz93, Melik-Alaverdian97}, which itself is a generalization of the standard DMC method~\cite{Foulkes01} to find ground states in the presence of broken time-reversal symmetry.  In this method,  we allow for LLM for electrons interacting with the effective interaction derived from LDA at zero magnetic field. We refer to this as ``2D-DMC." We find that, at this level of approximation, the phase diagram is substantially altered and neither the (3,3,1) nor the Pfaffian state is stabilized for a significant range of parameters (see Fig.~\ref{2D_DMC_BOUNDARY_ALL}). However, a conceptual difficulty with this method is an uncontrolled double-counting, because mixing with higher bands has already been incorporated through the modification of the transverse wave function, which, in a sense, is akin to LLM at a finite magnetic field. (At finite magnetic fields, it is LLM that leads to a modification of the form of the transverse wave function.) This study nonetheless shows the importance of LLM, indicating that the results from neither VMC nor 2D-DMC are fully reliable.

		\begin{figure}[H]
		\includegraphics[width=\columnwidth]{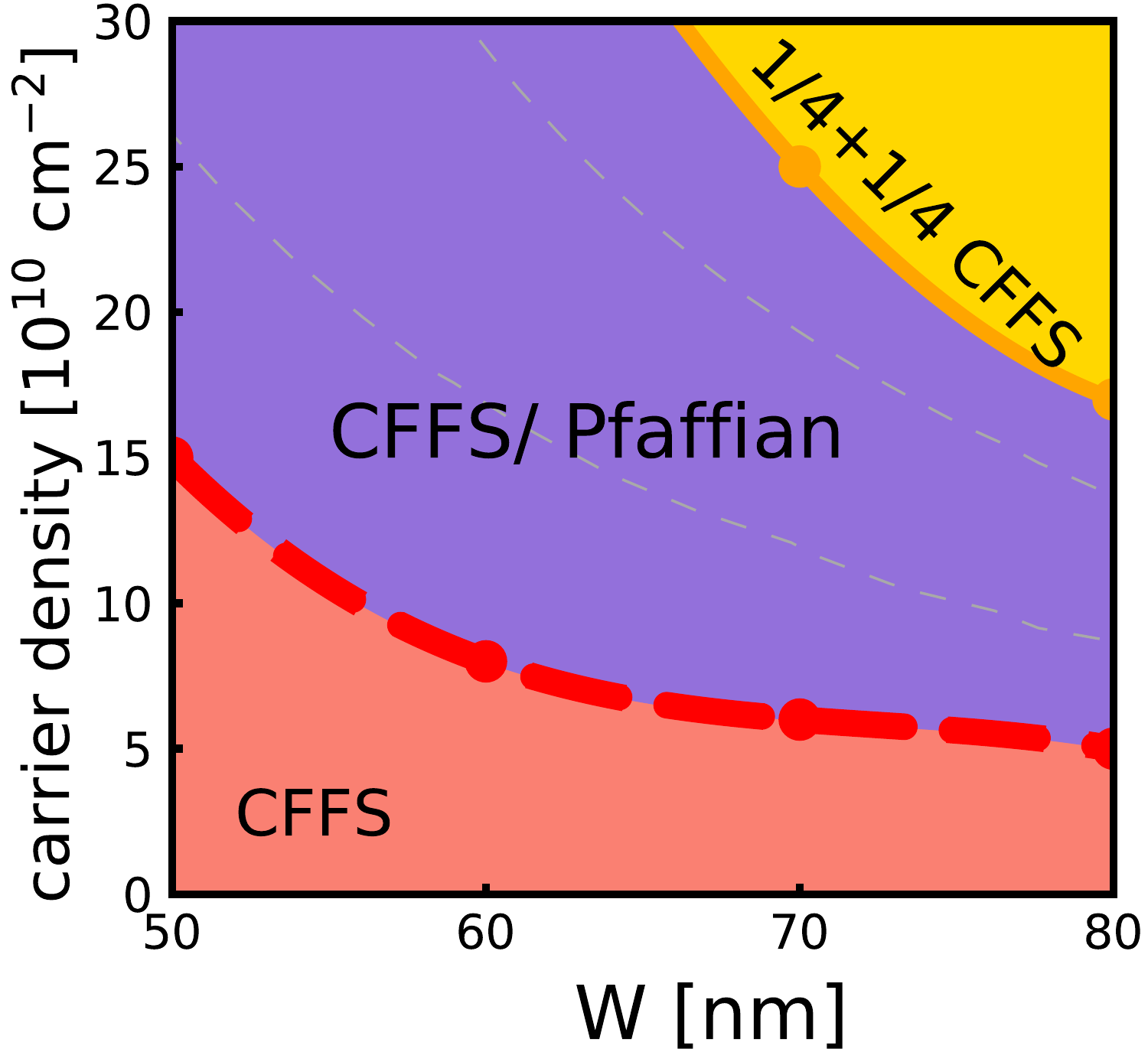}
		\caption{The phase diagram of states determined by 3D-DMC as a function of the quantum well width $W$ and the carrier density. Here both finite width and Landau level mixing are included in a DMC calculation directly in the presence of a magnetic field. The red region is the single-component CFFS state and the yellow region marks the $1/4+1/4$ CFFS state. In the purple region, the energies of the single-component CFFS and the single-component Pfaffian states are equal within numerical uncertainty. The uncertainty of the transition density from one-component state to the $1/4+1/4$ CFFS at each width is approximately $5\times 10^{10}\text{cm}^{-2}$. The region where experiments find an incompressible state~\cite{Shabani09b} is indicated by light dashed grey lines. }
		\label{3D_PHASE_2}
	\end{figure}

	The primary motivation of our work is to develop a technique that circumvents some of the above issues and treats finite width and LLM effects directly at a large magnetic field. Specifically, we use a three-dimensional (3D) version of the fixed phase DMC method, referred to below as ``3D-DMC," or simply as ``DMC." The most important advantage of the 3D-DMC method is that it directly gives the ground state energy (as well as the form of the transverse wave function) at a high magnetic field,  automatically including the effects of finite width and LLM. No reference is made to zero magnetic field in our calculation.  Of course, this method also makes an approximation, namely the choice of fixed phase, and all of our conclusions are subject to the validity of our choice of the phase. (We use the accurate lowest Landau level wave functions to fix the phase, which has been found to give good agreement with experiments in the past~\cite{Zhang16, Zhao18, Ma20}.) There are other practical difficulties with our method. One is that the required computation time does not allow treatment of very large systems; we have studied systems with up to about 25 particles. The second is that it does not allow treatment of two-component states with non-zero $\Delta_{\rm SAS}$. For two-component states, we assume that $\Delta_{\rm SAS}=0$, i.e. the wave function strictly vanishes at the center. This should be a decent approximation for sufficiently large widths and densities where $\Delta_{\rm SAS}$ is small.

		The phase diagram obtained from 3D-DMC calculations is shown in Fig.~\ref{3D_PHASE_2}. The light purple region shows the part of the phase diagram where the energies of the one-component CFFS and the one-component Pfaffian states are so close that we cannot distinguish between them within numerical uncertainty [although both of these energies are lower than the energy of the two-component $(3,3,1)$ state]. Given that experiments show an incompressible state here, we believe that the one-component Pfaffian state is the most likely possibility. Nonetheless, in light of the approximations made in the calculation, a definitive confirmation can come only from experiments, and we hope that our study will motivate further experimental studies of this state.
	
	We have also studied several other candidate wave functions at $\nu=1/2$ but found them not to be relevant for the issue at hand.
	
	Additionally, our 3D-DMC study yields the form of the transverse wave function directly in the presence of a high perpendicular magnetic field. Here, the double layer nature of the ground state for large widths or densities arises due to LLM. We find that, surprisingly, the form of the transverse wave function of the lowest symmetric band is not particularly sensitive to the nature of the 2D state; we find very similar forms for $\nu=1$, $1/3$, and $1/5$, as discussed later. Furthermore, also surprisingly, we find that the transverse wave function obtained from our 3D-DMC is also close to that obtained from LDA at zero magnetic field. Nonetheless, our phase diagram with 3D-DMC method is very different from that obtained from VMC.

	A recent work~\cite{Zhu16} has concluded that switching on tunneling in a bilayer favors the Pfaffian state. The model for quantum well considered in Ref.~\cite{Zhu16} is different from ours.
	
	The plan of our paper is as follows. In Sec.\,\ref{sec_cf_states}, we briefly review the fundamentals of the FQHE on a torus and give explicit forms of wave functions that are involved in our calculation. In Sec.\,\ref{sec_vmc} we report our VMC studies on the topic. We next introduce the general principles of the DMC method in Sec.\,\ref{sec_dmc_basics}. After that, we present our 2D-DMC and 3D-DMC investigations individually. We discuss our results in the end and more technical details can be found in the appendices.
	
	All calculations are performed in the torus geometry, except those presented in Appendix~\ref{VMC_SPHERE_SEC}. Throughout this work, we assume parameters appropriate for GaAs, with dielectric constant $\epsilon=12.6$ and band mass $m=0.067 m_e$, where $m_e$ is the electron mass in vacuum. The magnetic length is denoted $l_B=\sqrt{\hbar c/eB}$ where $B$ is the magnetic field.

	\section{Relevant states at half filling}\label{sec_cf_states}
	
	We shall include in our study several different states at filling factor $\nu=1/2$, which we now list. We primarily use the torus geometry for our study, because the CFFS can be constructed on a torus with explicit wave vector configuration. (On the sphere one must approach the CFFS by taking the limit $n\to \infty$ for Jain states at $\nu=\frac{n}{2n+1}$~\cite{Rezayi94,Balram15b,Balram17}, which requires going to very large systems that are not accessible to DMC.) We also give the VMC results in the spherical geometry in Appendix~\ref{VMC_SPHERE_SEC} for comparison. 
We start this section by reviewing some basics of FQHE on a torus\cite{Pu17, Bernevig12, Haldane85b, Haldane85, Greiter16}.
	
	\subsection{Basics of FQHE on a torus}
	
	We start by formulating the single-particle orbitals and we use them to construct the many-body wave functions. We map a torus to a parallelogram with quasi-periodic boundary conditions in the complex plane. The two edges of the parallelogram are given by $L$ and $L\tau$ in the complex plane, where $\tau$ is a complex number representing the modular parameter of the torus. We will take $L$ to be real. (We also use the symbol $\uptau$, with a different font, for the imaginary time in the introduction of the DMC algorithm; this should not cause any confusion, given that the two appear in very different contexts.) The location $\vec{z}=(x,y)$ of a particle in the complex plane is represented by the complex number $z=x+i y$. Later when we include the transverse dimension, the displacement vector in 3D space is labeled by $\vec{r}=(x, y, w)$. To make the quasi-periodic boundary conditions in $L$ and $L\tau$ directions compatible, the number of flux quanta through the torus, $N_\phi=BL^2{\rm Im}[\tau]/\phi_0$, must be an integer, where $\phi_0=hc/e$ is a single flux quantum. We will work with the symmetric gauge $\vec{A}=(B/2)(y, -x, 0)$, which corresponds to a uniform magnetic field $\vec{B}=-B\hat{z}$ perpendicular to the surface of the torus.  For simplicity, we choose a square torus with $\tau=i$. The magnetic translation operator is given by 
	\begin{align}
	t\left(\vec{\xi}\right)=e^{-\frac{i}{2l_B^2} \hat{z}\cdot (\vec{\xi}\times \vec z)}T\left(\vec{\xi} \right)
	\label{}
	\end{align}
	where $T\left(\vec{\xi}\right)$ is the usual translation operator.
	The single-particle orbitals are imposed with the quasi-periodic boundary conditions: 	\begin{equation}
	\begin{aligned}
	&t\left(L\right) \psi\left(z\right)=e^{i\phi_1}\psi\left(z\right)\\
	&t\left(L\tau\right) \psi\left(z\right)=e^{i\phi_\tau}\psi\left(z\right)
	\end{aligned}
	\label{PBC_single}
	\end{equation}
	where the phases $\phi_1$ and $\phi_\tau$ are the periodic boundary phases which define the Hilbert space. We have chosen $\phi_1=\phi_\tau=0$ because for our purpose, the calculation of the energy is independent of the choice of these phases.
	
	In general, the single-particle orbitals in the Lowest Landau level (LLL) in symmetric gauge can be written as:\cite{Greiter16, Pu17} 
	\begin{align}
	\psi^{(n)}\left(z\right)&=e^{\frac{z^2-|z|^2}{4 l_B^2}} f^{(n)}(z)\\
	\end{align}
	where $f\left(z\right)$ satisfies 
	\begin{equation}
	\begin{aligned}
	\frac{T\left(L\right)f\left(z\right)}{f\left(z\right)}=\frac{f\left(z+L\right)}{f\left(z\right)}&=1\\
	\frac{T\left(L\tau\right)f\left(z\right)}{f\left(z\right)}=\frac{f\left(z+L\tau\right)}{f\left(z\right)}
	&=e^{-i \pi N_\phi(2z/L+\tau)}
	\end{aligned}
	\label{PBC_eigenstate}
	\end{equation}
	The solutions to Eq. \ref{PBC_eigenstate} are given by\cite{Greiter16}
	\begin{equation}
	\begin{aligned}
	f^{(n)}\left(z\right) &= e^{i k^{(n)} z} \prod_{s=1}^{N_\phi}\theta\left( z/L-w_s^{(n)}|\tau\right)\\
	k^{(n)} &=\frac{-\pi N_\phi+2\pi n}{L}\\
	w_s^{(n)} &=\frac{1}{2\pi N_\phi}\left[-\pi N_\phi(2-\tau)-2\pi n \tau+\pi+2\pi(s-1)\right]
	\end{aligned}
	\end{equation}
	where $\theta\left( z|\tau\right)$ is the odd Jacobi theta function\cite{Mumford07} (see Appendix~\ref{theta_function_definition} for its definition and properties). Here we have $n=0,1,2,\cdots N_\phi-1$; $w_s^{(n)} L$ give the positions of zeros; and $k^{(n)}$ is a real number labeling the eigenvalues of magnetic translation $t(L/N_\phi)$:
	\begin{equation}
	\label{T1}
	t\left(L/N_\phi\right)\psi^{(k)}(z,\bar{z}) =e^{\i{{2\pi} k\over N_\phi}}\psi^{(k)}(z,\bar{z}).
	\end{equation}

		Starting from single-particle wave functions, one can construct many-body wave functions that preserve the quasi-periodic boundary conditions. In general, the many-body wave function at filling $p/q$, where $p$ and $q$ are co-primes, has a $q$ fold center-of-mass (CM) degeneracy\cite{Haldane85b}. The Laughlin wave function at $\nu=1/m$ is given by\cite{Haldane85, Greiter16, Haldane85b}
	\begin{equation}
	\begin{aligned}
	\label{Laughlin_wf}
	\Psi_{1/m}^{(n)}(\{z_i\})=e^{\sum_i\frac{z_i^2-|z_i|^2 }{4l_B^2}} F_\frac{1}{m}^{(n)}\left( Z\right)\prod_{i<j}\left[\theta\left(\frac{z_i-z_j}{L}|\tau\right)\right]^{m}
	\end{aligned}
	\end{equation}
	where $F_\frac{1}{m}^{(n)}(Z)$ describes the CM part with $Z=\sum_{i=1}^N z_i$:
	\begin{equation}
	\begin{aligned}
	F_\frac{1}{m}^{(n)}(Z)=&e^{i K^{(n)}Z}\prod_{s=1}^{m} \theta\left(Z/L-W_s^{(n)}|\tau\right),\\
	K^{(n)}
	=&(-\pi N_\phi+2 \pi n)/L\\
	W_s^{(n)}
	=&\frac{ N_\phi \tau-N_\phi-2n \tau-(m-1)+2(s-1)}{2m} 
	\end{aligned}
	\end{equation}
	where $n=0, 1, 2,\dots,m-1$ labels the $m$-fold CM degeneracy\cite{Haldane85,Greiter16}. In the special case $m=1$, Eq.~\ref{Laughlin_wf} gives the wave function $\Psi_1$ for filled LLL. For the filled LLL wave function, we drop the superscript $n$ for $F_1(Z)$, since $n$ can take only one value $n=0$. 

The Jain state at $\nu=\frac{s}{2ps+1}$ is constructed as
	\begin{equation}
	\begin{aligned}
	\Psi_{s\over 2ps+1}=\mathcal{P}_\text{LLL}\Psi_s\Psi_1^2
	\end{aligned} \label{Jainwf}
	\end{equation}
	where $\Psi_s$ stands for the wave function of electrons filling the lowest $s$ LLs, $\Psi_1^2$ attaches $2p$ vortices to each electron to composite-fermionize it, and $\mathcal{P}_\text{LLL}$ projects the wave function into the LLL. This form is valid for both the spherical and the torus geometries. On torus, the wave function in Eq.~\ref{Jainwf} does not have a well defined CM momentum, but $2ps+1$ degenerate CM eigenstates can be constructed as discussed by Pu {\it et al.}\cite{Pu17} Ref.~\onlinecite{Pu17} also shows how LLL projection can be conveniently accomplished for the Jain states in the torus geometry.
	
	\subsection{One-component CFFS state}
	
An important state involved for our purposes is the one-component CFFS. As mentioned above, this state thrives in narrow quantum wells. 
	The construction of the CFFS wave function at $\nu=1/2p$ in torus geometry is accomplished by attaching 2p flux quanta to an electron fermi sea state and projecting it into the LLL \cite{Rezayi94,Shao15,Geraedts18,Wang19,Pu18}: 
		\begin{equation}
\Psi_\text{CFFS, 1/2p}\left(\left\{ z_i\right\}\right)=\mathcal{P}_\text{LLL} \Psi_{FS} \Psi_1^{2p}
\end{equation}
where $\Psi_\text{FS}={\rm det}[e^{i\vec{k}_n\cdot\vec{r}_i}]$ stands for fermi sea wave function. It can be projected into the LLL to produce 
	\begin{equation}
	\begin{aligned}
	  & \Psi_\text{CFFS, 1/2p}\left(\left\{ z_i\right\}\right) 
	=e^{\frac{\sum_i z_i^2-|z_i|^2}{4l_B^2}}   F_1\left( Z+i\ell_B^2K\right)^{2p} \\
	&\times \det{\left[G_{k_n}\left(z_m\right)\right]} \left[\prod_{i<j}\theta\left( \frac{z_i-z_j}{L}|i\right)\right]^{2p-2}
	\end{aligned}
	\label{CFFS WF}
	\end{equation}
	where
	\begin{equation}
	\begin{aligned}
	G_{k_n}\left(z_m\right)&=e^{-\frac{k_n l_B^2}{4}(k_n+2\bar{k}_n)}e^{\frac{i}{2}(\bar{k}_n+k_n)z_m}\cdot\\ &\cdot\prod_{j, j\neq m}\theta\left(\frac{z_m+2pik_n l_B^2-z_j}{L}|i\right).
	\end{aligned}
	\end{equation}
Here $k_n$ stand for the magnetic momenta occupied by the CFFS, with the CM momentum given by $K=\sum_n k_n$. The empirical rule is that the configuration of $k_n$'s that produces the ground state is  as compact as possible, i.e. minimizes $\sum_n\left(k_n-K/N\right)^2$. More details can be found in References \onlinecite{Rezayi94,Pu18,Fremling18,Pu20b}.

	\subsection{Pfaffian state}
	
	Three distinct Pfaffian wave functions on the torus are given by \cite{Greiter91,Greiter92a} 
	\begin{equation}
	\label{Pfaffian_wfn}
	\begin{aligned}
	&\Psi_\text{Pf, 1/2}\left( \left\{ z_i\right\}\right)\\
	=&Pf\left( M_{ij} \right)F_1^2\left(Z\right)  \prod_{i<j}\theta^2\left(\frac{z_i-z_j}{L}|i\right) e^{\frac{\sum_i z_i^2-|z_i|^2}{4l_B^2}}.
	\end{aligned}
	\end{equation}
	Here $Pf\left(M_{ij}\right)$ is the Pfaffian of the matrix $M_{ij}=\frac{\theta_a\left(\frac{z_i-z_j}{L}|i\right)}{\theta_1\left(\frac{z_i-z_j}{L}|i\right)}$, and the choices $a=2, 3, 4$ produce three distinct Pfaffian wave functions.
	The definition of $\theta_a\left(z|\tau\right)$ can be found in Appendix\,\ref{theta_function_definition}. These three states are degenerate for a three-body Hamiltonian for which the Pfaffian state is exact and are believed to become degenerate for Coulomb interaction in the thermodynamic limit \cite{Peterson08}. Our calculations also show that the energy difference between them is negligible because: (1) for VMC calculation the difference is much smaller than the difference between the Pfaffian state and the CFFS; and (2) for DMC calculation the energy differences are smaller than the statistical uncertainty. (See Appendix\,\ref{PF_DEGENERACY}) Due to these reasons and the limit of our computational resources, we choose $a=2$ below.
	
	\subsection{Uncoupled $1/4+1/4$ two-component CFFS state}
	
	In the limit of very wide quantum wells, we expect the system to form two uncoupled $1/4$ CFFSs, which is referred to as $1/4+1/4$ CFFS. The wave function of this two-component state is the product of the two $1/4$ CFFSs defined in Eq.\,\ref{CFFS WF}:
	\begin{equation}
	\Psi_\text{CFFS, 1/4+1/4}=
\Psi_\text{CFFS, 1/4}\left(\left\{ z_i\right\}\right) \Psi_\text{CFFS, 1/4}\left(\left\{ z_{[j]}\right\}\right)
	\end{equation}
	where $i=1, 2, \dots, N_e/2$ denote the electrons belonging to the first layer and $[j]\equiv N_e/2+j=N_e/2+1, N_e/2+2,\dots,N_e$ denote the electrons belonging to the second layer. 
	
	\subsection{The pseudo-spin singlet CFFS states}
	
	We also consider the pseudo-spin singlet CFFS states, which is compressible and it is constructed by attaching flux quanta to the pseudo-spin-singlet fermi sea wave function. Here the term "pseudo-spin" refers to the layer index. The pseudo-spin singlet CFFS state has interlayer correlations, in contrast to the $1/4+1/4$ CFFS state. One can write its wave function by simply replacing in Eq.\,\ref{CFFS WF} the determinant in the wave function of the pseudo-spin polarized 1/2 CFFS by the product of determinants of the two pseudo-spins \cite{Hossain20a}:
	\begin{equation}
	\begin{aligned}
	\det \left[G_{k_n}\left( z_m \right)\right] \to \det \left[G_{k_n}\left( z_i \right)\right] \det \left[ G_{k_l}\left( z_{[j]} \right)\right]
	\end{aligned}
	\end{equation}
	where $i=1, 2, \dots, N_e/2$ and $[j]=N_e/2+1, N_e/2+2,\dots,N_e$ denote the electrons belonging to two pseudo-spin components. The Jastrow factor remains the same as in Eq.\,\ref{CFFS WF} which includes both intra-layer and inter-layer correlations. To make sure that the state is a singlet, one also needs to make the momentum distribution identical for both pseudo-spins.
	
	\subsection{The Halperin $(3,3,1)$ state}
	
	The Halperin $(3,3,1)$ state reads
	\begin{equation}
	\begin{aligned}
	&\Psi_\text{$(3,3,1)$}\left( \left\{ z_i\right\} \right)=\\
	&e^{\frac{\sum_i z_i^2-|z_i|^2}{4l_B^2}} F_{(3,3,1)}\left(Z\right)\prod_{1\leq i<j\leq N_e/2}\theta^3\left( \frac{z_i-z_j}{L}|i\right)\cdot\\
	\cdot&\prod_{N_e/2<[i]<[j]\leq N_e}\theta^3\left( \frac{z_{[i]}-z_{[j]}}{L}|i\right) \prod_{\substack{1\leq i\leq N_e/2,\\N_e/2<[j]\leq N_e}}\theta\left( \frac{z_i-z_{[j]}}{L}|i\right).
	\end{aligned}
	\end{equation}
	Here 
	\begin{equation}
	\begin{aligned}
	F_{(3,3,1)}(Z)&=F^{(0)}_\frac{1}{2}\left(Z_L\right)F^{(0)}_\frac{1}{2}\left(Z_R\right)F_1\left(Z\right)
	\end{aligned}
	\end{equation}
	where $Z_L=\sum_{i=1}^{N_e/2}z_i$, $Z_R=\sum_{[j]=N_e/2+1}^{N_e}z_{[j]}$, and $Z= Z_L+Z_R$ (here $L$ and $R$ denote the left and right layers).
	
	\section{VMC calculation of the phase diagram}\label{sec_vmc}
	\begin{figure}[H]
		\includegraphics[width=\columnwidth]{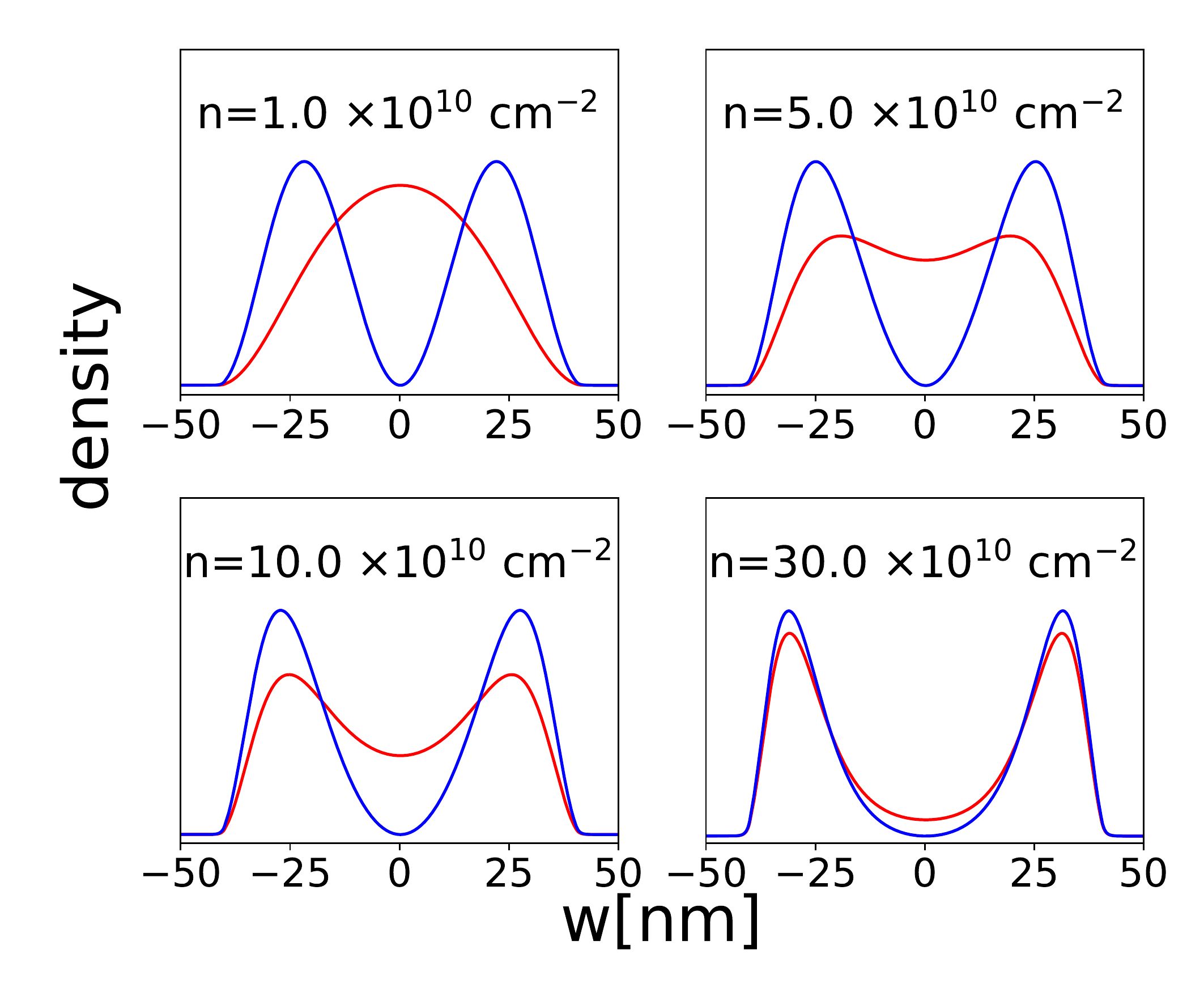}
		\caption{The transverse density profile for the lowest (red) and the first excited (blue) subbands in the quantum well of width $W=80$ nm calculated by LDA.}
		\label{LDA_band}
	\end{figure}
	
\begin{figure}[H]
		\includegraphics[width=\columnwidth]{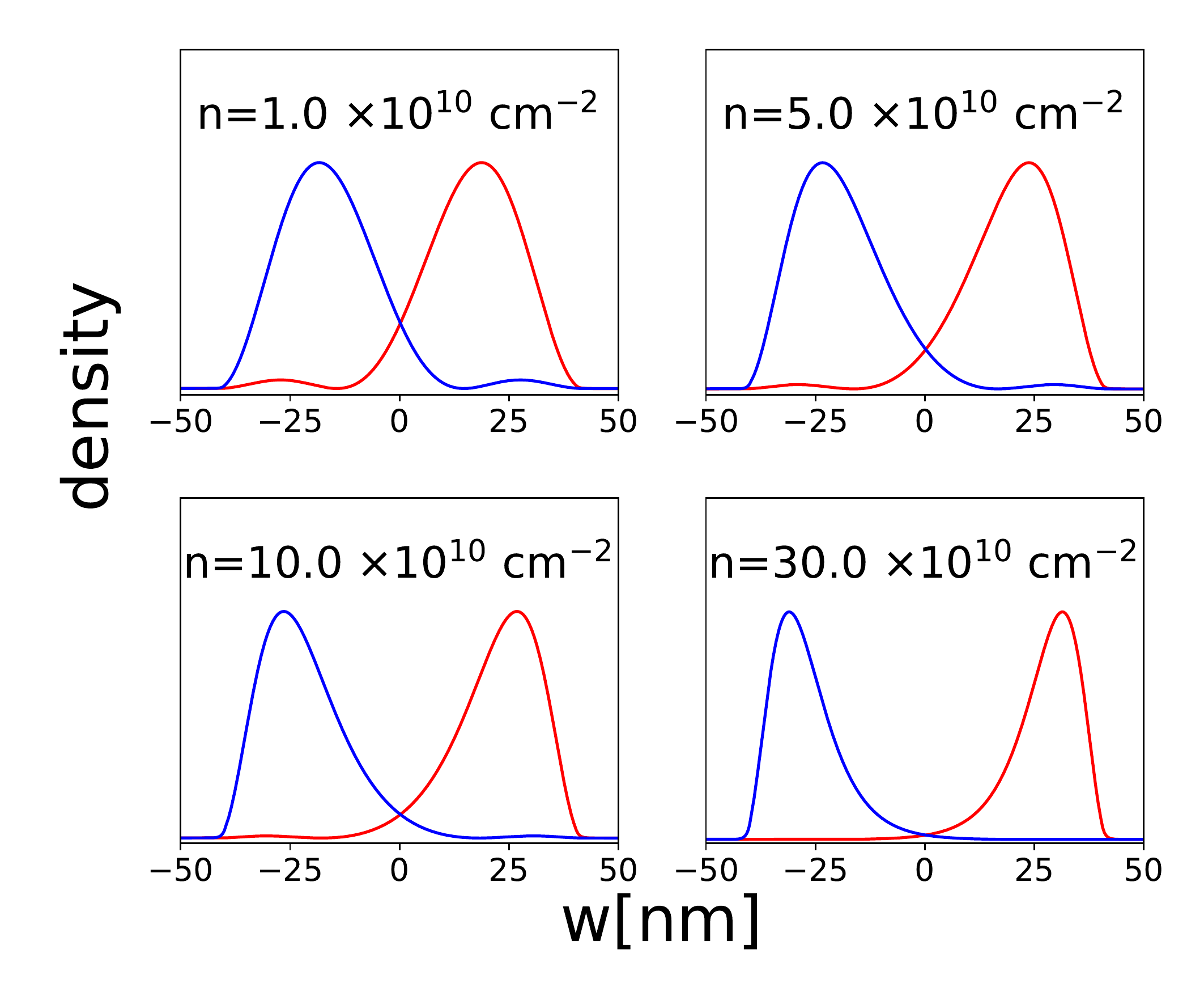}
		\caption{The density profiles of the left (blue) and right (red) bases in the quantum well of $W=80$ nm calculated by LDA.}
		\label{LDA_LR_COMPONENT}
	\end{figure}

	We shall model the confinement potential as a quantum well with the infinite depth and a width of $W$. In some circumstances the finite depth is also considered, but in general this does not cause any significant difference because the GaAs quantum wells we discuss in this article (and also those in experiments) are  generally very deep. The problem is modeled via a VMC calculation which includes an effective two-dimensional interaction, defined as follows:
	\begin{equation}
	V_{\text{eff}}\left( \vec{r}\right)=\frac{e^2}{\epsilon}\int dw_1\int dw_2 \frac{|\psi(w_1)|^2|\psi(w_2)|^2}{\sqrt{|\vec{r}|^2+(w_1-w_2)^2}}.
	\label{V_eff}
	\end{equation}
	Here $\psi(w)$ is the transverse wave function, $w$ is the transverse coordinate, and $\vec{r}$ is a two-dimensional vector. In the simplest approximation, the subband wave functions are taken as the single-particle solutions of a quantum well problem $\psi_S(w)=\sqrt{\frac{2}{W}}\cos\left(\frac{\pi w}{W}\right)$ and $\psi_A(w)=\sqrt{\frac{2}{W}}\sin\left(\frac{2\pi w}{W}\right)$, where $S$ and $A$ refer to symmetric and antisymmetric. In this approximation the subband gap is $\Delta_{\rm SAS}=\frac{3\pi^2}{2}\frac{\hbar^2}{mW^2}$, where $m$ is the band mass of the electron.  A better approximation for $\psi(w)$ is obtained by LDA at zero magnetic field, where one assumes a Fermi liquid state in the 2D plane~\cite{Park99b,Faugno19}. In this and the next section that introduces the 2D fixed-phase DMC, we use the LDA form for $\psi(w)$. We denote the lowest two subbands as $\psi_S$ and $\psi_A$, in which $S$ represents the symmetric subband and $A$ represents the anti-symmetric subband. The typical LDA density profiles of the lowest two subbands are shown in Fig.~\ref{LDA_band}. Before going further, let us discuss how the occupation of the subbands changes as one tunes the subband gap.  When $\Delta_{\rm SAS}$ is much larger than  the Fermi energy, as is the case for either very small $W$ or small densities, only the lowest subband is occupied.  In the limit when the lowest two bands are approximately degenerate ($\Delta_{\rm SAS}\approx 0$), which happens at large $W$ or at large densities, two-component states are possible, where the two components are linear combinations of the two subbands. Because the system tends to form two layers at large widths, we choose the left-right bases as (Fig.~\ref{LDA_LR_COMPONENT}):
	\begin{equation}
	\begin{aligned}
	\psi_L=\frac{1}{\sqrt{2}}(\psi_S +\psi_A)\\
	\psi_R=\frac{1}{\sqrt{2}}(\psi_S -\psi_A)\\
	\end{aligned}
	\label{vmc_bases}
	\end{equation}
	More generally, we can choose $\psi_\theta=\frac{1}{\sqrt{2}}(\psi_S +e^{i \theta}\psi_A)$ and $\psi'_\theta=\frac{1}{\sqrt{2}}(\psi_S -e^{-i \theta}\psi_A)$. However, because the systems becomes a bilayer  for sufficiently wide quantum wells or large densities, we expect that $\theta=0$ will produce the lowest energy.

Similarly to Eq.\,\ref{V_eff}, we define the effective interactions as follows. For one-component states, only the lowest symmetric subband is used for defining the effective interactions whereas for two-component states, both intra-component interaction and inter-component interaction are needed:
	\begin{equation}
	\begin{aligned}
	V_{\text{SS}}\left( \vec r\right)=\frac{e^2}{\epsilon}\int dw_1\int dw_2 \frac{\rho_\text{S}(w_1)\rho_\text{S}(w_2)}{\sqrt{|\vec{r}|^2+(w_1-w_2)^2}}\\
	V_{\text{LL}}\left( \vec r\right)=\frac{e^2}{\epsilon}\int dw_1\int dw_2 \frac{\rho_\text{L}(w_1)\rho_\text{L}(w_2)}{\sqrt{|\vec{r}|^2+(w_1-w_2)^2}}\\
	V_{\text{RR}}\left( \vec r\right)=\frac{e^2}{\epsilon}\int dw_1\int dw_2 \frac{\rho_\text{R}(w_1)\rho_\text{R}(w_2)}{\sqrt{|\vec{r}|^2+(w_1-w_2)^2}}\\
	V_{\text{LR}}\left( \vec r\right)=\frac{e^2}{\epsilon}\int dw_1\int dw_2 \frac{\rho_\text{L}(w_1)\rho_\text{R}(w_2)}{\sqrt{|\vec{r}|^2+(w_1-w_2)^2}}\\
	\end{aligned}
	\label{V_eff_explicit}
	\end{equation} 
The densities are defined as $\rho_\text{S}=\left|\psi_S\right|^2$ and $\rho_\text{L,R}=\left|\psi_\text{L,R}\right|^2$.
	
\begin{figure}[H]
		\includegraphics[width=\columnwidth]{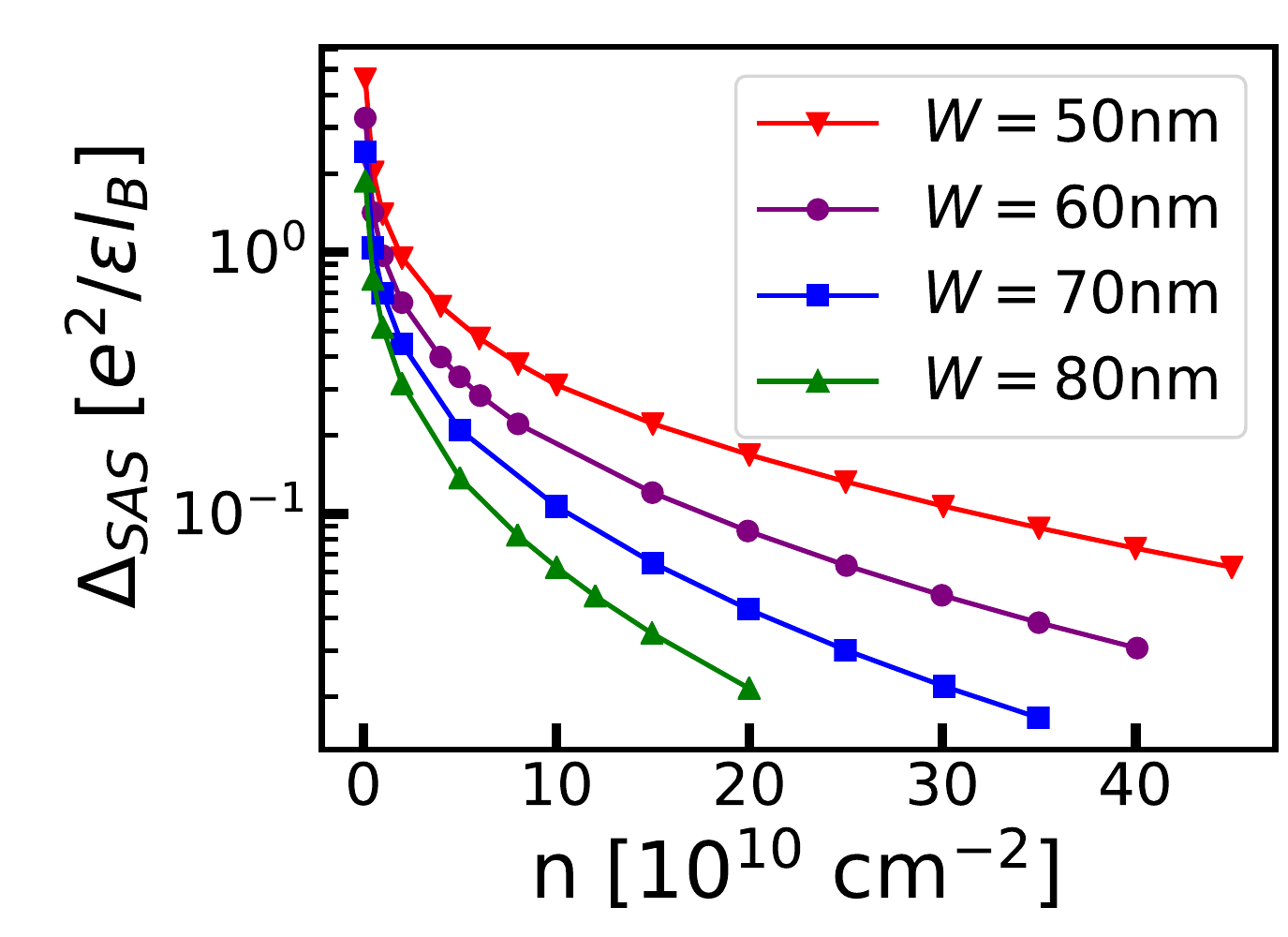}
		\caption{Subband gap $\Delta_{\rm SAS}$ calculated in LDA for various quantum well widths as a function of the density.}
		\label{Delta_SAS}
	\end{figure}

We mention here two caveats. First of all, we consider states for which either only the lowest subband is occupied, or the two lowest subbands are equally occupied. All of our trial wave functions, namely the single-component CFFS, the single component-Pfaffian, the pseudo-spin singlet CFFS, the uncoupled $1/4+1/4$ CFFS, and the $(3,3,1)$ satisfy this requirement. In principle, we can also consider a partially polarized CFFS, which will have an unequal occupation of two subbands, but we have not done so (because it significantly enhances the calculational difficulty). All other states considered here cannot be partially polarized. Second, the value of $\Delta_{\rm SAS}$ is relevant for transitions from a single-component to a two-component state.  $\Delta_{\rm SAS}$ is typically very large compared to the Coulomb energy differences between the relevant states. We determine the value of  $\Delta_{\rm SAS}$ from the LDA calculation (Fig.~\ref{Delta_SAS}).

\begin{figure}[H]
		\includegraphics[width=\columnwidth]{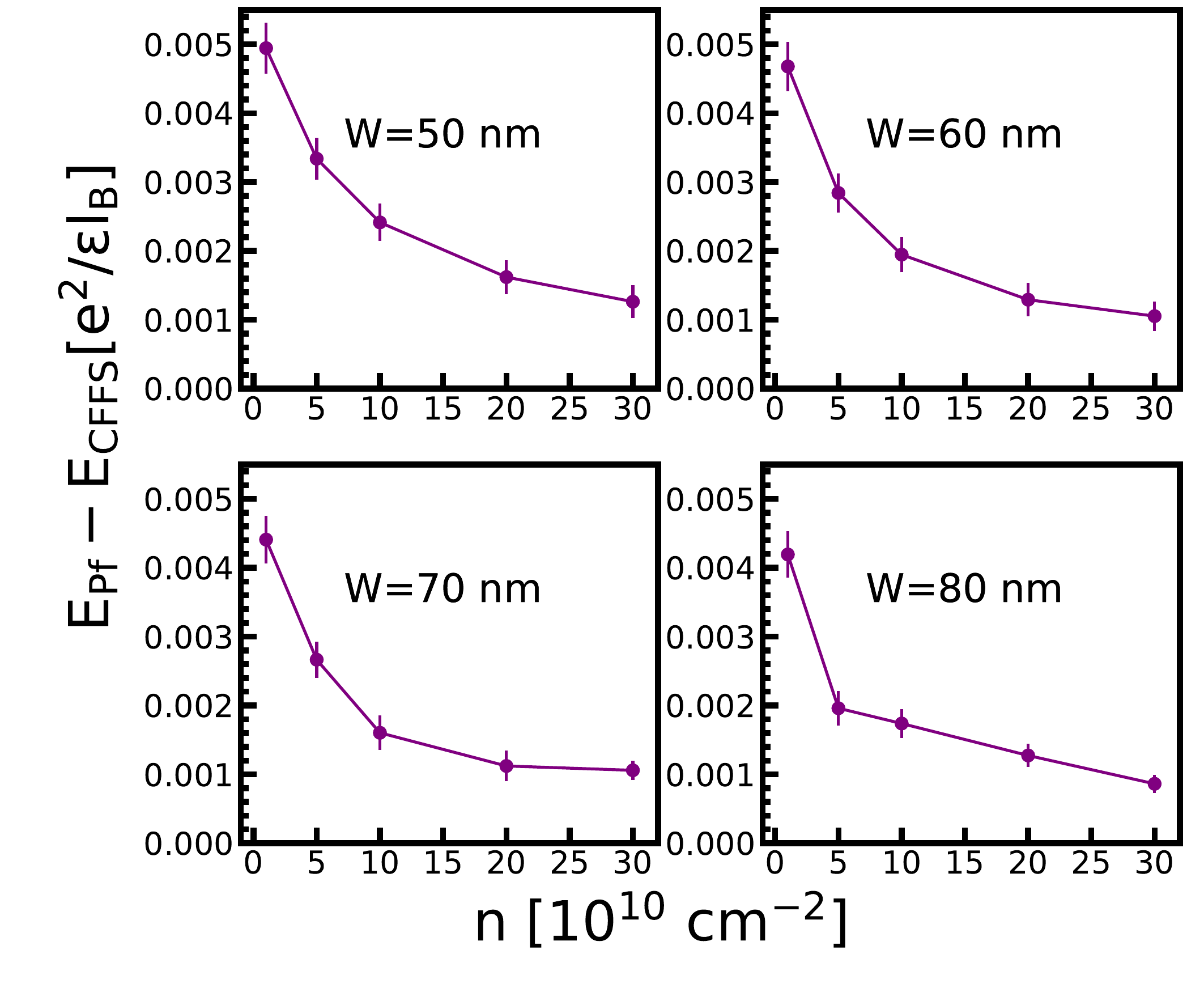}
		\caption{The VMC calculation of the energy difference per particle between the Pfaffian and the one-component CFFS state in the thermodynamic limit. The well widths are shown on the plots.}
		\label{VMC_E_1}
	\end{figure}

	The energies of one-component states relative to the CFFS are shown in Fig.~\ref{VMC_E_1}. 
	As one can see, the CFFS remains the lowest energy state for all parameters, although the Pfaffian comes as close as $0.001 \frac{e^2}{\epsilon l_B}$ at densities greater than $2\times 10^{11} \text{cm}^{-2}$.  This conclusion is also supported by exact diagonalization studies of finite systems in the spherical geometry. In Appendix~\ref{ED_Ajit}, we show the overlap between the exact ground state of the LDA interaction with the one-component CFFS and the Pfaffian states, and find that in the entire region of parameter space that we considered, the one-component CFFS always has a very high overlap with the exact ground state and thus is superior to the Pfaffian.
[Note: We have also performed the energy comparison in the spherical geometry, where we see a different result, namely that the Pfaffian state has lower energy in the thermodynamic limit for some parts of the phase diagram.  We believe that the torus results are more reliable because the thermodynamic extrapolation on the sphere is less accurate for finite widths. See Appendix\,\ref{VMC_SPHERE_SEC} for further discussion.]

\begin{figure}[H]
		\includegraphics[width=\columnwidth]{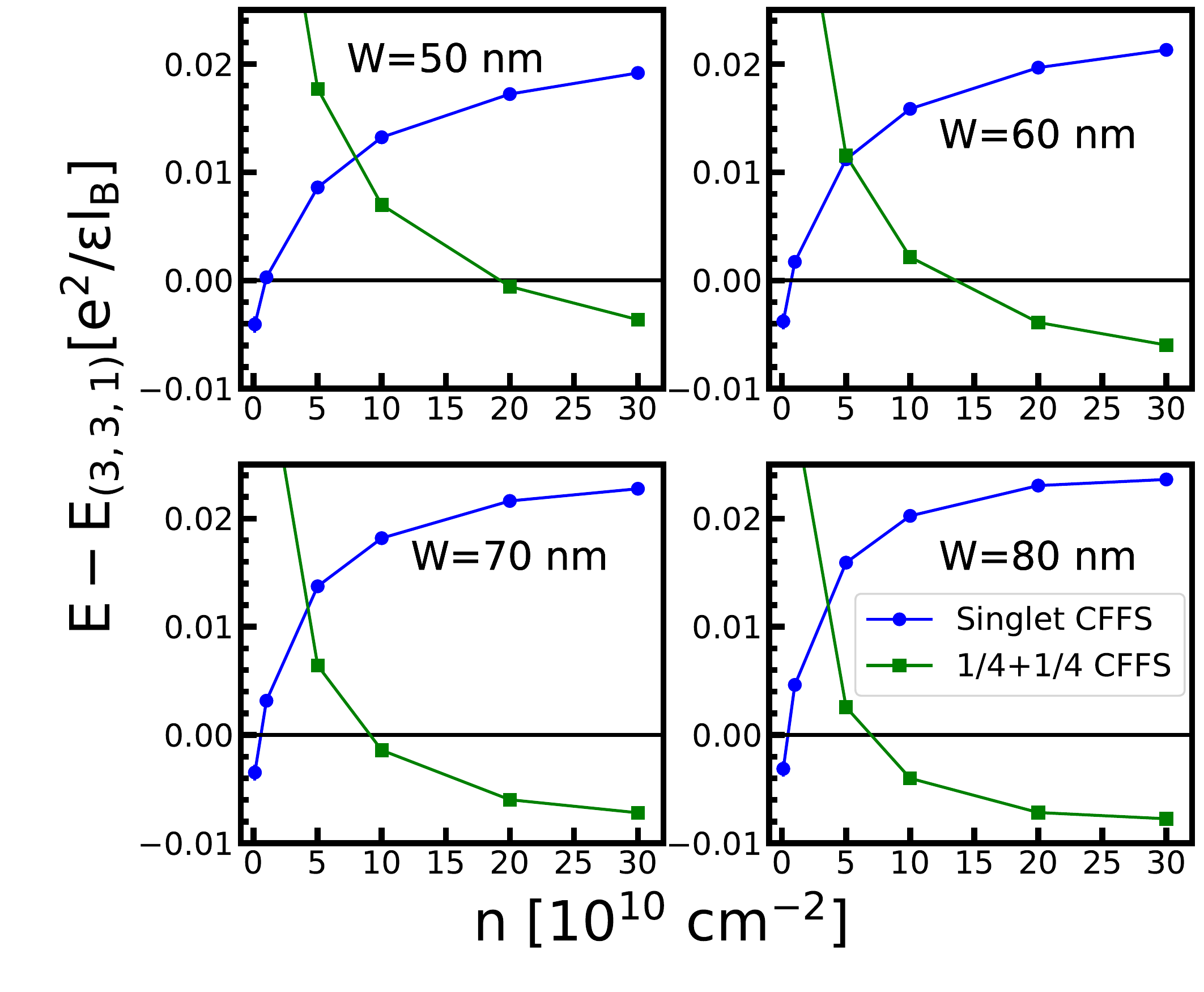}
		\caption{The VMC calculation of the energy per particle of the $1/4+1/4$ CFFS state and the singlet CFFS state relative to the $(3,3,1)$ state in the thermodynamic limit. The well widths are shown on the plots. The statistical errors are smaller than the symbol sizes.}
		\label{VMC_E_2}
	\end{figure}

\begin{figure}[H]
		\includegraphics[width=\columnwidth]{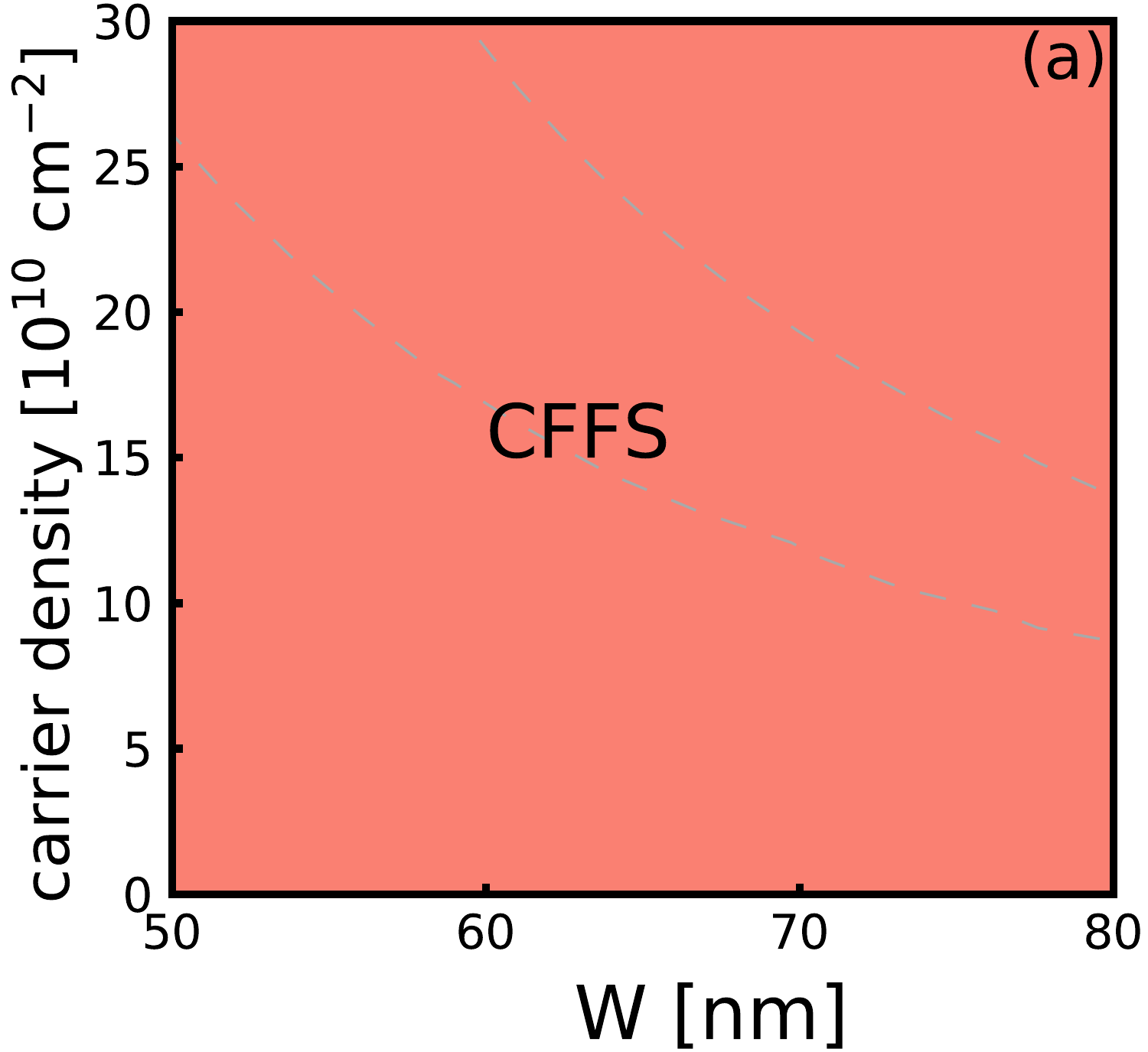}
		\includegraphics[width=\columnwidth]{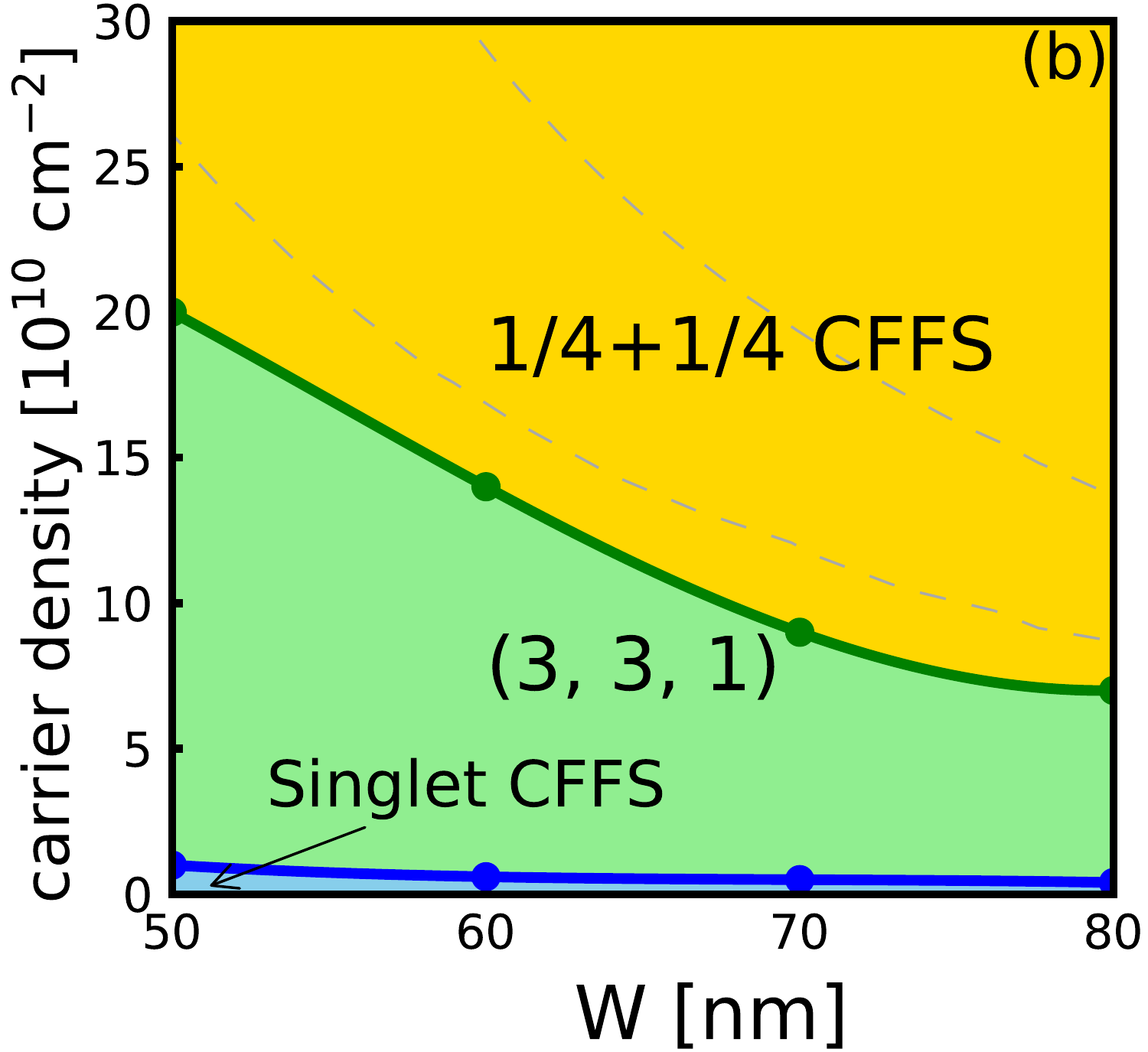}
		\caption{(a) The phase diagram of one component states, including CFFS (red) and the Pfaffian state (purple). The Pfaffian state is not stabilized for the parameters considered. (b) The phase diagram of two component states, including the $(3, 3, 1)$ state (green), pseudo-spin singlet (blue), and $1/4+1/4$ CFFS state (yellow). The region where experiments find an incompressible state~\cite{Shabani09b} is indicated by light dashed grey lines. At each width, the uncertainty of the transition density is about $1\times 10^{10} \text{cm}^{-2}$. The overall phase boundary is obtained by smoothly joining the transition points at $W=50,60,70,80$ nm.}
		\label{VMC_PHASE_DIAGRAM_1}
	\end{figure}

	The energies of the two-component states, namely the Pseudo-spin singlet CFFS, $(3,3,1)$ and the uncoupled $1/4+1/4$ CFFS, relative to the $(3,3,1)$ are shown in Fig.~\ref{VMC_E_2}.
	A transition from the singlet CFFS to the Halperin $(3,3,1)$ occurs at very low densities, followed by a second transition into the uncoupled $1/4+1/4$ CFFS (Fig.~\ref{VMC_E_2}). This behavior is similar to that found in earlier VMC calculations on the zero-width bilayer systems\cite{Scarola01b}. The phase diagrams for one and two-component states separately are shown in Fig.~\ref{VMC_PHASE_DIAGRAM_1}.
	

	\begin{figure}[H]
		\includegraphics[width=\columnwidth]{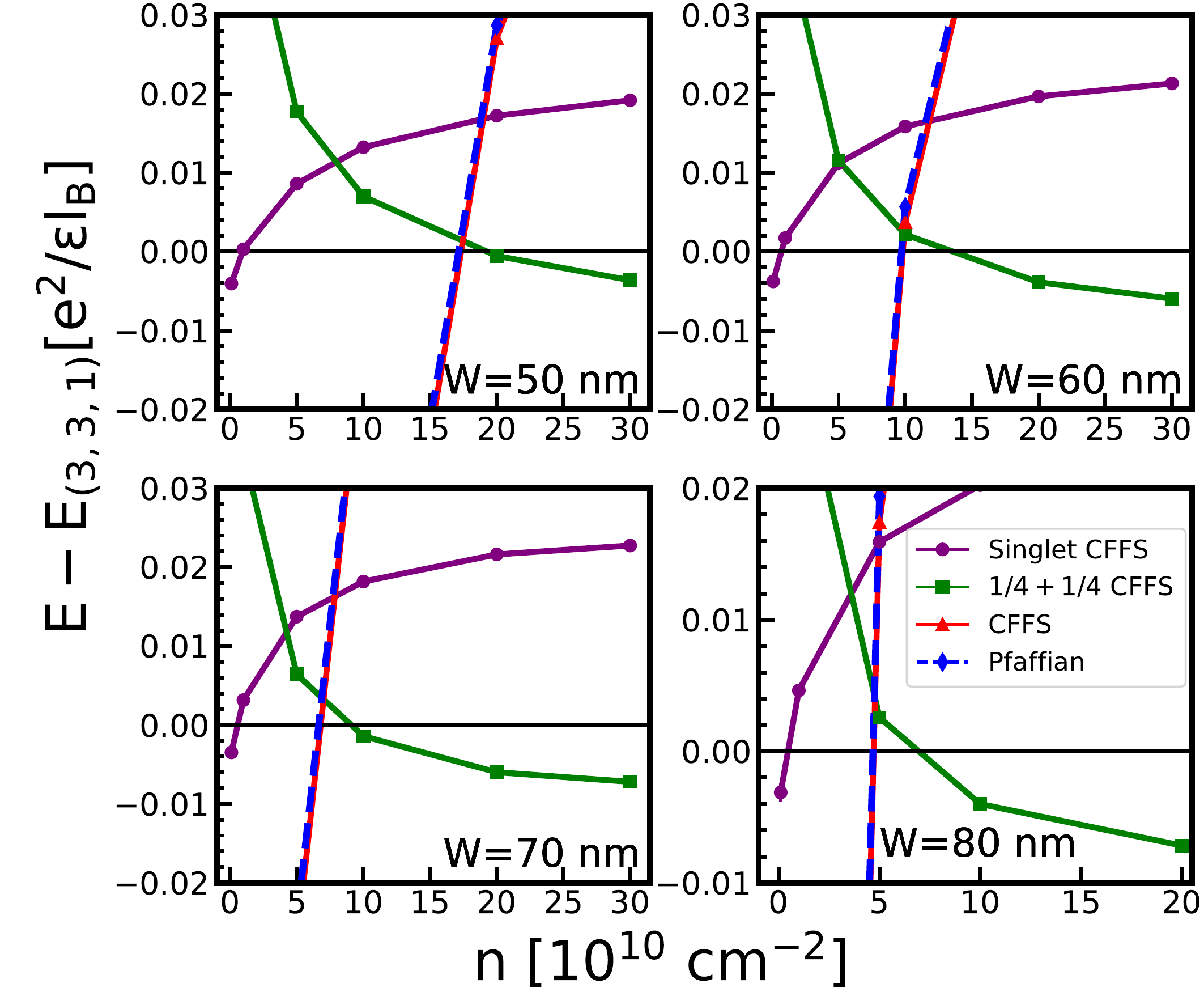}
		\caption{The VMC calculation of the energy per particle of the one-component CFFS state, the Pfaffian state, the $1/4+1/4$ CFFS state, and the singlet CFFS state in the thermodynamic limit. All energies are measured relative to the energy of the $(3, 3, 1)$ state. The well widths are shown on the plots. The energies of the one-component states change rapidly relative to the $(3, 3, 1)$ state due to the $\Delta_{\rm SAS}$ component. The statistical errors are smaller than the symbol sizes.}
		\label{VMC_E_3}
	\end{figure}

	Fig.~\ref{VMC_E_3} shows the energies of all states.  We add $\frac{1}{2} \Delta_\text{SAS}$ to the energy per particle for each two-component state, because half of all particles occupy the second subband. We quote all energies relative to the $(3,3,1)$ state in Fig.~\ref{VMC_E_3}, which yields the phase diagrams in Fig.~\ref{VMC_PHASE_DIAGRAM_2}. In general, one can see the ground state of the system is in a one-component state when the carrier density is low and the system makes a transition into a two-component state at high density. This result is qualitatively consistent with the experiments~\cite{Shabani09b, Suen92, Luhman08}, and favors the possibility that the observed incompressible state is the $(3,3,1)$ state. 
	
	There are significant differences, however. As noted earlier, the lower phase boundary is very sensitive to $\Delta_{\rm SAS}$. However, the upper theoretical phase boundary ought to be more reliable, and its deviation from the experimental phase boundary is thus significant. We note, however, that the calculation, so far, does not include LLM or disorder.

	\section{Fixed-phase Diffusion Monte Carlo Method}\label{sec_dmc_basics}
	
	In the following sections, we will use the fixed-phase DMC method to evaluate the phase diagram. The general DMC is a standard Monte Carlo method designed to obtain the ground state of the many-body Schr\"{o}dinger equation\cite{Reynolds82, Foulkes01} by a stochastic method. 
	By setting time to an imaginary variable $(t\to t=-i \uptau)$, the Schr\"odinger equation takes the form
	\begin{equation}
	\begin{aligned}
	-\hbar \partial_\uptau \Psi\left(\vec{R}, \uptau\right)= \left( H\left(\vec{R}\right)-E_T\right)\Psi\left(\vec{R}, \uptau\right)
	\end{aligned}
	\end{equation}
	where $\vec{R}=\left( \vec{r}_1, \vec{r}_2, \dots, \vec{r}_{N_e}\right)$ is the collective coordinate of the system and $E_T$ is a constant energy offset. When $\Psi\left(\vec{R}, \uptau\right)$ is real and non-negative, one can interpret the above equation as a diffusion equation, with $\Psi\left(\vec{R}, \uptau\right)$ interpreted as the density distribution of randomly moving walkers. The energy offset $E_T$ 
controls the population of random walkers. Starting from an initial trial wave function $\Psi_T$, as the walkers diffuse stochastically, the distribution gradually converges to a stable distribution that represents the ground state (provided $\Psi_T$ has a non-zero overlap with the ground state). 
More details can be found in Ref.~[\onlinecite{Foulkes01},\onlinecite{Mitas98}]	.
	
	The applicability of the DMC method relies on the assumption that the ground state is real and non-negative. However, this condition is not satisfied in a system with broken time-reversal symmetry, which is  the case in the presence of a magnetic field. To overcome this difficulty, the fixed-phase DMC method has been proposed\cite{Ortiz93, Melik-Alaverdian97}. The key idea is to 	write the wave function as 
	\begin{equation}
	\Psi(\vec{R})= \left| \Psi(\vec{R}) \right | \exp \left[ i \phi(\vec{R})\right]
	\label{amp_phase}
	\end{equation}
and determine the $\left| \Psi(\vec{R})\right |$ that gives the lowest energy for a fixed phase $\phi(\vec{R})$ by DMC method. This amounts to solving the Schr\"odinger equation
	\begin{equation}
	\begin{aligned}
	&H_\text{DMC} \left| \Psi\left(\vec{R}, \uptau\right)\right|=\\
	& \left( -\sum_{i=1}^N \frac{\hbar^2\nabla_i^2}{2m} +V_{\text{DMC}}\left(\vec{R}\right)-E_T\right) \left|\Psi\left(\vec{R}, \uptau\right)\right|=E \left|\Psi\left(\vec{R}, \uptau\right)\right|\\
	\end{aligned}
	\label{eqn_amp}
	\end{equation}
	with
	\begin{equation}
	V_{\text{DMC}}\left( \vec{R} \right)=V\left( \vec{R} \right)+\frac{1}{2 m}\sum_{i=1}^N
	\left[ \hbar \nabla_i \phi \left(\vec{R}\right)+\frac{e}{c} {\mathbf A}\left({\mathbf r }_i\right) \right]^2 \label{Vdmc}.
	\end{equation}

	The diffusion equation is often efficiently solved by an importance sampling method. The so-called guiding function is defined as
	\begin{equation}
	f\left( \vec{R}, \uptau \right)=\left|\Psi_T \left( \vec{R}\right) \right| \left|\Psi \left( \vec{R}, \uptau\right) \right|
	\end{equation}
	where $\Psi_T$ is the trial wave function.
	Instead of solving Eq.\,\ref{eqn_amp}, we have an equivalent equation:
	\begin{equation}
	\begin{aligned}
	-\hbar\partial_{\uptau}f({\bf R},\uptau)
	&=-\frac{\hbar^2}{2m}\nabla^2 f({\bf R},\uptau)+\frac{\hbar^2}{m}\nabla \cdot \left({\bf v}_D  f({\bf R},\uptau)\right)\\
	&+\left(E_L({\bf R})-E_T\right) f({\bf R},\uptau)
	\end{aligned}
	\label{equiv_schrodingerr}
	\end{equation}
	where $\nabla=\left(\nabla_1, \nabla_2, \dots, \nabla_N\right)$ is the dN-dimensional (in d space dimensions) gradient operator, $\vec{v}_D\left(\vec{R}\right)$ is the dN-dimensional drift velocity defined by
	\begin{equation}
	\vec{v}_D\left(\vec{R}\right)=\nabla \ln \left|\Psi_T \left(\vec{R}\right)\right|,
	\end{equation}
	and
	\begin{equation}
	E_L\left(\vec{R}\right)=|\Psi_T|^{-1}H_\text{DMC} |\Psi_T|
	\end{equation}
	is the local energy. We give their explicit forms in Appendix~\ref{DMC_algorithm} based on Ref.~[\onlinecite{Ortiz93}]. 
	
	The accuracy of the DMC energy depends on the choice of the phase $\phi(\vec{R})$. In this paper, our initial DMC trial wave functions will be our candidate trial wave functions described earlier. (In the case of 3D-DMC, these will also include the transverse wave function.)  
	Each trial wave function identifies a specific phase $\phi_T$. The DMC algorithm then produces the lowest energy state for each choice of the trial wave function.

We stress that the DMC calculation automatically includes LLM. In fact, it is a non-perturbative method for treating LLM, which has been shown in past studies to give rather accurate results\cite{Guclu05a,Ortiz93,Melik-Alaverdian95, Bolton96, Melik-Alaverdian97, Melik-Alaverdian99, Zhao18, Zhang16, Hossain20}. 	
	
	\section{2D fixed-phase DMC study with effective interaction}\label{sec_2d_dmc}
	
We implement a 2D fixed-phase DMC study of the problem where we obtain the lowest energy using DMC while setting $V(\vec{R})$ in Eq.~\ref{Vdmc} to $V_{\rm eff}(\vec{R})$ introduced in Eq.~\ref{V_eff}. This allows for LLM in a model where electrons confined to 2D are interacting via $V_{\rm eff}(\vec{R})$. As noted above, the phase is fixed by the trial wave functions described above.
	
\begin{figure}[H]
		\includegraphics[width=\columnwidth]{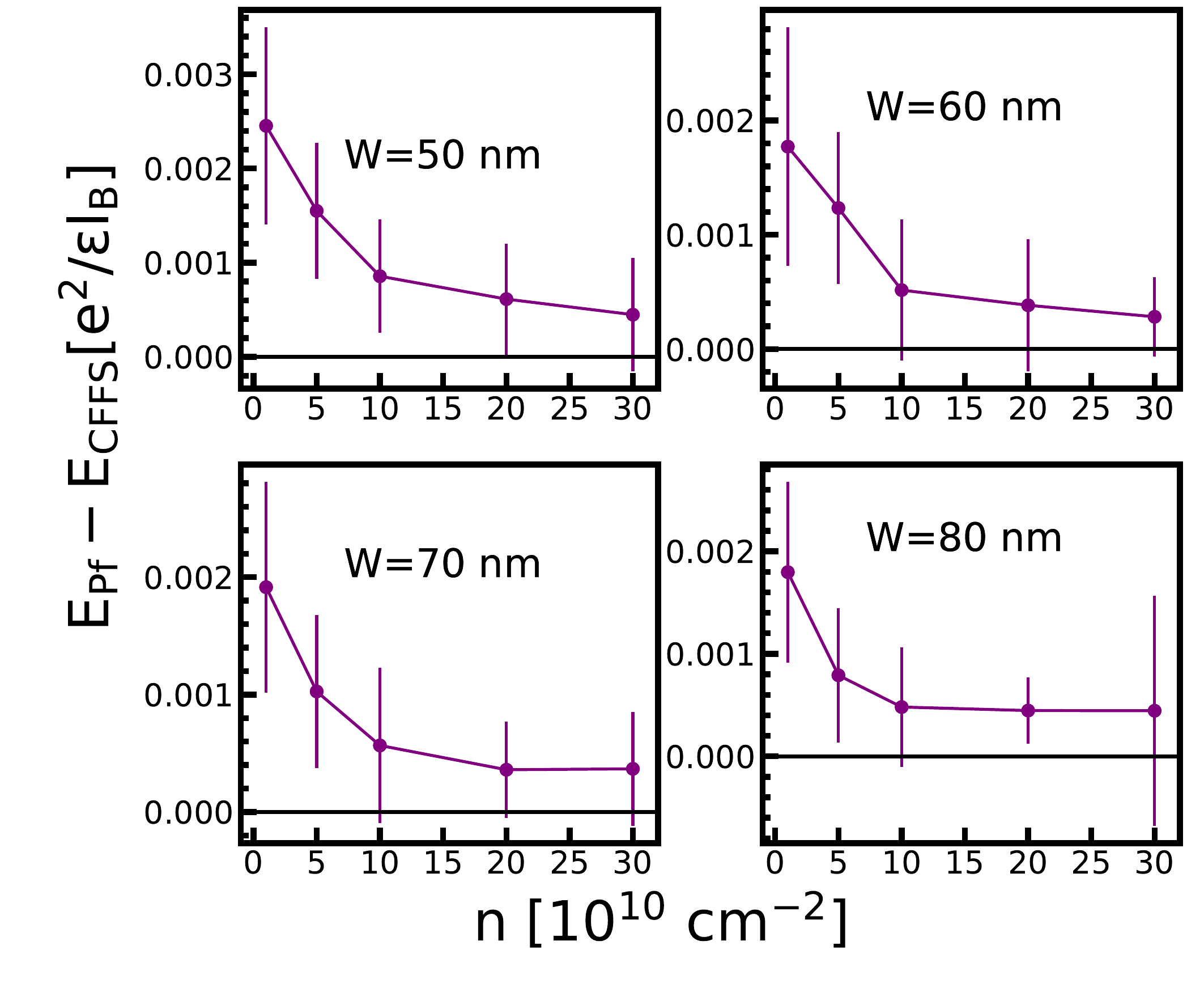}
		\caption{2D-DMC calculation of the energy difference per particle between the one-component CFFS and the Pfaffian state in the thermodynamic limit.}
		\label{DMC_2d_1}
	\end{figure}

\begin{figure}[H]
		\includegraphics[width=\columnwidth]{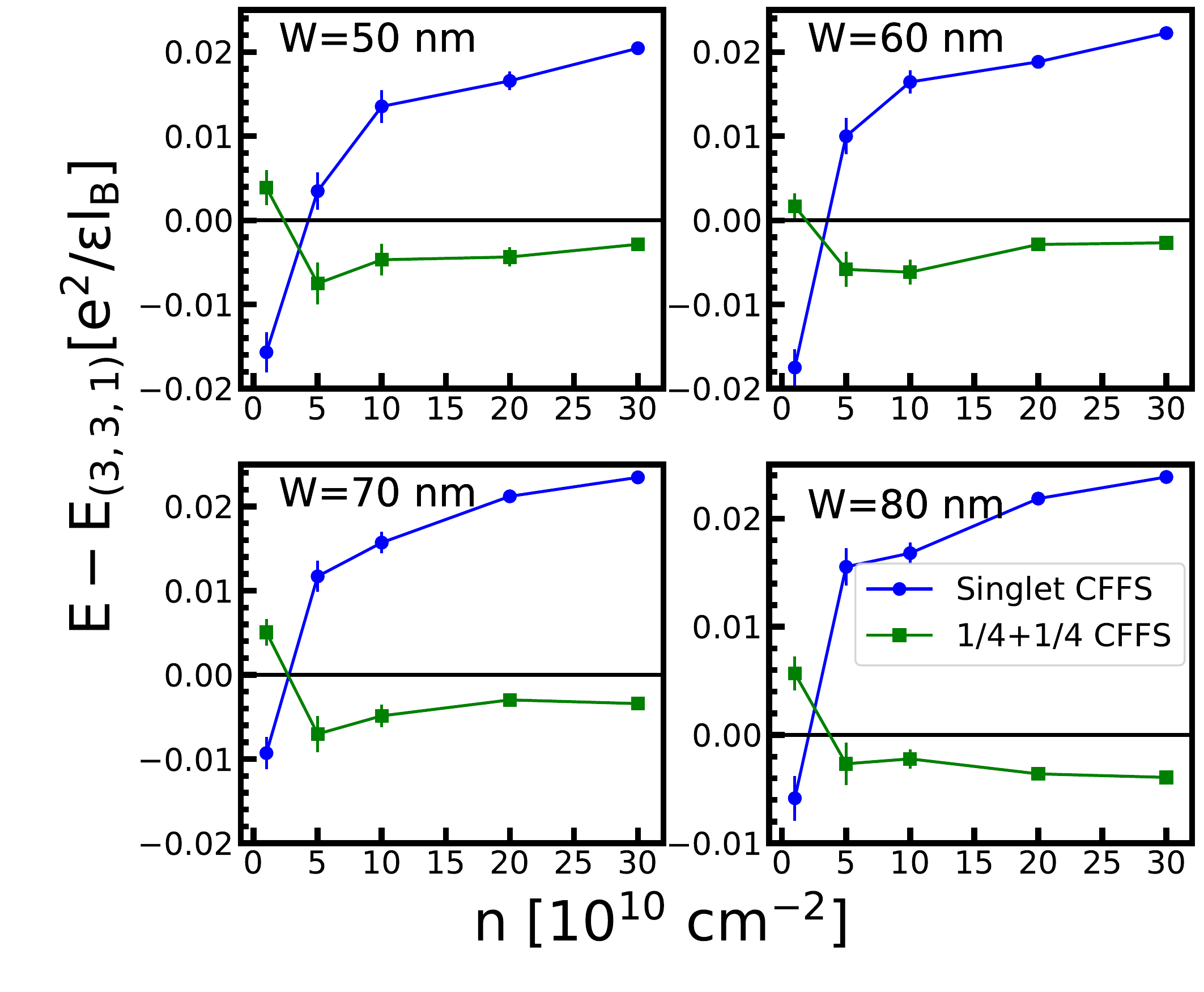}
		\caption{2D-DMC calculation of the energy per particle of the  $1/4+1/4$ CFFS and the singlet CFFS state in the thermodynamic limit relative to the $(3,3,1)$ state. The well widths are indicated on the plots.}
		\label{DMC_2d_2}
	\end{figure}

\begin{figure}[H]
		\includegraphics[width=\columnwidth]{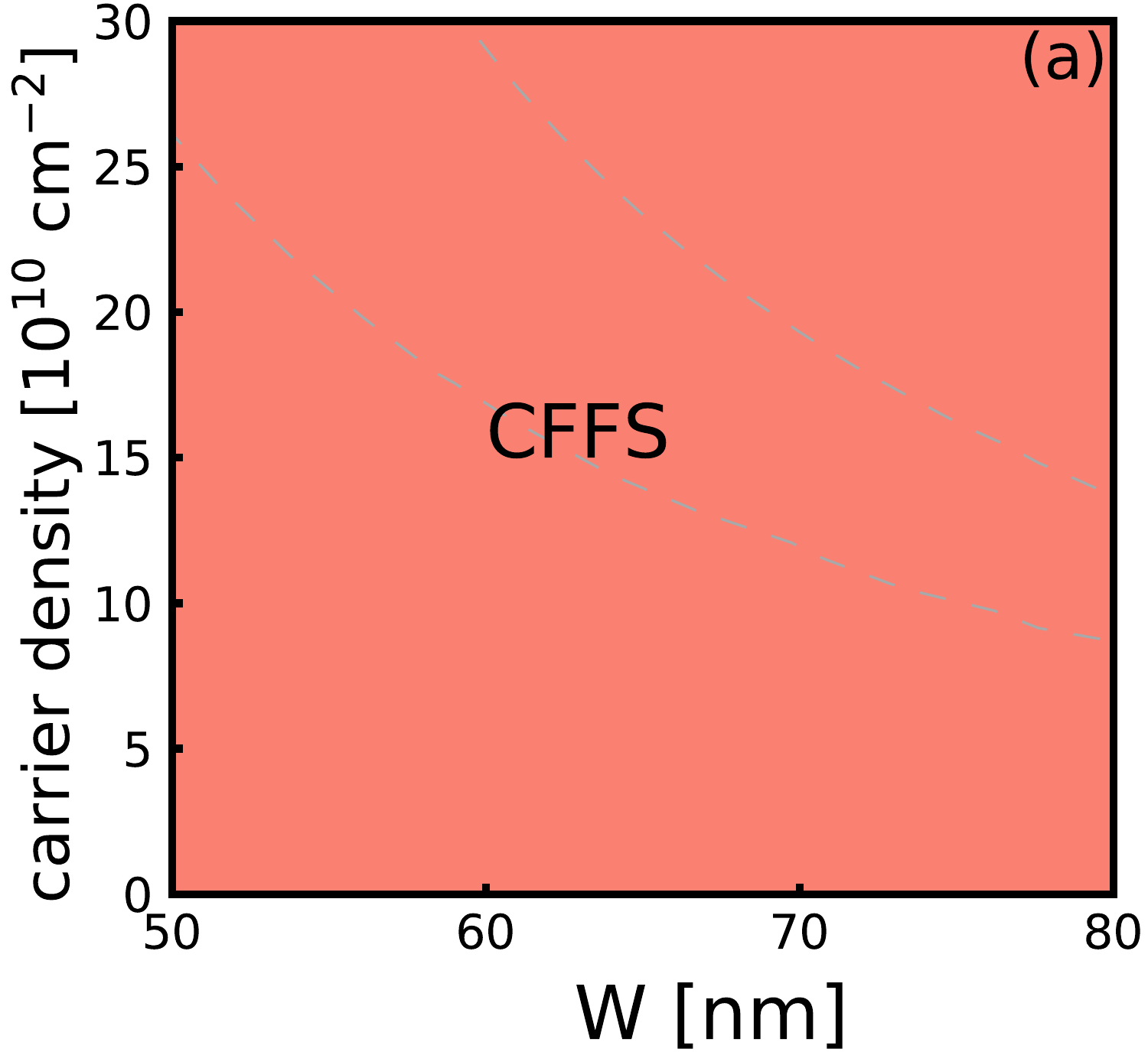}
		\includegraphics[width=\columnwidth]{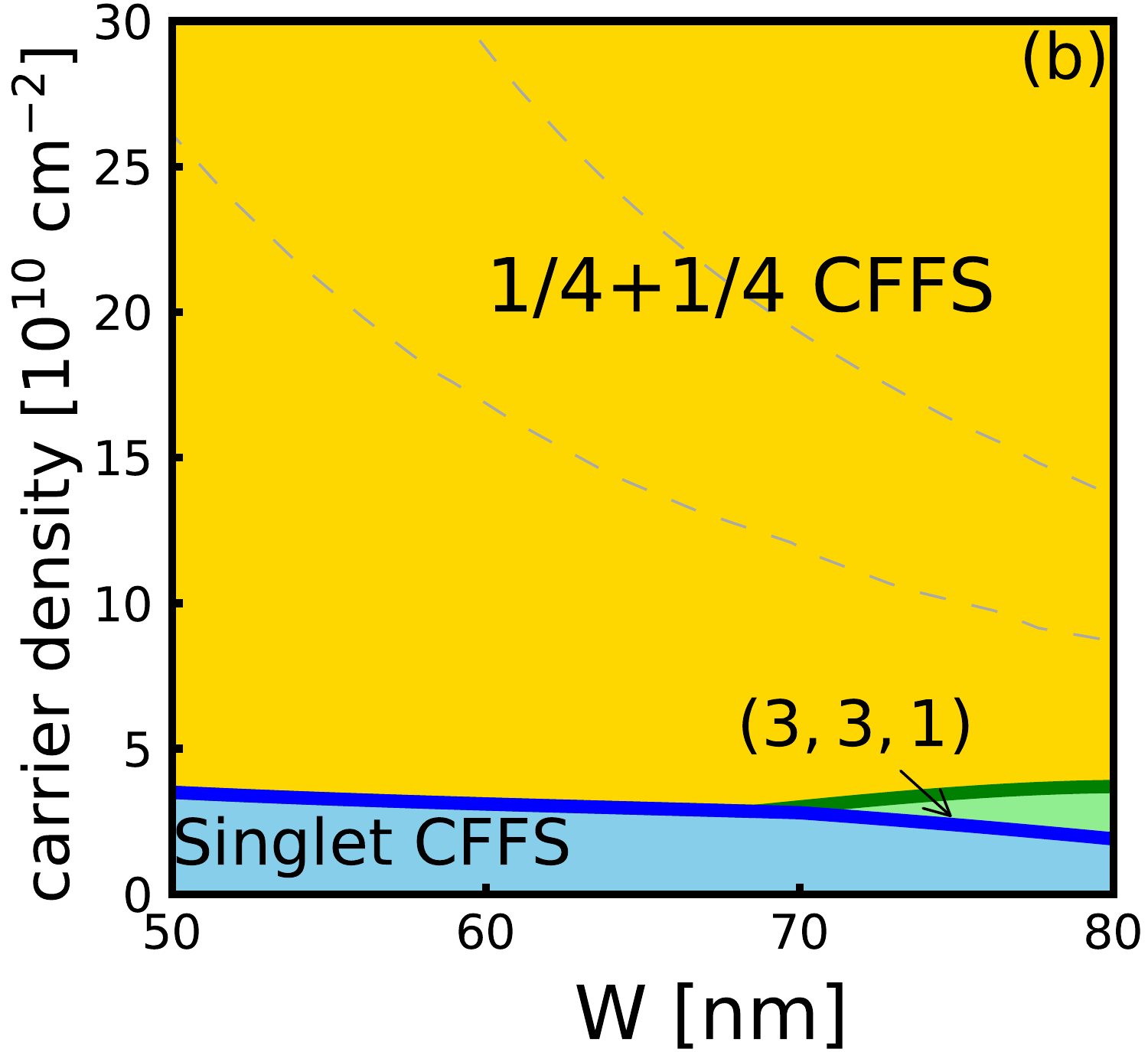}
		\caption{(a) The phase diagrams of one component states obtained by 2D-DMC on torus. (b) The phase diagrams of two component states. The states considered are the CFFS state (red), the $(3, 3, 1)$ state (green), the singlet CFFS state (blue), the $1/4+1/4$ CFFS state (yellow), and the Pfaffian state (purple). In the lower panel, the uncertainty of the transition density from the singlet CFFS state to $(3,3,1)$ is about $1\times 10^{10} \text{cm}^{-2}$, while the transition density from $(3, 3, 1)$ or singlet CFFS to $1/4+1/4$ CFFS has an uncertainty of about $2\times 10^{10} \text{cm}^{-2}$.The region where experiments find an incompressible state~\cite{Shabani09b} is indicated by light dashed grey lines. The overall phase boundary is obtained by smoothly joining the transition points at $W=50,60,70,80$ nm.}
		\label{2D_DMC_BOUNDARY}
	\end{figure}

	As shown in Fig.~\ref{DMC_2d_1}, the comparison between the one-component CFFS and the Pfaffian state is very similar to that from VMC calculation and no transition occurs into the Pfaffian state (Fig.~\ref{2D_DMC_BOUNDARY}(a)). Meanwhile, the result for two-component states is quite different from the VMC result (Fig.~\ref{DMC_2d_2}). We find that the uncoupled $1/4+1/4$  CFFS is very efficient in lowering its energy in the presence of the LLM. In contrast to the VMC result, the system makes a transition from the pseudo-spin singlet CFFS directly into the $1/4+1/4$ CFFS state for most parameters (Fig.~\ref{2D_DMC_BOUNDARY}(b)). For very large widths and low densities, we find a small region of $(3,3,1)$ state.

	When both one-component and two-component states are considered, the resulting phase diagram is shown in Fig.~\ref{2D_DMC_BOUNDARY_ALL}. The one-component CFFS makes a transition into the uncoupled two-component $1/4+1/4$ without going through an incompressible state, except in a small region where the well-width is large. We note here that the extrapolation of the 2D-DMC is less linear for the two-component states, which leads to a larger statistical error of about $2\times 10^{10} \text{cm}^{-2}$ for the density where the phase transition occurs.

	\begin{figure}[H]
		\includegraphics[width=\columnwidth]{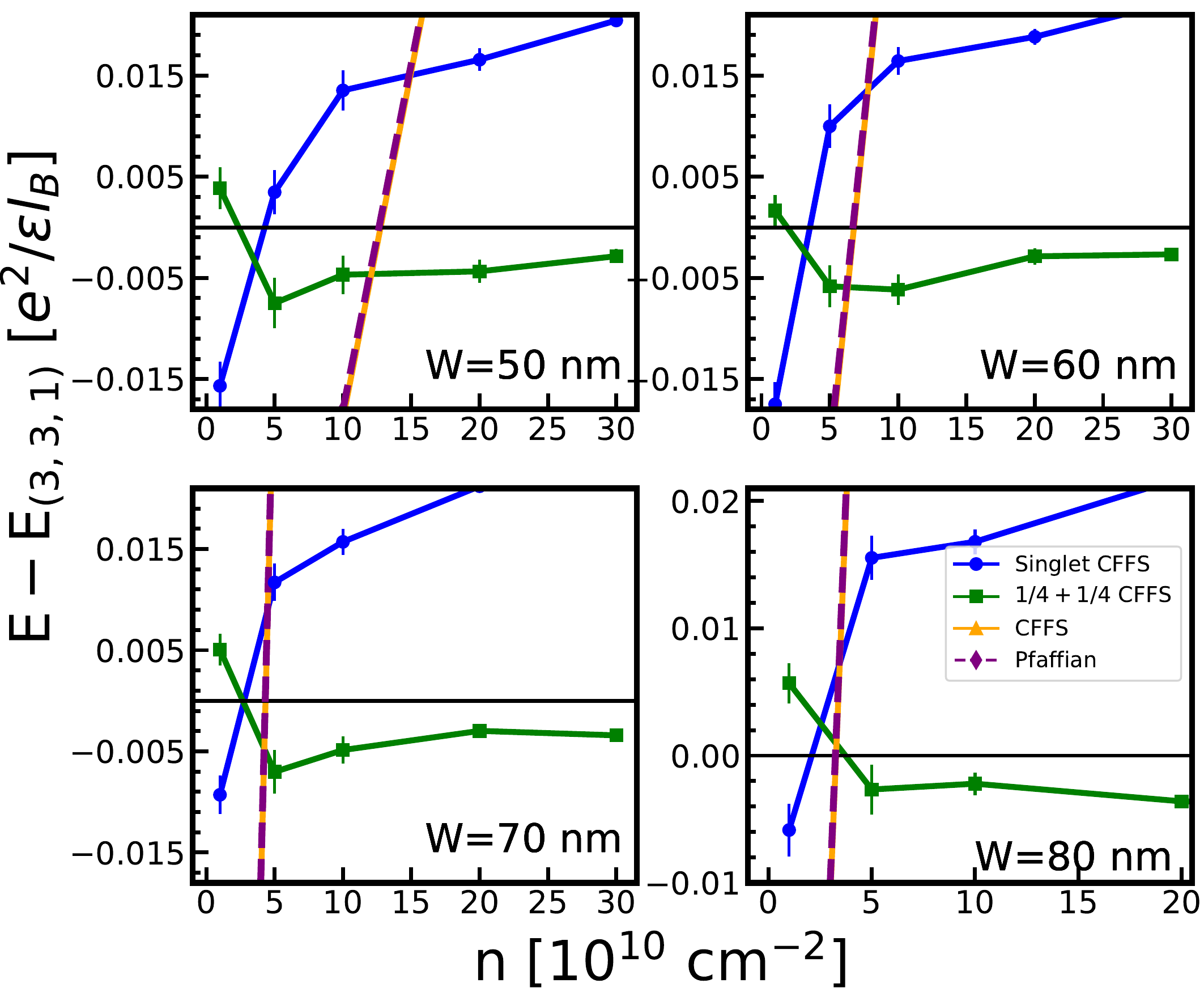}
		\caption{2D-DMC calculation of the energy per particle of the one-component CFFS state, the Pfaffian state,  the $1/4+1/4$ CFFS and the singlet CFFS state relative to the $(3, 3, 1)$ state in the thermodynamic limit. The well widths are labeled on the plots. The energies include contribution from $\Delta_{\rm SAS}$.} 
		\label{DMC_2d_3}
	\end{figure}
	
	\section{3D fixed-phase DMC study of the $1/2$ FQHE}\label{sec_dmc_3d}
	The transverse trial wave function for the one-component states are chosen to be:
	\begin{equation}
	\begin{aligned}
	\Psi_\text{trans}\left( \left\{ w_i\right\}\right)&=\prod_{i=1}^{N_e}\psi_S(w_i)\\
	&=\prod_{i=1}^{N_e} \left[\cos\left( \frac{\pi w_i}{W} \right)-\alpha \cos\left( \frac{3 \pi w_i}{W} \right)\right],
	\end{aligned}
	\label{trans_wf_sing}
	\end{equation}
	where $W$ is the width of the quantum well and $\alpha$ is a parameter introduced to improve the converging speed. Empirically we find the program to be most efficient and stable when $\alpha$ is tuned from $0.2$ to $0.8$ when the well width ranges from $2 l_B$ to $10 l_B$. However, one should keep in mind that the choice of $\alpha$ is a technical matter; as long as the number of iterations is large enough, any choice of $\alpha$ leads to the same result because the fixed-phase DMC solves for the lowest energy state within a given phase sector independent of the initial wave function.

	For two-component states, before coming to 3D-DMC, 
	 it is necessary to address a significant difficulty. In general, one needs to evaluate the energy expectation of a given two-component state [e.g. Halperin $(3,3,1)$ state] by fully anti-symmetrizing the wave function. For two-component states in a single layer with real spin, the Coulomb interaction does not depend on the spin index, and all the cross-terms produced by anti-symmetrization vanish, and one can treat the two components as two sets of distinguishable particles, which greatly simplifies the calculation. This, however, is not true for the present case since the Coulomb interaction explicitly depends on the transverse coordinates and the cross-terms are nonzero. Here, one must include all the permutation-terms to fully anti-symmetrize the wave function. This is impractical for systems with greater than 10 particles because there are $\frac{N_e!}{(N_e/2)! (N_e/2)!}$ inter-component permutations. A special case is when the two transverse bases have no overlap. In this case, all cross-terms vanish and one can calculate the energy expectation as if the two components were two distinguishable sets of particles. We, therefore, use a transverse trial wave function for the two components to be strictly spatially separated, i.e. one basis function is strictly confined in the left half of the quantum well while the other component in the other half. In other words, our basis is given by:
	
	\begin{equation}
	\Psi_\text{trans} \left( \left\{ w_i, w_{[j]}\right\}\right)=\prod_{i=1}^{N_e/2}  \prod_{\left[j\right]=N_e/2+1}^{N_e} \psi_L\left(w_i \right) \psi_R \left( w_{[j]}\right)
	\label{trans_wf_left}
	\end{equation}
	where
	\begin{equation}
	\begin{aligned}
	\psi_L\left( w_i \right)&=\left\{\begin{array}{ll}
	-\sin(\frac{2\pi w_i}{W}), & \text{if } -W/2<w_i<0\\
	0, & \text{if } 0\leqslant w_i<W/2
	\end{array}
	\right.\\
	\psi_R\left( w_{[j]} \right)&=\left\{\begin{array}{ll}
	0, & \text{if } -W/2<w_{[j]}<0\\
	\sin(\frac{2\pi w_{[j]}}{W}), & \text{if } 0\leqslant w_{[j]}< W/2
	\end{array}
	\right.
	\end{aligned}
	\end{equation}
	represents the left- and right-components.
		The 3D trial wave function for the $(3,3,1)$ state is constructed as:
	\begin{equation}
	\Psi_{(3,3,1)}^{\text{3D}}\left({\mathbf{R}}\right) =\Psi_{(3,3,1)}\left( \left\{ z_i, z_{[j]}\right\}\right) \Psi_\text{trans} \left( \left\{ w_i, w_{[j]}\right\}\right),
	\end{equation}
	The other two-component states are constructed similarly, with the in-plane part replaced by the corresponding wave functions.
	
	In Appendix~\ref{Full_Antisymmtrized} we test the regime of validity of our approximation (that the right and left components are non-overlapping) for a system of four particles, for which we can implement full antisymmetrization. We find that our approximation becomes excellent near the upper phase boundary in Fig.~\ref{3D_PHASE_2}.

	\subsection{Transverse density profile evaluated by 3D-DMC}
	\begin{figure}[H]
		\includegraphics[width=\columnwidth]{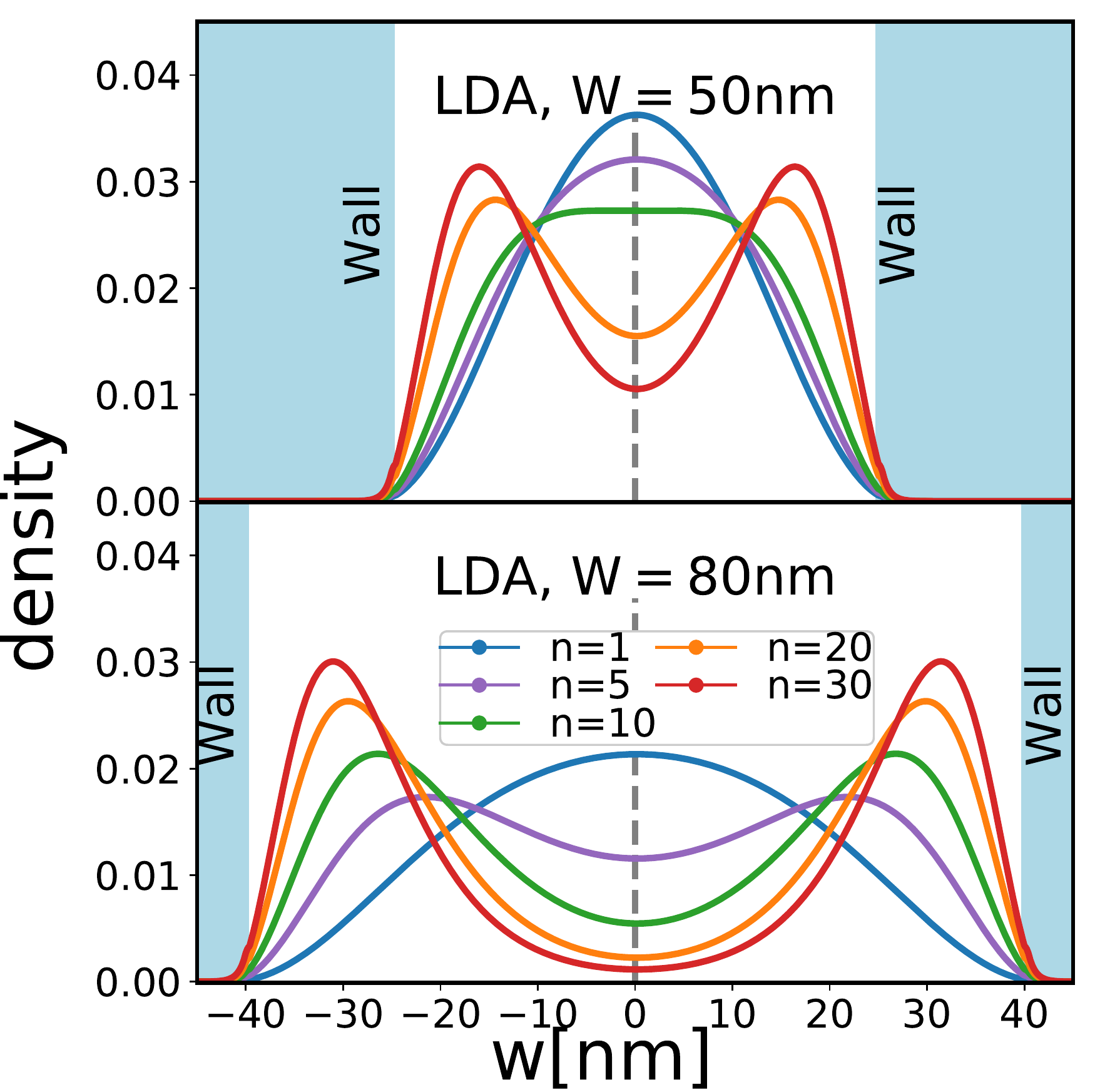}
		\caption{The transverse density calculated by LDA, which assumes a finite depth quantum well with the realistic parameters of the GaAs. The carrier densities are shown in units of $10^{10}\text{cm}^{-2}$. The area under each profile is normalized to unity.}
		\label{LDA_density}
	\end{figure}
	
	\begin{figure}[H]
		\includegraphics[width=\columnwidth]{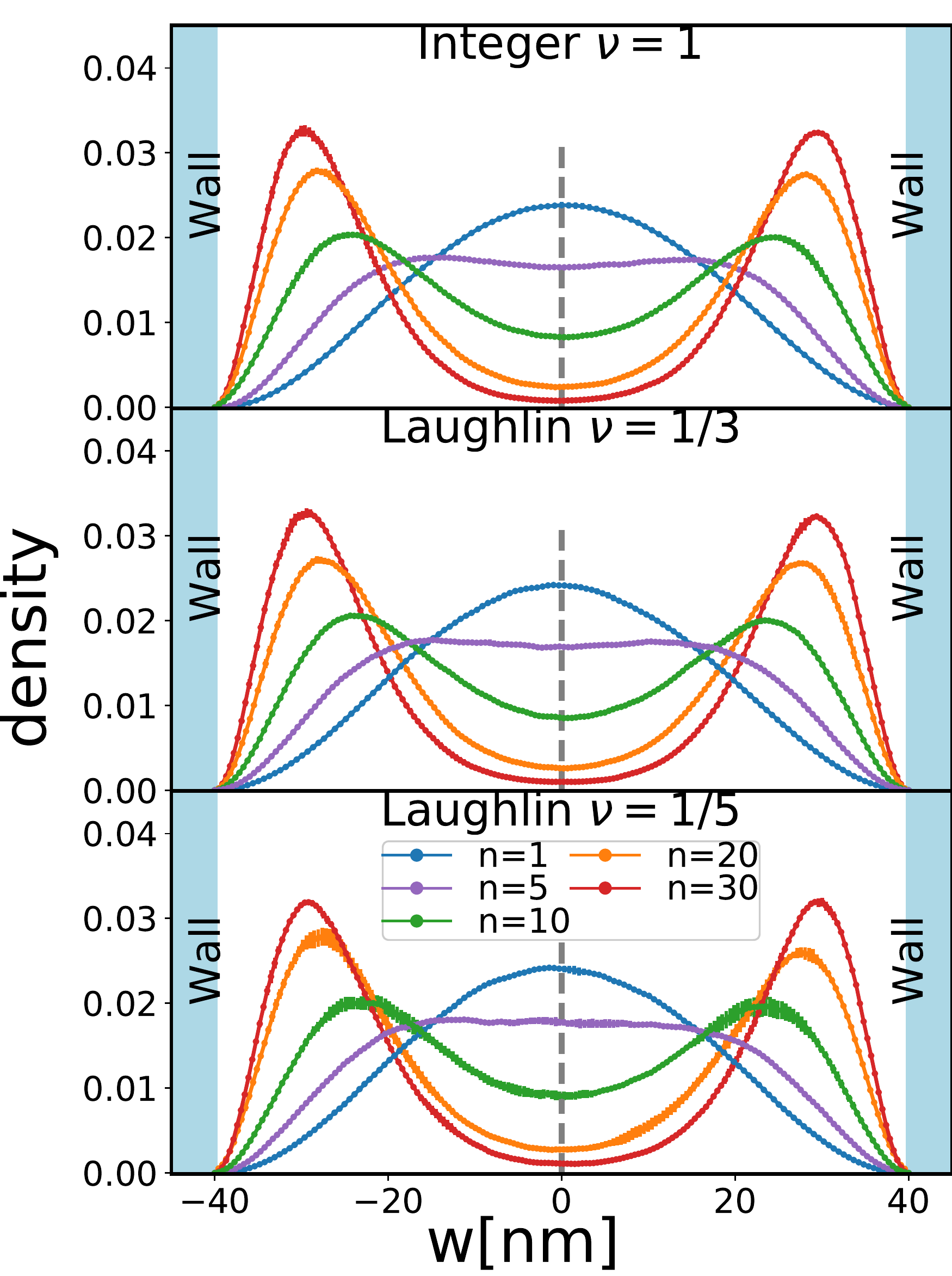}
		\caption{3D-DMC calculation of transverse densities for different FQHE states. The calculations are performed for $N_e=16$ and quantum well width $W=80 \text{nm}$. The legend shows the carrier densities corresponding to each color, measured in units of $10^{10}\text{cm}^{-2}$}
		\label{DMC_density_others}
	\end{figure}
	
	It is essential to quantitatively understand how the transverse distribution of electrons evolves as the well-width increases. We first show the transverse density calculated by LDA (Fig.~\ref{LDA_density}). The LDA package \cite{Martin20} is for realistic parameters with finite well-depth, and the transverse density extends outside the well by about $3\sim 4$ nm or less on each side. 
	This justifies our infinite-depth approximation. (In principle our method can also deal with finite depth quantum well, but technically that makes the form of the transverse trial wave function and the local energy more complicated.) In our approach, we implement the 3D-fixed-phase DMC and explicitly calculate the transverse density profiles of different candidate states. 
	
	Let us first consider one component states. Fig.~\ref{FINITE_SIZE_DEN} in Appendix~\ref{transverse density} shows that the transverse density for the CFFS is insensitive to the system size. We have found similar behavior at other filling factors. Hence, we believe that the density profiles shown in our work represent the thermodynamic limit.
	
	We have also studied the transverse profile for several filling factors, e.g. for $\nu=1$, 1/3, 1/5 FQHE states. As shown in Fig.~\ref{DMC_density_others}, we find that the transverse densities are not sensitive to the filling factor. 

	\begin{figure}[H]
	\includegraphics[width=\columnwidth]{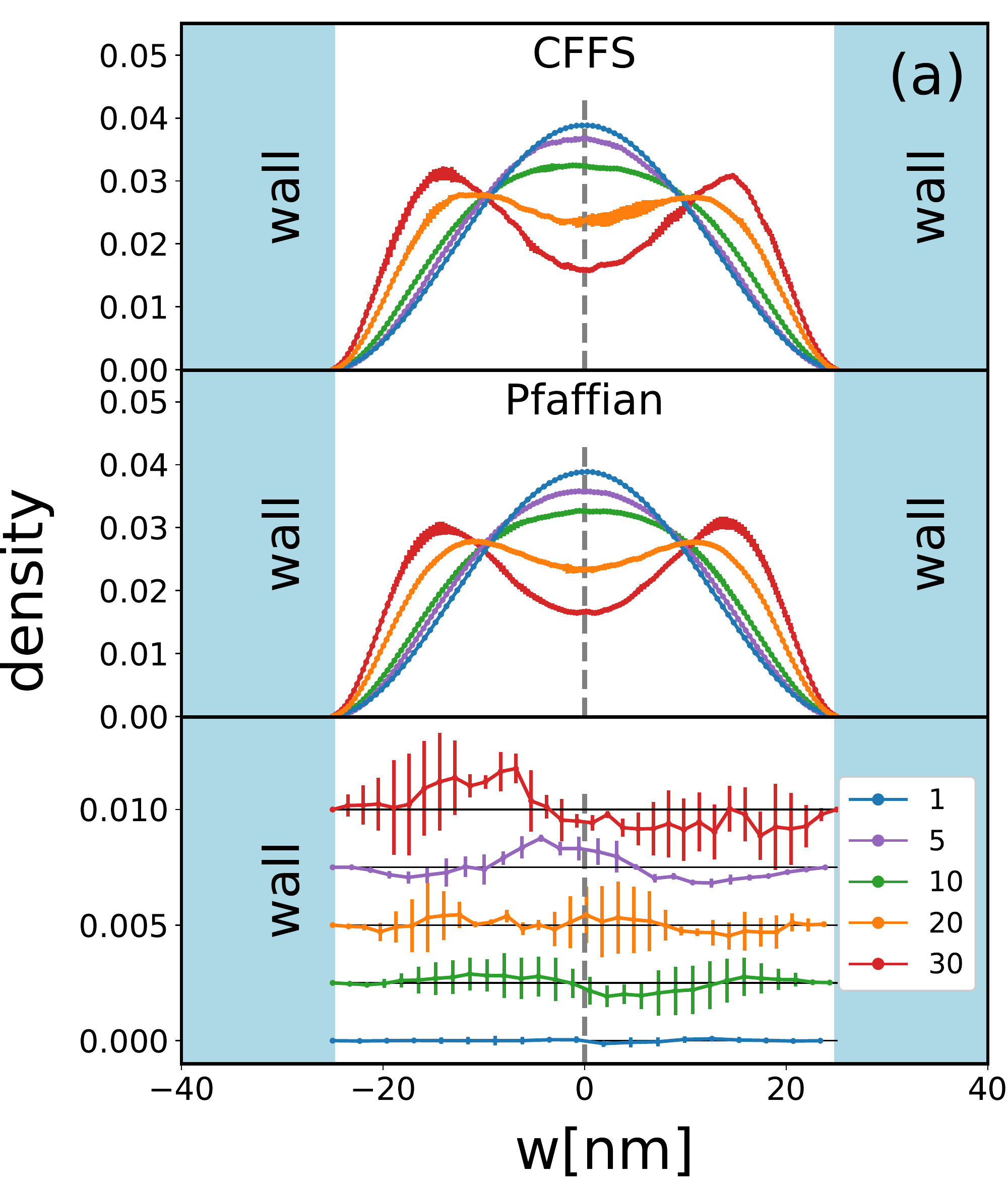}
	\includegraphics[width=\columnwidth]{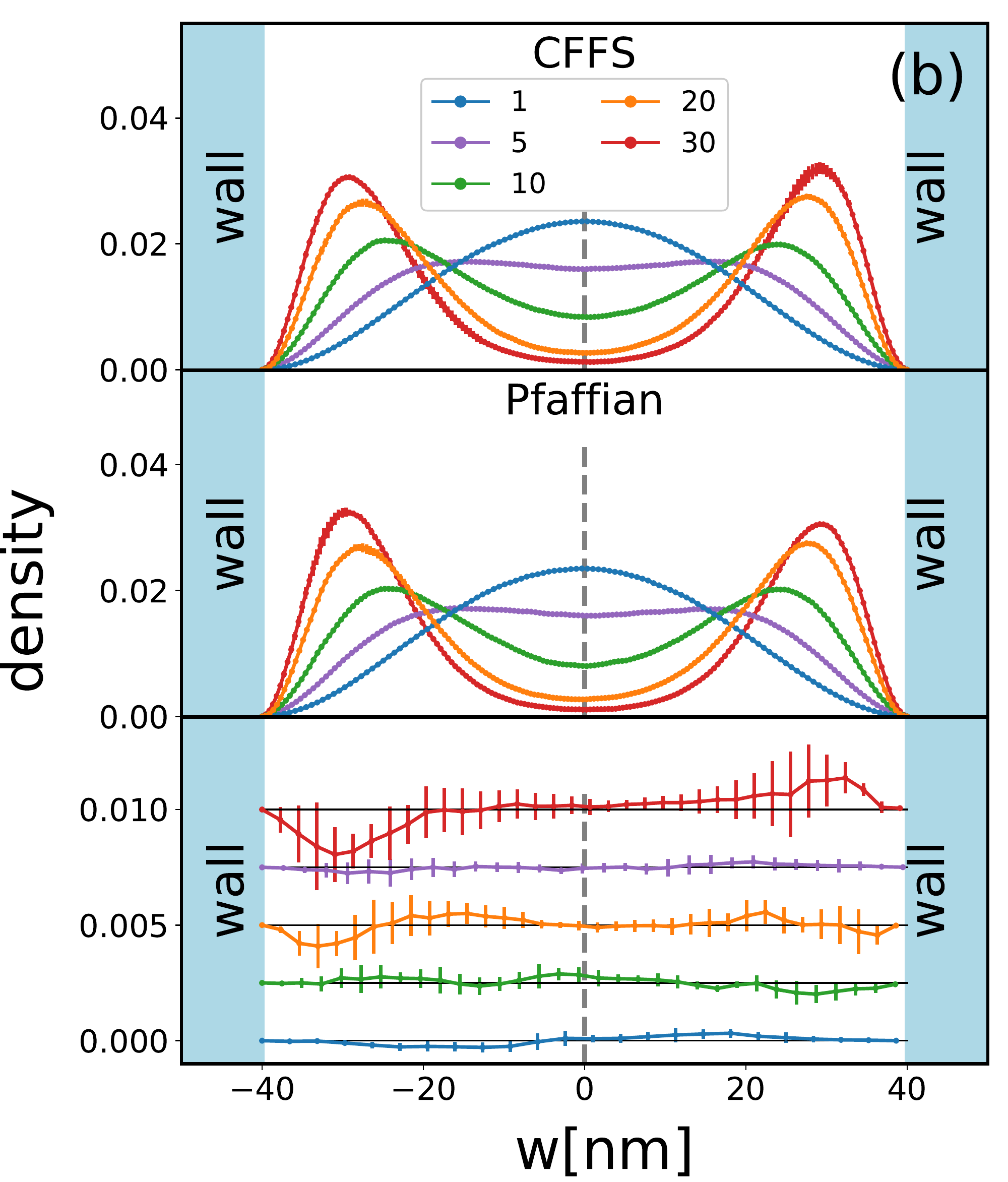}
	\caption{Transverse densities for the one-component CFFS and the  Pfaffian state calculated by 3D-DMC. The legend shows the carrier density in units of $\mathrm{10^{10} cm^{-2}}$. The results are shown for quantum well widths $W=50$nm (a) and $W=80$ nm WQW (b). At each width, the density profiles of one-component CFFS the Pfaffian state are shown individually in the upper two panels of (a) and (b). The lowest panel shows the differences between the two densities $\rho_\text{CFFS}-\rho_\text{Pfaf}$; the scale on the left corresponds to the lowest plot, and the rest are shifted up by 0.0025 units; also only 1 out of every 5 data points in the calculation are shown for clarity. The system size is $N_e=16$.} 
	\label{Density_1L}
	\end{figure}	
	
	In Figure\,\ref{Density_1L} we show the transverse density for the one-component CFFS and Pfaffian states at $\nu=1/2$ in a $\mathrm{50 nm}$ and a $\mathrm{80 nm}$ well width, with the areal density ranging from $\mathrm{n=1\times10^{10} cm^{-2}}$ to $\mathrm{3\times 10^{11} cm^{-2}}$. Other widths we consider in this article are between $\mathrm{50 nm}$ and $\mathrm{80 nm}$ and the profiles of the transverse density are similar (not shown).  As one can see, the system becomes more and more two-component-like with increasing carrier density.
	If one compares the 3D-DMC results with the LDA results, one can see that the two methods give very similar predictions, although the two ``humps," which indicate the onset of bilayer-like physics, appear at somewhat smaller densities in the LDA results.

\begin{figure}[H]
		\includegraphics[width=\columnwidth]{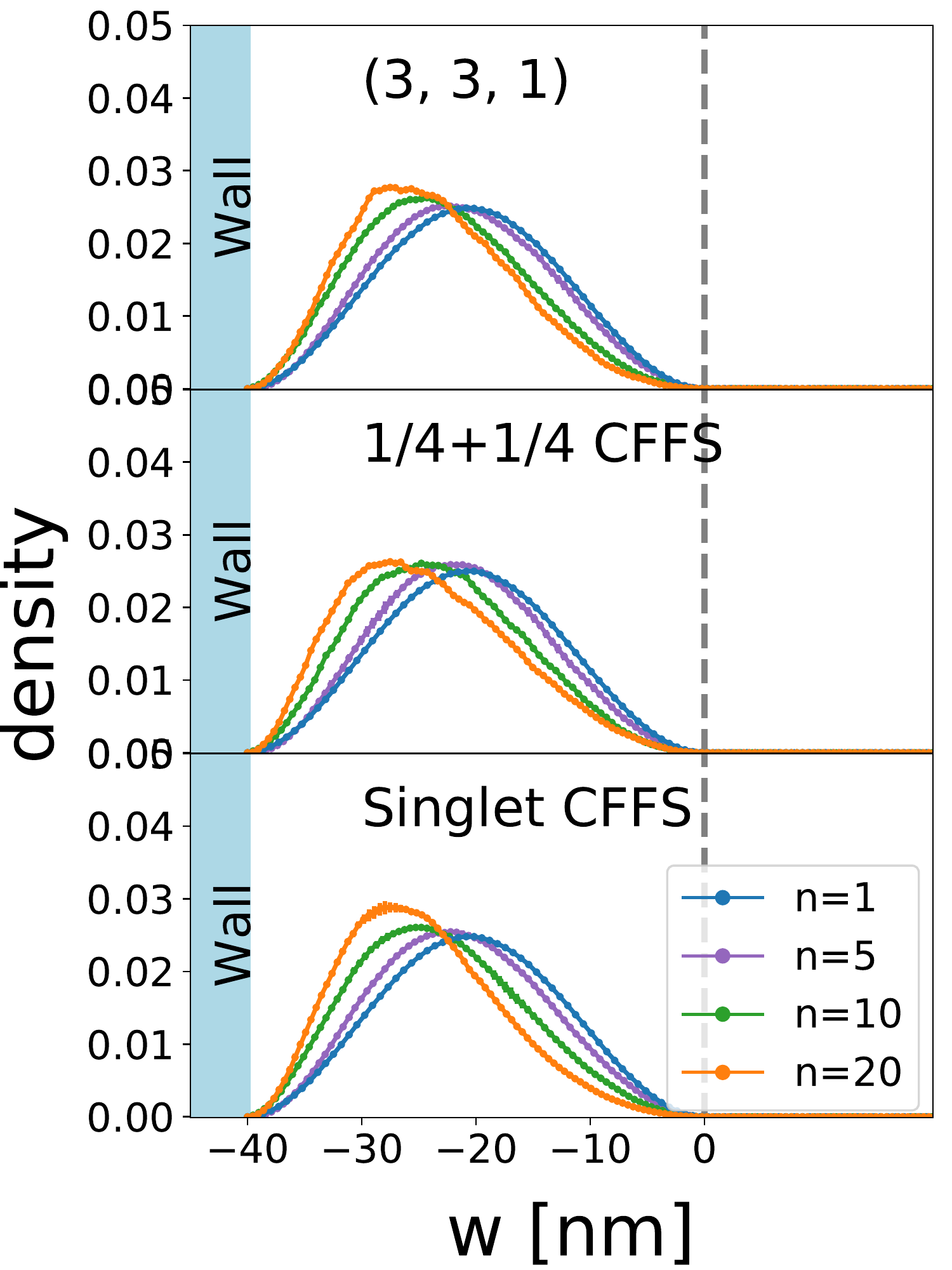}
		\caption{Transverse density of the left-component obtained by 3D-DMC for two-component states. The right component is analogous. The legend shows the carrier density in units of $\mathrm{10^{10} cm^{-2}}$. The system size is $N_e=8$.}
		\label{Density_2L}
	\end{figure}

	We next show in Fig.~\ref{Density_2L} the transverse density profiles for two-component states, assuming that the density vanishes at the center point (for reasons discussed above). The transverse wave function is insensitive to the state in 2D, and, as expected, the system becomes more bilayer-like with increasing carrier density.

	\subsection{Energy calculation and phase diagram by 3D-DMC}
	
\begin{figure}[H]
		\includegraphics[width=\columnwidth]{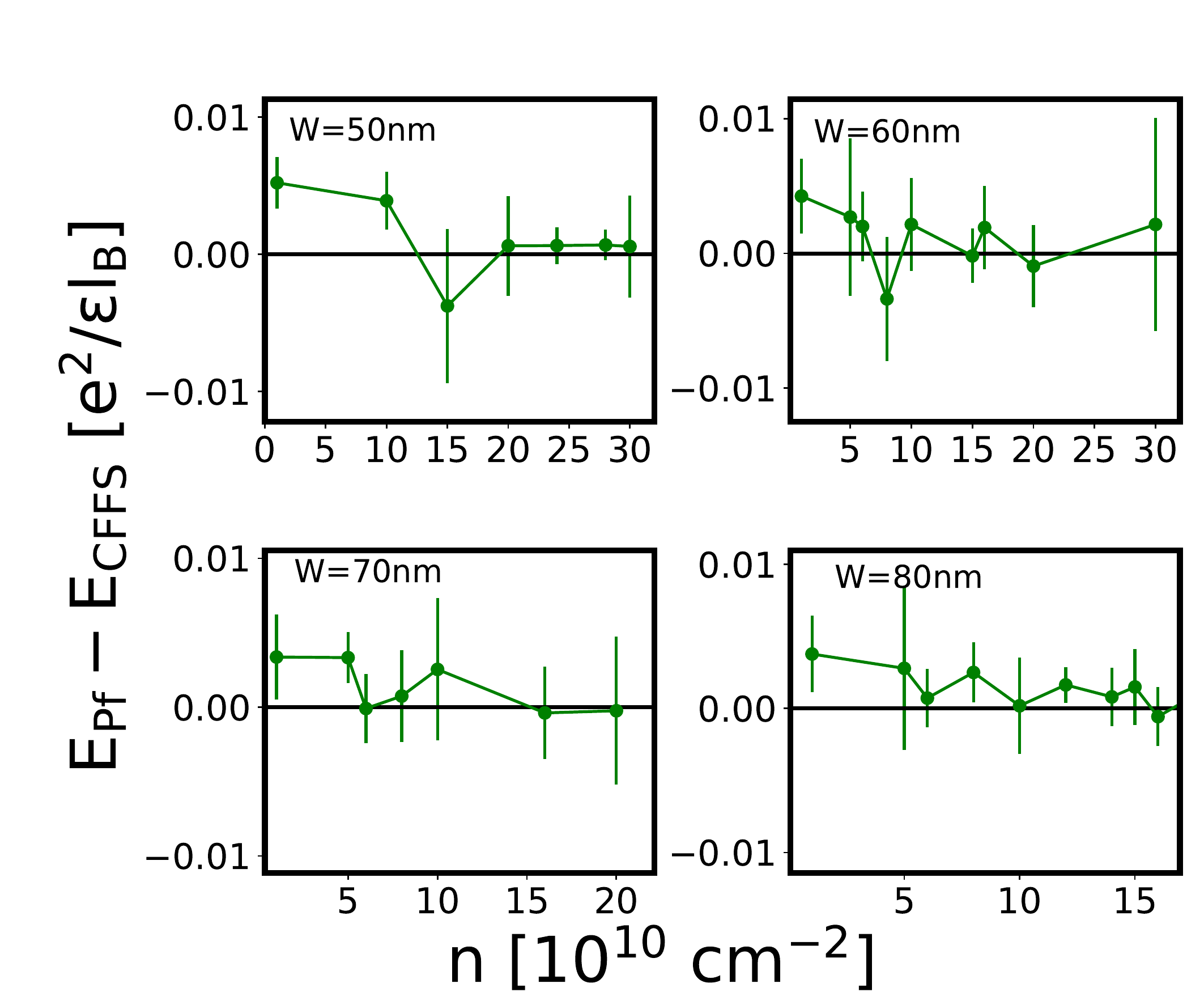}
		\caption{The energy difference between the Pfaffian state and the one-component CFFS state in the thermodynamic limit as a function of the carrier density at different well widths calculated by 3D-DMC in the torus geometry.}
		\label{3D_1L_energy}
	\end{figure}

\begin{figure}[H]
		\includegraphics[width=\columnwidth]{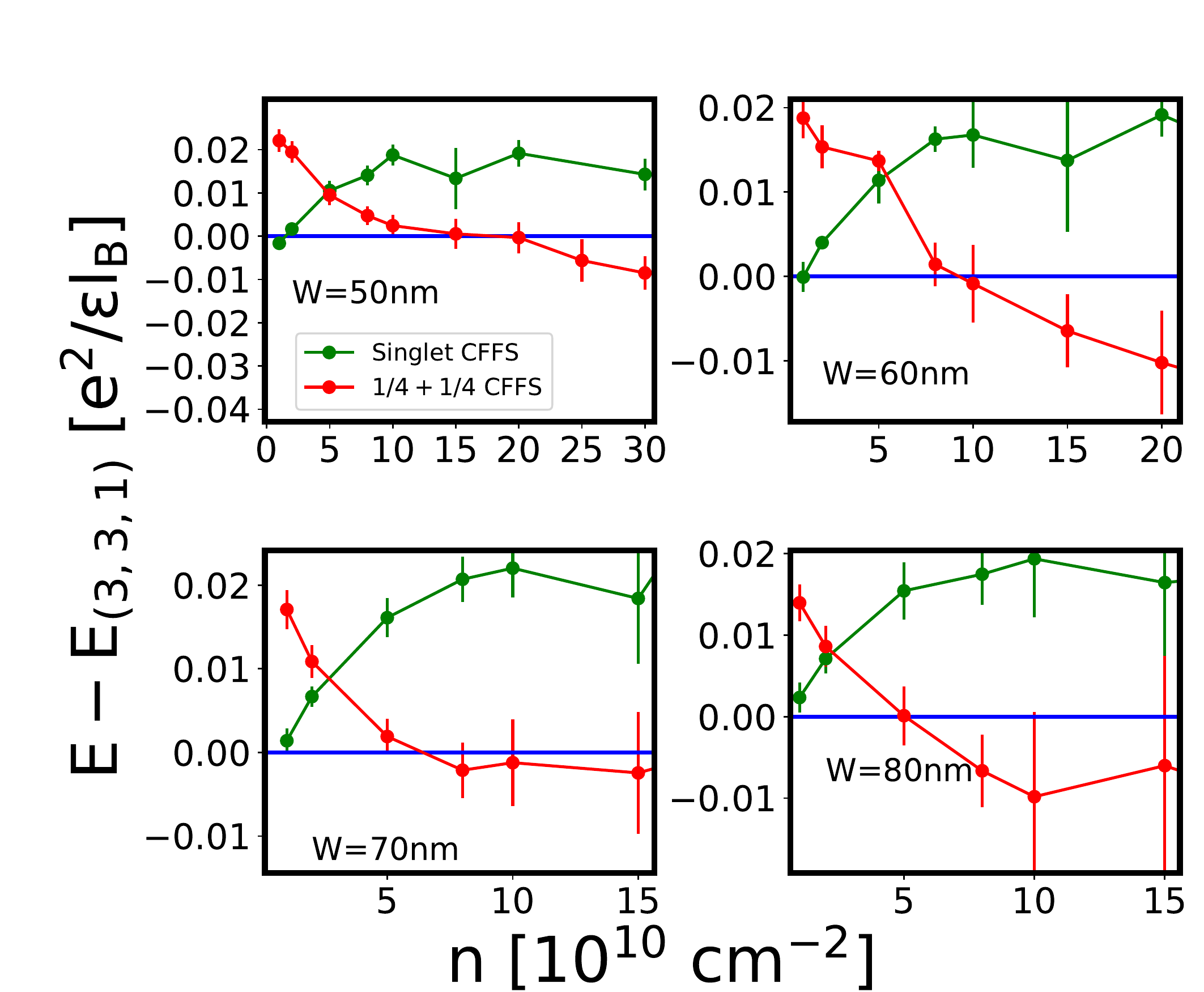}
		\caption{The energy of several two-component states relative to the Halperin $(3, 3, 1)$ state in the thermodynamic limit as functions of the carrier density at different well widths calculated by 3D-DMC in the torus geometry.}
		\label{3D_strict2L_energy}
	\end{figure}

\begin{figure}[H]
		\includegraphics[width=\columnwidth]{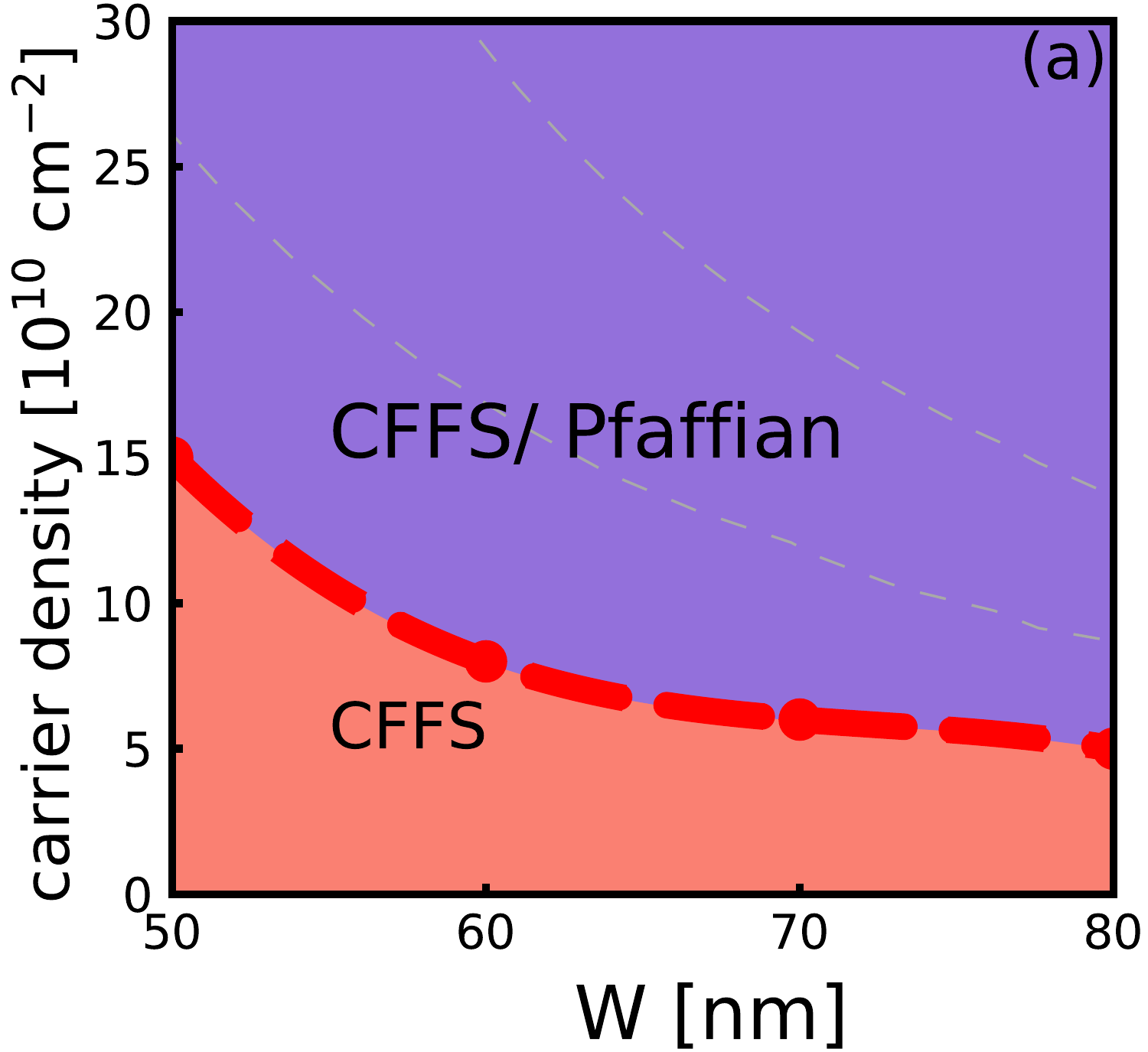}
		\includegraphics[width=\columnwidth]{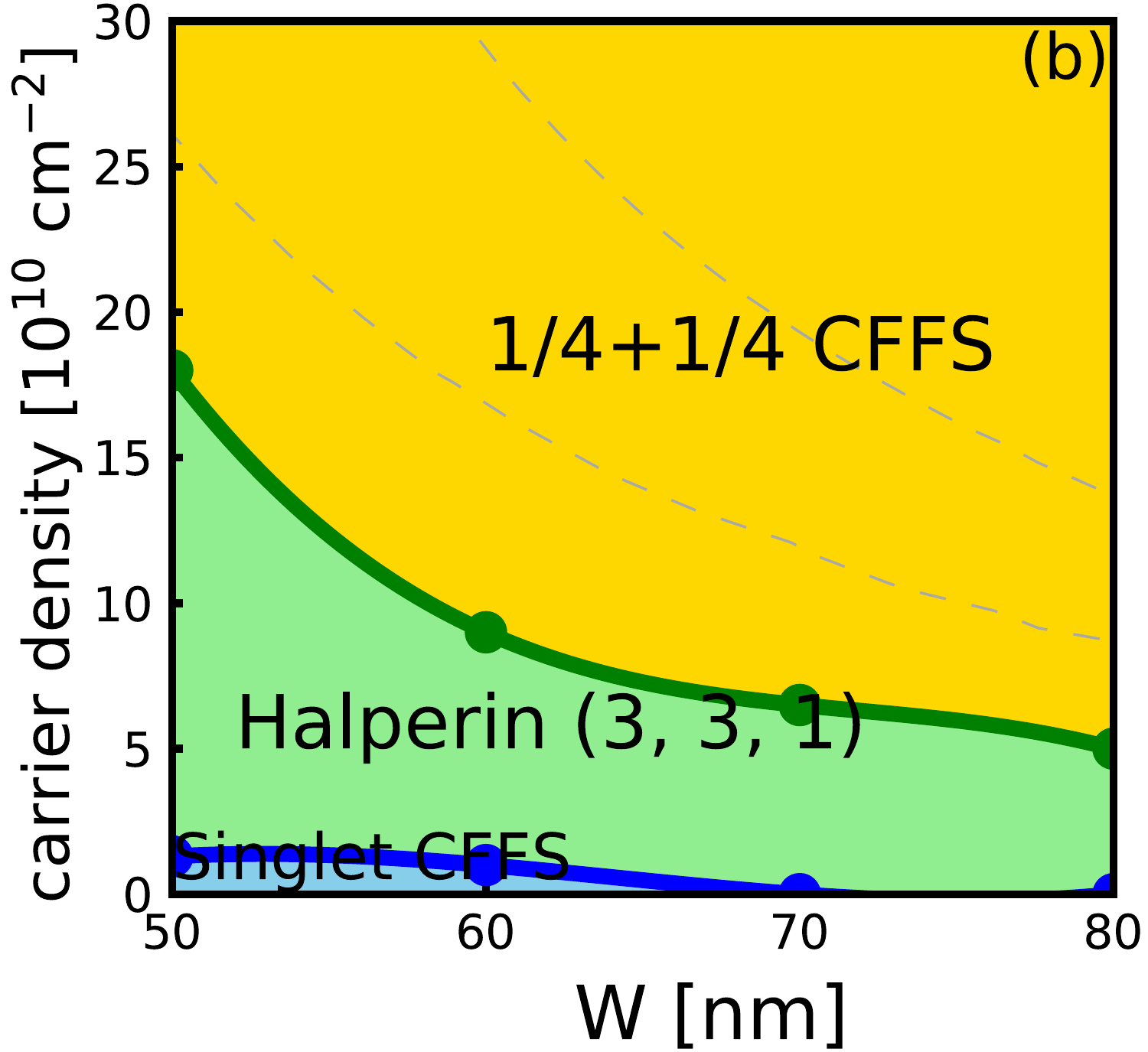}
		\caption{
			(a) The phase diagrams of one component states obtained by 3D-DMC on the toroidal geometry. Above the dashed boundary, the uncertainty is greater than the energy difference between the one-component CFFS and the Pfaffian state so we suggest it is either the one-component CFFS or the Pfaffian state (purple). (b) The phase diagrams of two-component states. The states considered are the CFFS state (red), the $(3, 3, 1)$ state (green), the singlet CFFS state (blue), and the $1/4+1/4$ CFFS state (yellow).In the lower panel, The uncertainty of the transition density at each width is about $2\times10^{10}\text{cm}^{-2}$. The region where experiments find an incompressible state~\cite{Shabani09b} is indicated by light dashed grey lines. The overall phase boundary is obtained by smoothly joining the transition points at $W=50,60,70,80$ nm.}
		\label{3D_PHASE_1}
	\end{figure}

	In this section, we show our calculation of the energy expectations of different states considered in this article. We first show in Fig.~\ref{3D_1L_energy} the energy comparison between the one-component CFFS and the  Pfaffian state. The energy of the Pfaffian state gets closer and closer to the CFFS as the carrier density increases for each well-width. In fact, their difference becomes so small that it is comparable to the statistical error and we are not able to determine which one is lower.  Within the two-component states, the energies are shown in Fig.~\ref{3D_strict2L_energy} and the theoretical phase diagram is shown in Fig.~\ref{3D_PHASE_1} (b). As the density increases, the system first makes a transition from the pseudo-spin singlet CFFS state into the Halperin $(3,3,1)$ state, and finally into the uncoupled $1/4+1/4$ state. This is qualitatively similar to the behavior found in the VMC calculation. The resulting phase diagrams for one- and two-component states (separately) are shown in Fig.~\ref{3D_PHASE_1} (a).

	\begin{figure}[H]
		\includegraphics[width=\columnwidth]{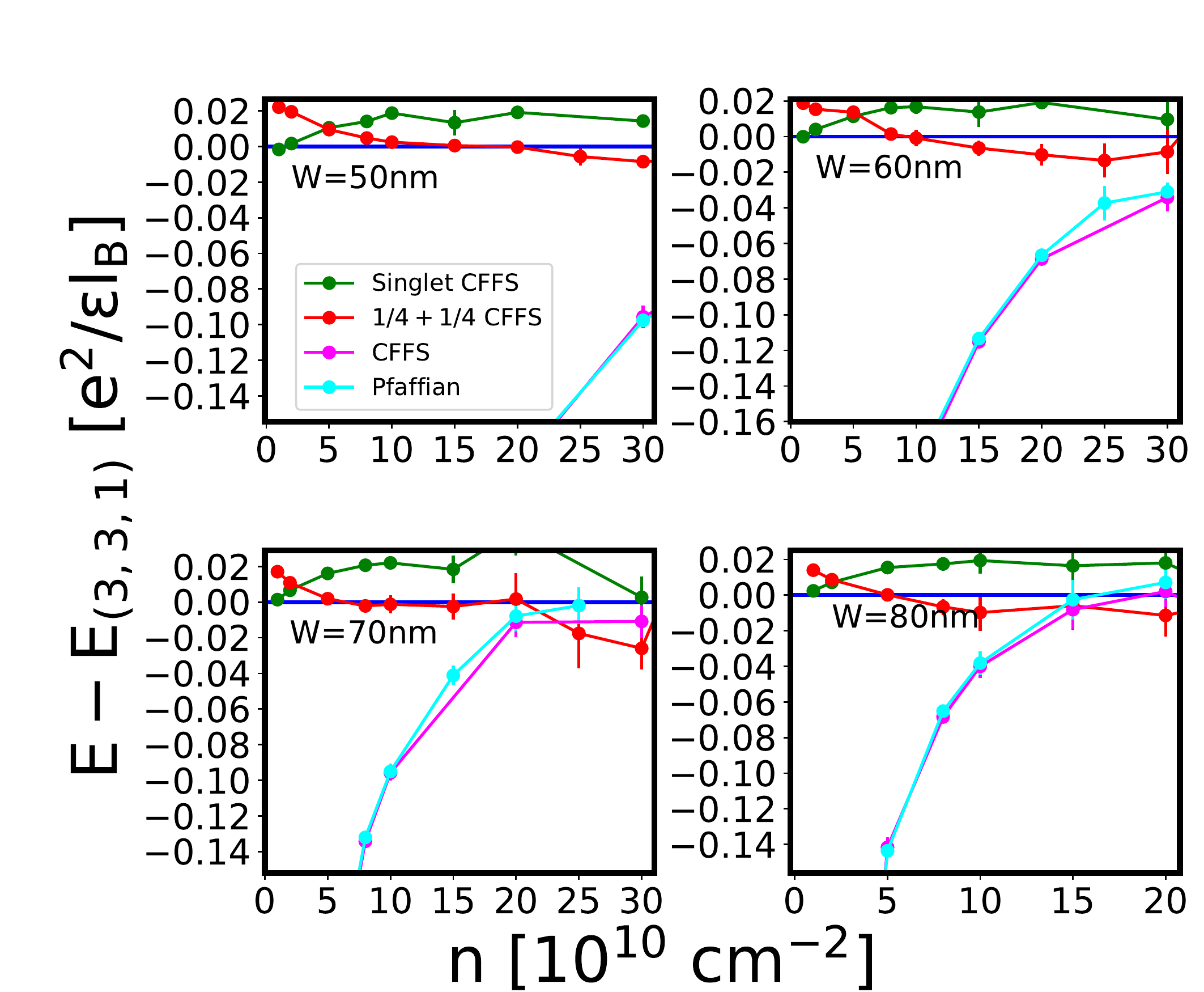}
		\caption{The energy of all the states calculated by 3D-DMC  on the torus in the thermodynamic limit.  The energy of each state is measured relative to that of the $(3,3,1)$ state.}
		\label{3D_1L_strict2L_energy}
	\end{figure}
	
	Because the 3D-DMC automatically includes $\Delta_{\rm SAS}$, we can directly compare all the states. Their energies are given in Fig.~\ref{3D_1L_strict2L_energy}, and the resulting phase diagram is shown in Fig.~\ref{3D_PHASE_2}. This phase diagram is different from that found by VMC or 2D-DMC and suggests that the experimentally observed incompressible state is likely to be the one-component Pfaffian state.

	\section{Discussion}

This work concerns the nature of the FQHE at $\nu=1/2$ in wide quantum wells. We have evaluated the phase diagram of states at $\nu=1/2$ as a function of the quantum well width and the carrier density at three different levels of approximation. 

Figure~\ref{VMC_PHASE_DIAGRAM_2} shows the phase diagram obtained by a variational Monte Carlo calculation. In this calculation, we evaluate an effective 2D interaction with the help of a transverse wave function calculated by LDA at zero magnetic field. A shortcoming of this method is the assumption that the transverse wave function and $\Delta_{\rm SAS}$ evaluated at zero magnetic field remain valid at finite magnetic fields as well. This is of particular concern for the phase boundaries separating one- and two-component states, because these phase boundaries depend sensitively on $\Delta_{\rm SAS}$, which is a relatively large energy, and also a rapidly varying function of the quantum well width and density. For that reason, in Fig.~\ref{VMC_PHASE_DIAGRAM_2} the phase boundary separating the $(3,3,1)$ and $1/4+1/4$ CFFS states is more reliable than that separating the  one-component CFFS and the $(3,3,1)$ states. 

The VMC calculation also does not include the effect of LLM directly.  Figure~\ref{2D_DMC_BOUNDARY_ALL} includes LLM for electrons interacting with an effective 2D interaction within a fixed phase DMC calculation. 

The principal result of the present work is given in Fig.~\ref{3D_PHASE_2}, which is the phase diagram obtained from a 3D fixed phase DMC. This method produces the ground state energy directly at a finite magnetic field, including, in principle, the effect of finite width and LLM. This suggests, although does not prove, that the incompressible state observed in experiments is the one-component Pfaffian state.

A technical difficulty of the 3D-DMC method is that for two-component states we must assume that the transverse wave function vanishes at the center of the quantum well. One may question if this affects comparisons between one- and two-component states. Fortunately, this is an excellent approximation near the upper phase boundary of Fig.~\ref{3D_PHASE_2}, which separates the single component ``CFFS/Pfaffian" state from the two-component ``$1/4+1/4$ CFFS" state. That gives us some degree of confidence that the transition from the two-component $1/4+1/4$ CFFS occurs into the one-component Pfaffian state. 
Nonetheless, a definitive confirmation must await further experimental studies. In particular, thermal Hall measurements, which have shown half-quantized value at $5/2$~\cite{Banerjee18}, can convincingly reveal whether the FQHE state here has a non-Abelian origin.

We also note that we do not consider the anti-Pfaffian state, which is the hole partner of the Pfaffian state~\cite{Levin07, Lee07, Balram18}. These two are degenerate in energy in the absence of LLM, but LLM is expected to select one of them. We have not investigated this issue here, both because the anti-Pfaffian is harder to deal with numerically, and because the energy differences are expected to be small compared to the Monte Carlo uncertainty.

Before ending we list other assumptions made in our study. We do not consider the crystal phase. Previous theoretical studies of possible states in an ideal bilayer~\cite{Faugno18, Faugno20} (i.e. two 2D layers separated by a distance $d$) did not find any crystal states, but a crystal may occur in wide quantum wells~\cite{Thiebaut15}. Such a crystal might be responsible for the fact that the experiments see an insulator on the either side of the FQHE state, rather than the compressible $1/4+1/4$ CFFS state. Of course, an alternative possibility is that disorder, omitted in our study, may turn the $1/4+1/4$ CFFS into an insulator.  Experimental studies in better quality samples can clarify the situation.

\textit{Acknowledgement}: We thank Mansour Shayegan for many insightful discussions. The work was made possible by financial support from the U.S. Department of Energy under Award No. DE-SC0005042. The VMC and DMC calculations were performed using Advanced CyberInfrastructure computational resources provided by The Institute for CyberScience at The Pennsylvania State University.  A. C. B.  acknowledges the Science and Engineering Research Board (SERB) of the Department of Science and Technology (DST) for funding support via the Start-up Grant SRG/2020/000154. Computational portions of exact diagonalization research work were conducted using the Nandadevi supercomputer, which is maintained and supported by the Institute of Mathematical Science's High-Performance Computing Center. Some of the numerical diagonalizations were performed using the DiagHam package, for which we are grateful to its authors.

	\begin{appendix}
		\counterwithin{figure}{section}
		\section{VMC results from the spherical geometry}
		\label{VMC_SPHERE_SEC}
		
\begin{figure}[H]
			\includegraphics[width=\columnwidth]{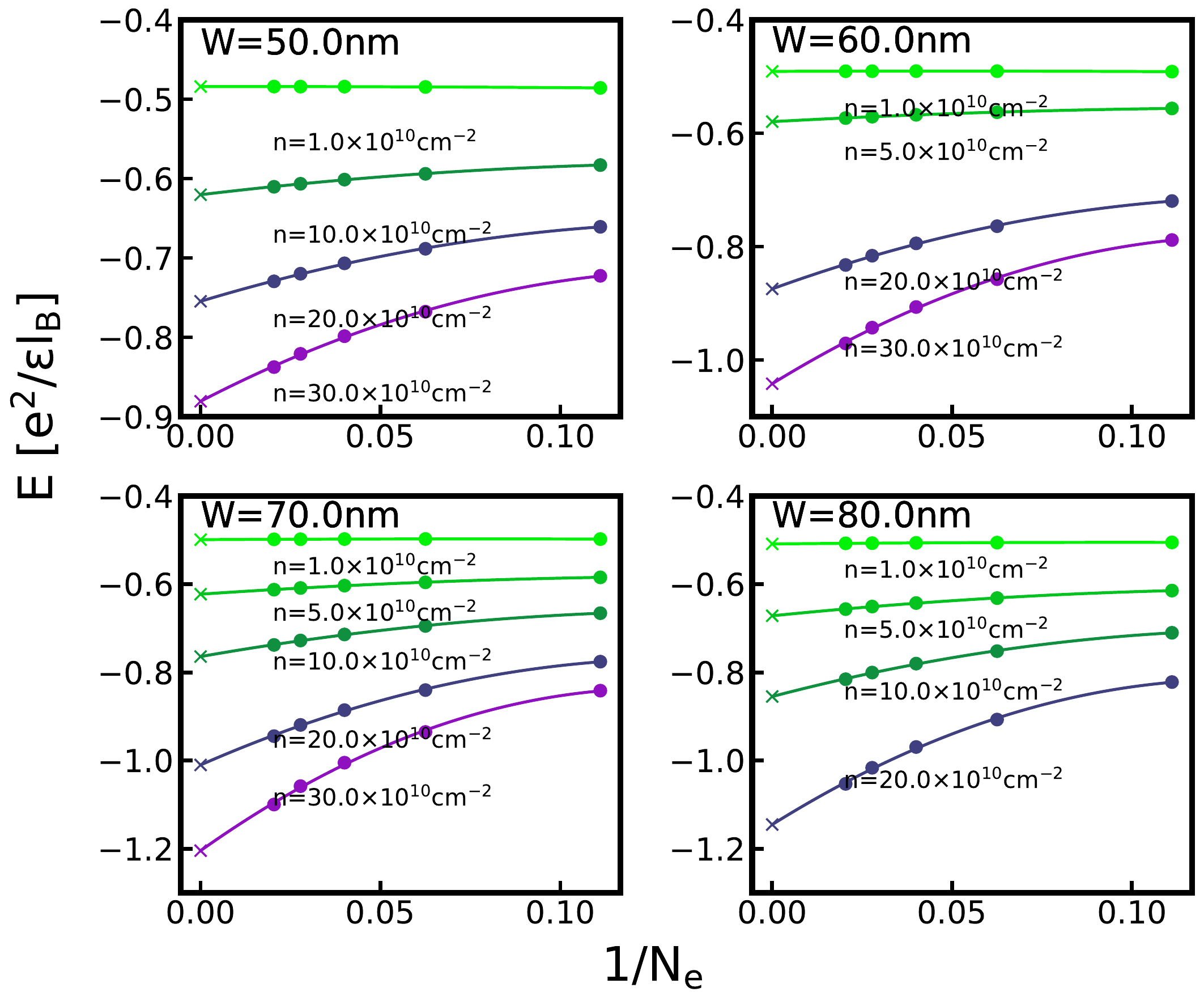}
			\caption{Finite-size extrapolation of the energy  for the one-component CFFS state for different widths and carrier densities. The calculation is done by VMC on the sphere. The well widths are shown on the plots.}
			\label{VMC_extrap_CFFS_sphere}
		\end{figure}

\begin{figure}[H]
			\includegraphics[width=\columnwidth]{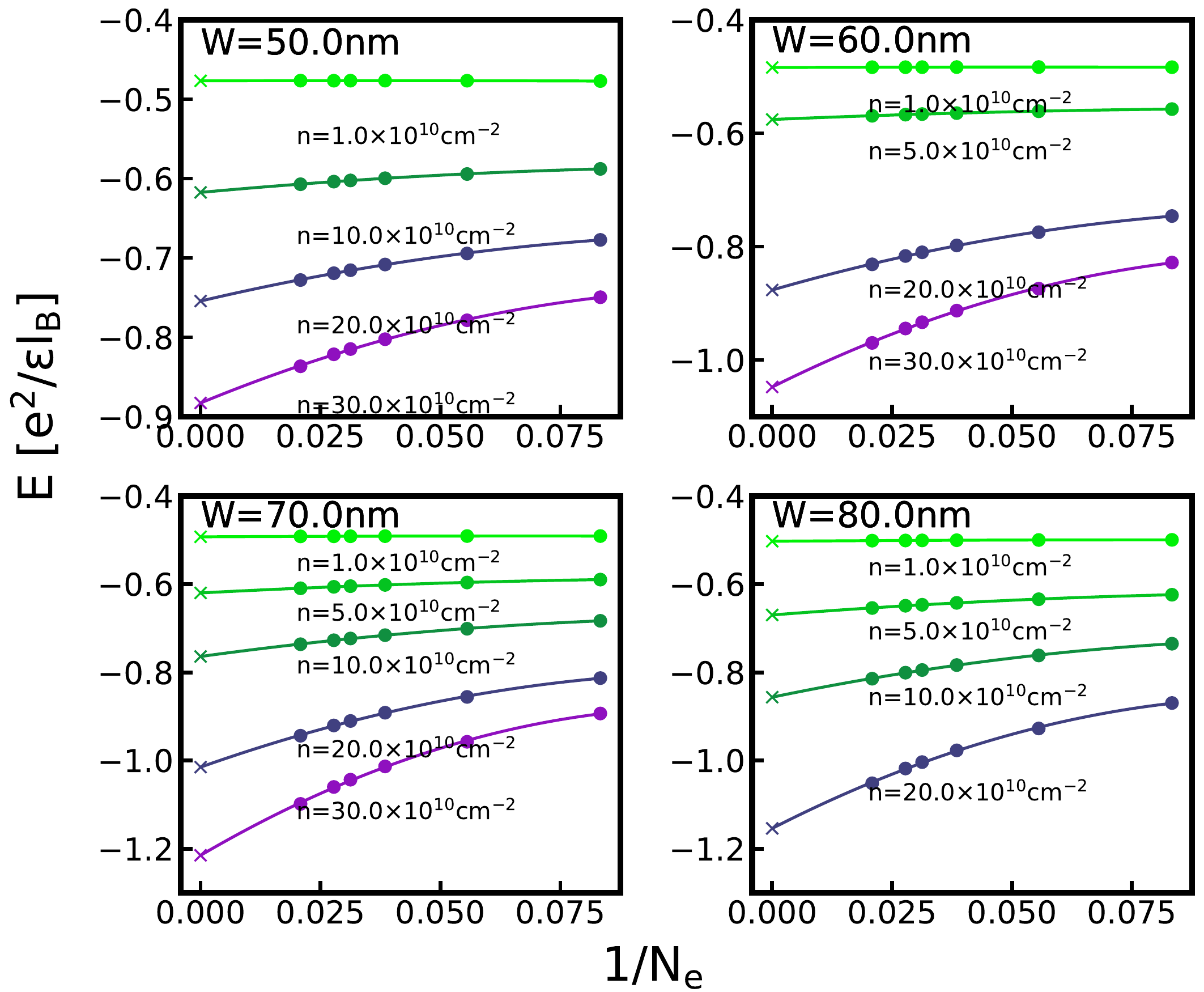}
			\caption{Finite-size extrapolation of the energy  for the Pfaffian state for different widths and carrier densities. The calculation is done by VMC on the sphere. The well widths are shown on the plots.}
			\label{VMC_extrap_Pfaffian_sphere}
		\end{figure}

\begin{figure}[H]
			\includegraphics[width=\columnwidth]{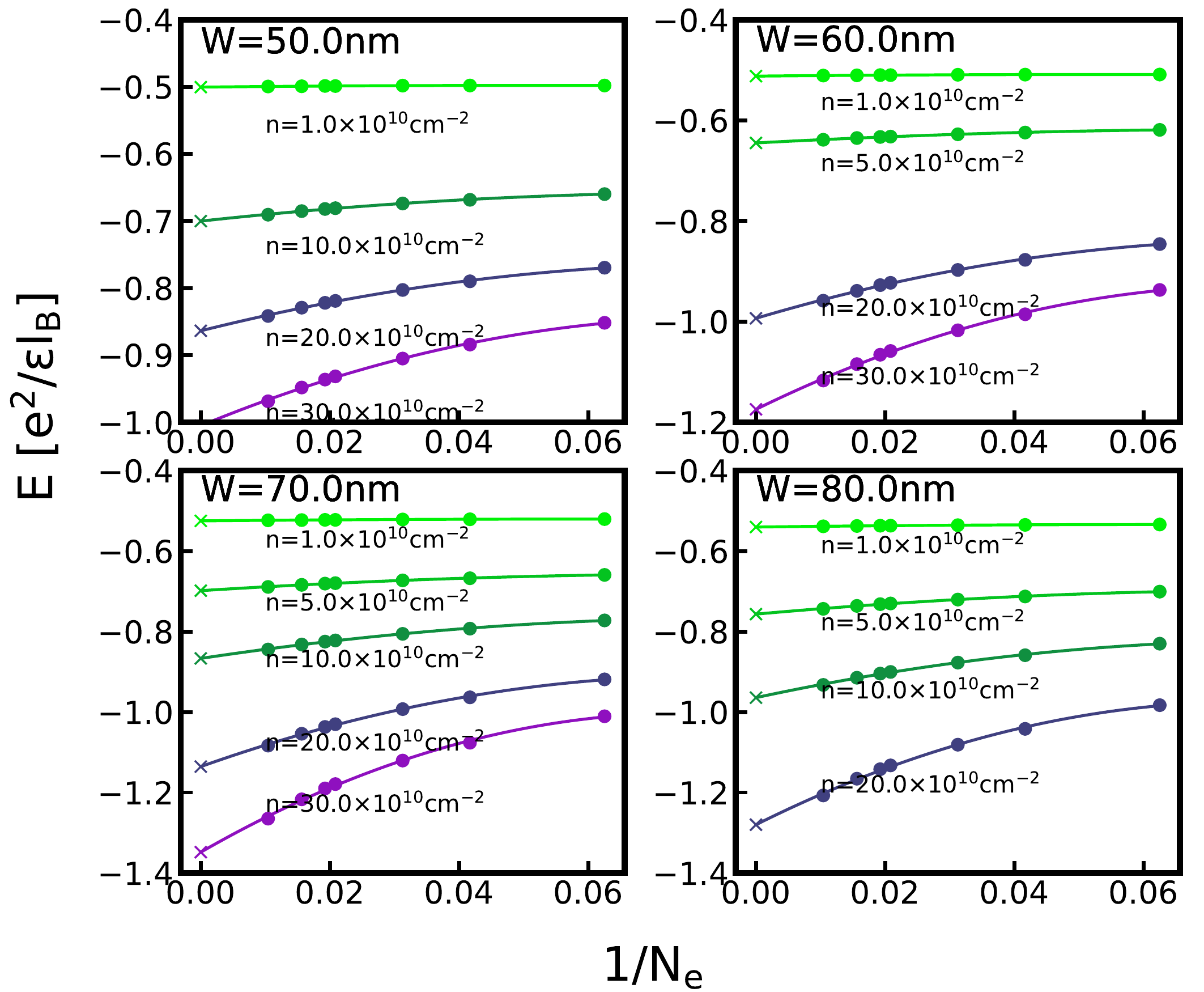}
			\caption{Finite-size extrapolation of the energy  for the $(3,3,1)$ state for different widths and carrier densities. The calculation is done by VMC on the sphere. The well widths are shown on the plots.}
			\label{VMC_extrap_331_sphere}
		\end{figure}

\begin{figure}[H]
			\includegraphics[width=\columnwidth]{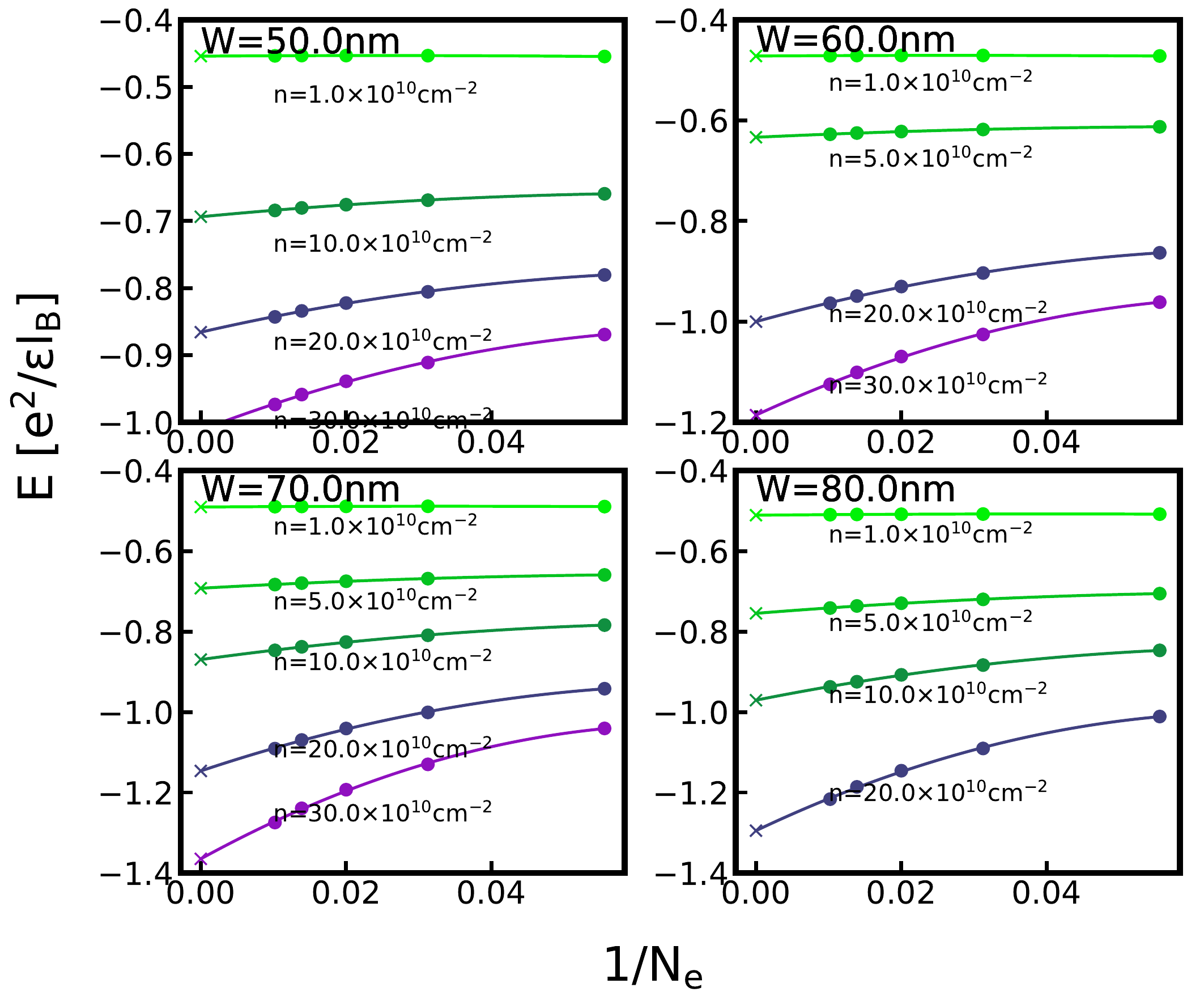}
			\caption{Finite-size extrapolation of the energy  for the $1/4+1/4$ CFFS state for different widths and carrier densities. The calculation is done by VMC on the sphere. The well widths are shown on the plots.}
			\label{VMC_extrap_2CFFS_sphere}
		\end{figure}

\begin{figure}[H]
			
			\includegraphics[width=\columnwidth]{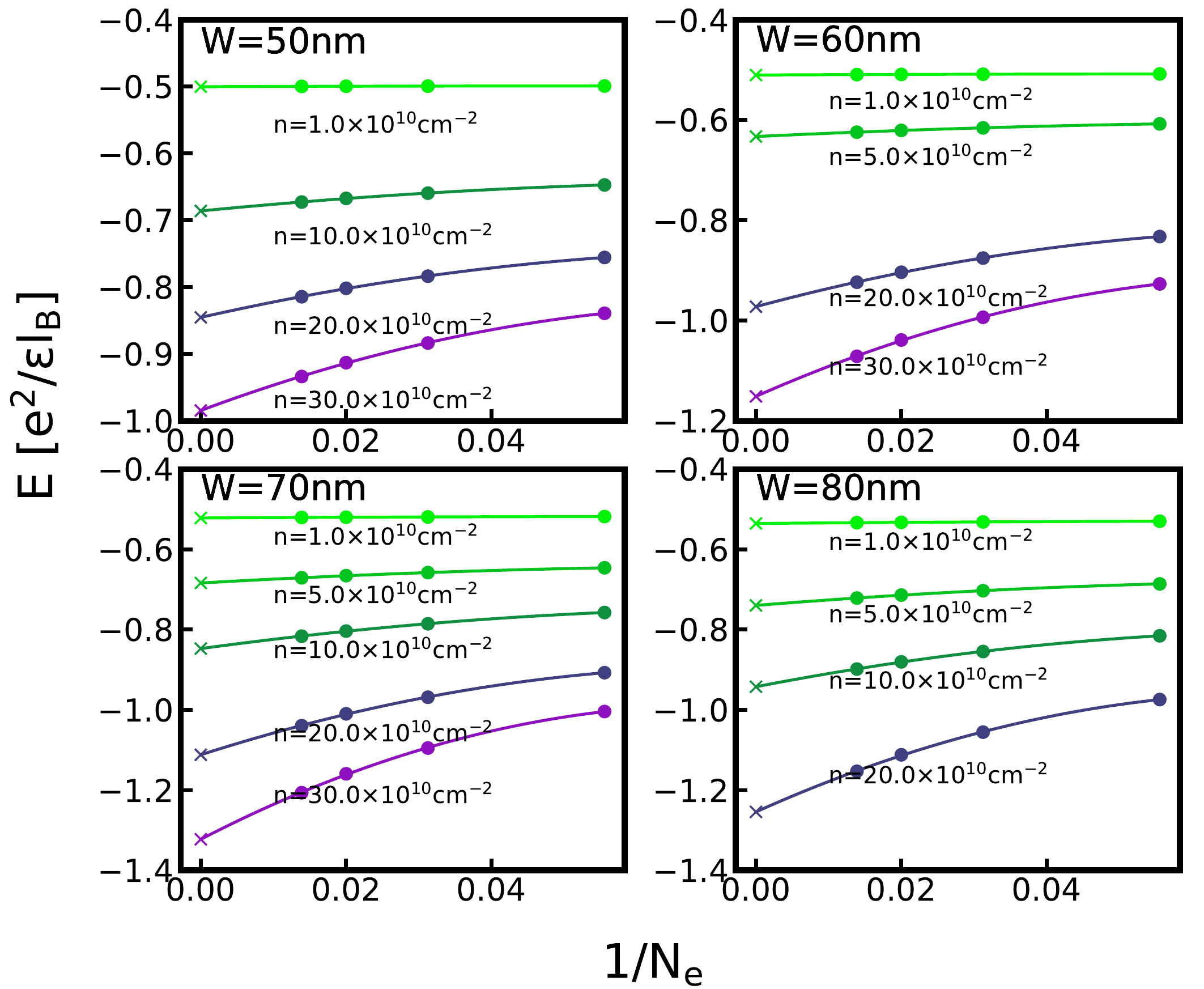}
			\caption{Finite-size extrapolation of the energy  for the pseudo-spin singlet CFFS state for different widths and carrier densities. The calculation is done by VMC on the sphere. The well widths are shown on the plots.}
			\label{VMC_extrap_singlet_sphere}
		\end{figure}

\begin{figure}[H]
			\includegraphics[width=0.95\columnwidth]{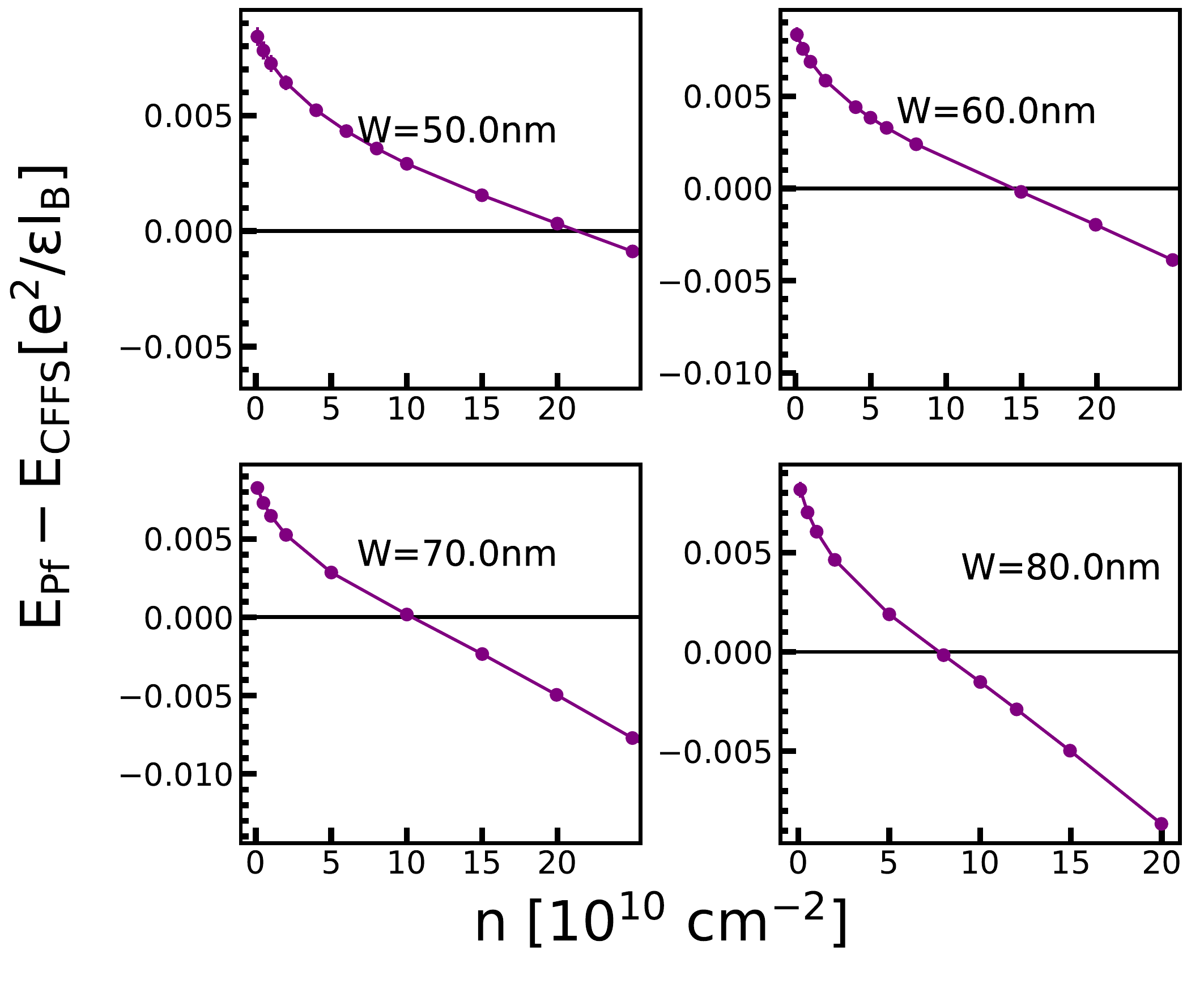}
			\includegraphics[width=0.95\columnwidth]{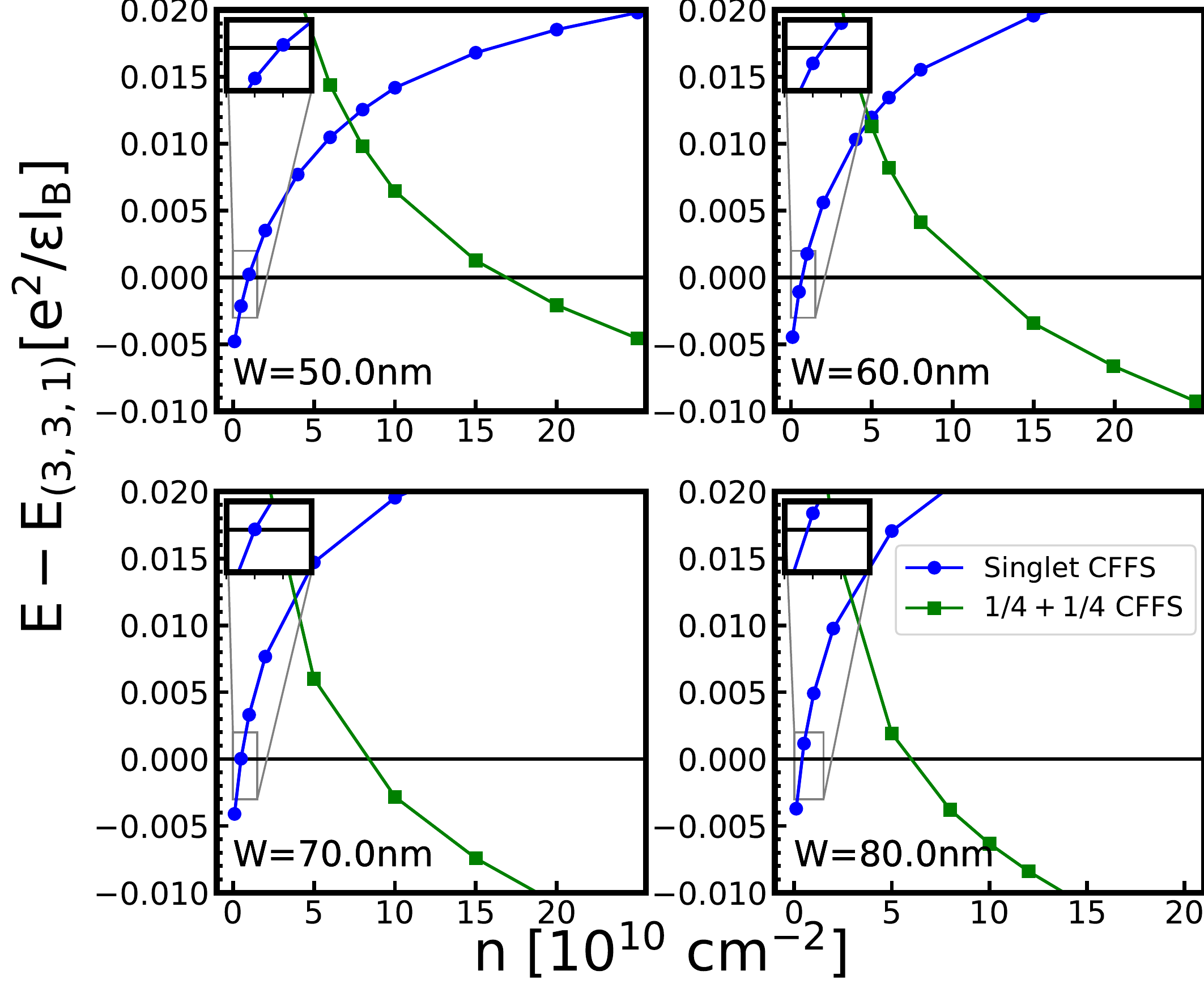}
			\caption{The VMC energies of different states relative to either the CFFS state  or the $(3,3,1)$ state, as labeled in each figure. All energies are thermodynamic limits evaluated on the sphere. The statistical errors are smaller than the symbol sizes.}
			\label{sphere_VMC_FW_1}
		\end{figure}

\begin{figure}[H]
			\includegraphics[width=0.95\columnwidth]{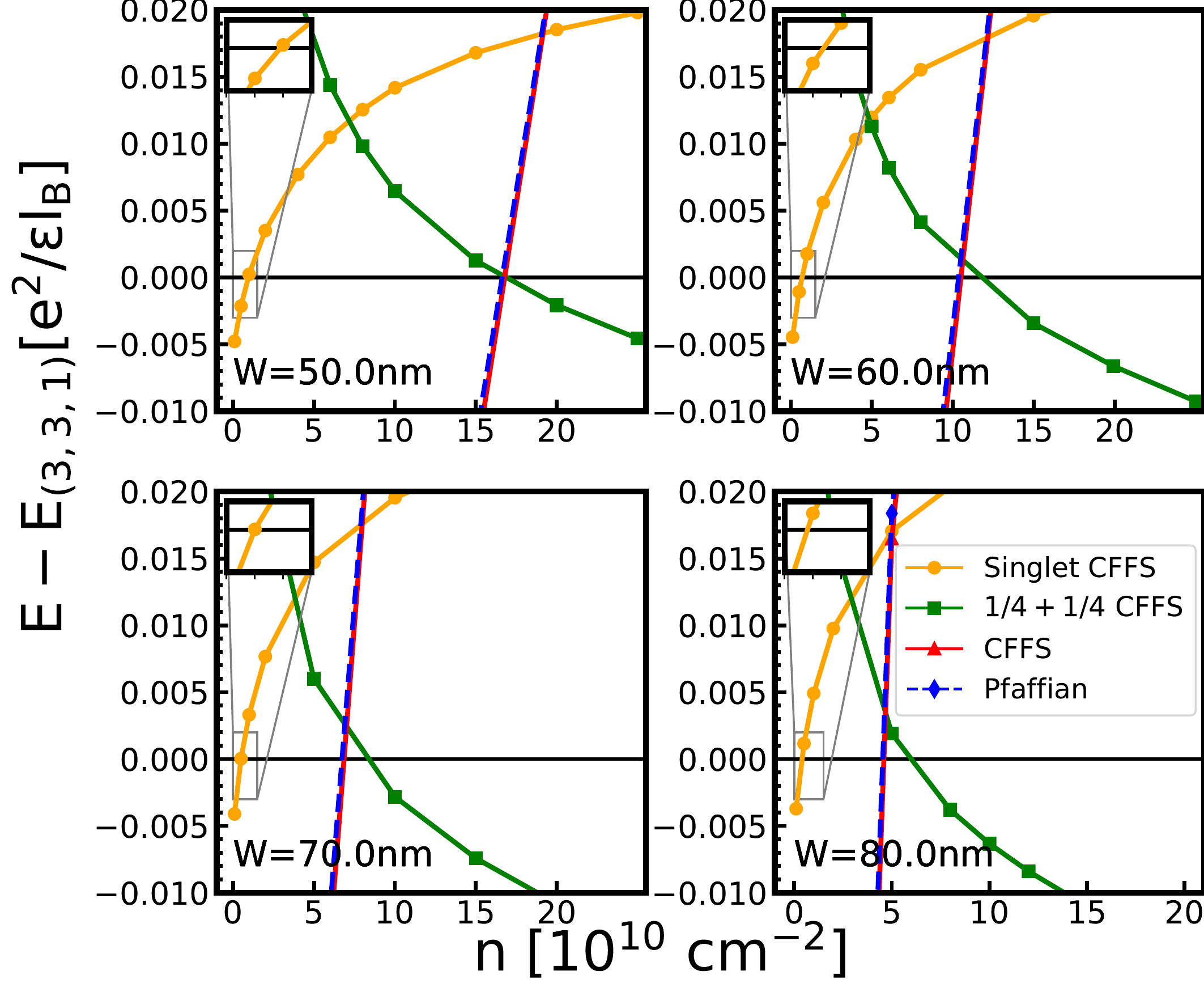}
			\caption{The VMC energies of all states relative to the $(3,3,1)$ state. All energies are evaluated on the sphere, and represent the thermodynamic limit. An offset of $\frac{1}{2}\Delta_{\rm SAS}$ per particle is included for the two-component states. The statistical errors are smaller than the symbol sizes.}
			\label{sphere_VMC_FW_2}
		\end{figure}

\begin{figure}[H]
			\includegraphics[width=\columnwidth]{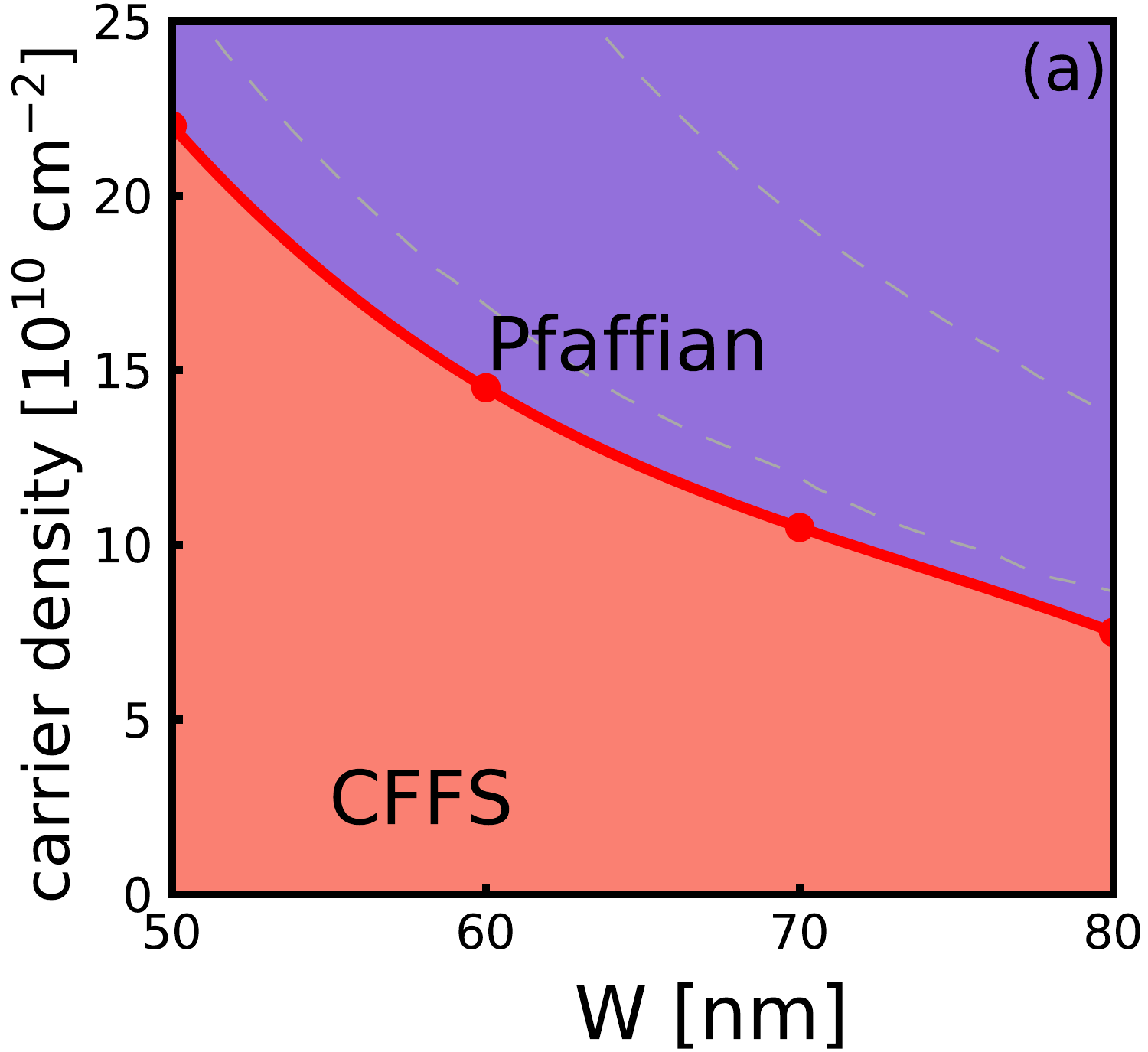}
			\includegraphics[width=\columnwidth]{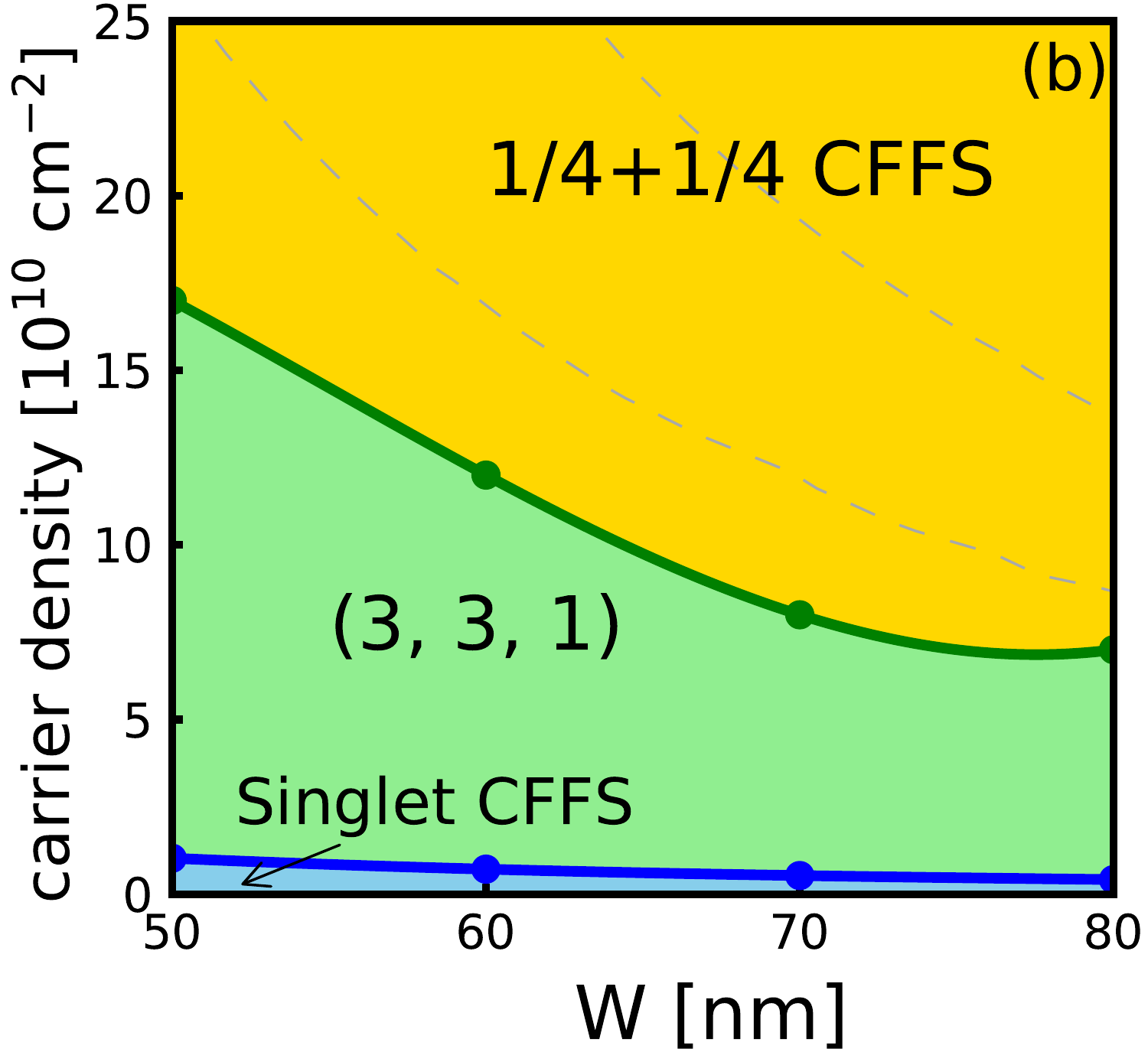}
			\caption{(a) The phase diagram of one-component states. (b) The phase diagram of two component states. The phase boundaries are obtained from VMC calculation in the spherical geometry. 
				The region where experiments find an incompressible state~\cite{Shabani09b} is indicated by light dashed grey lines. The uncertainty in the density of the transition point is approximately $1\times 10^{10} \text{cm}^{-2}$}
			\label{sphere_VMC_PT_1}
		\end{figure}

\begin{figure}[H]
			\includegraphics[width=0.95\columnwidth]{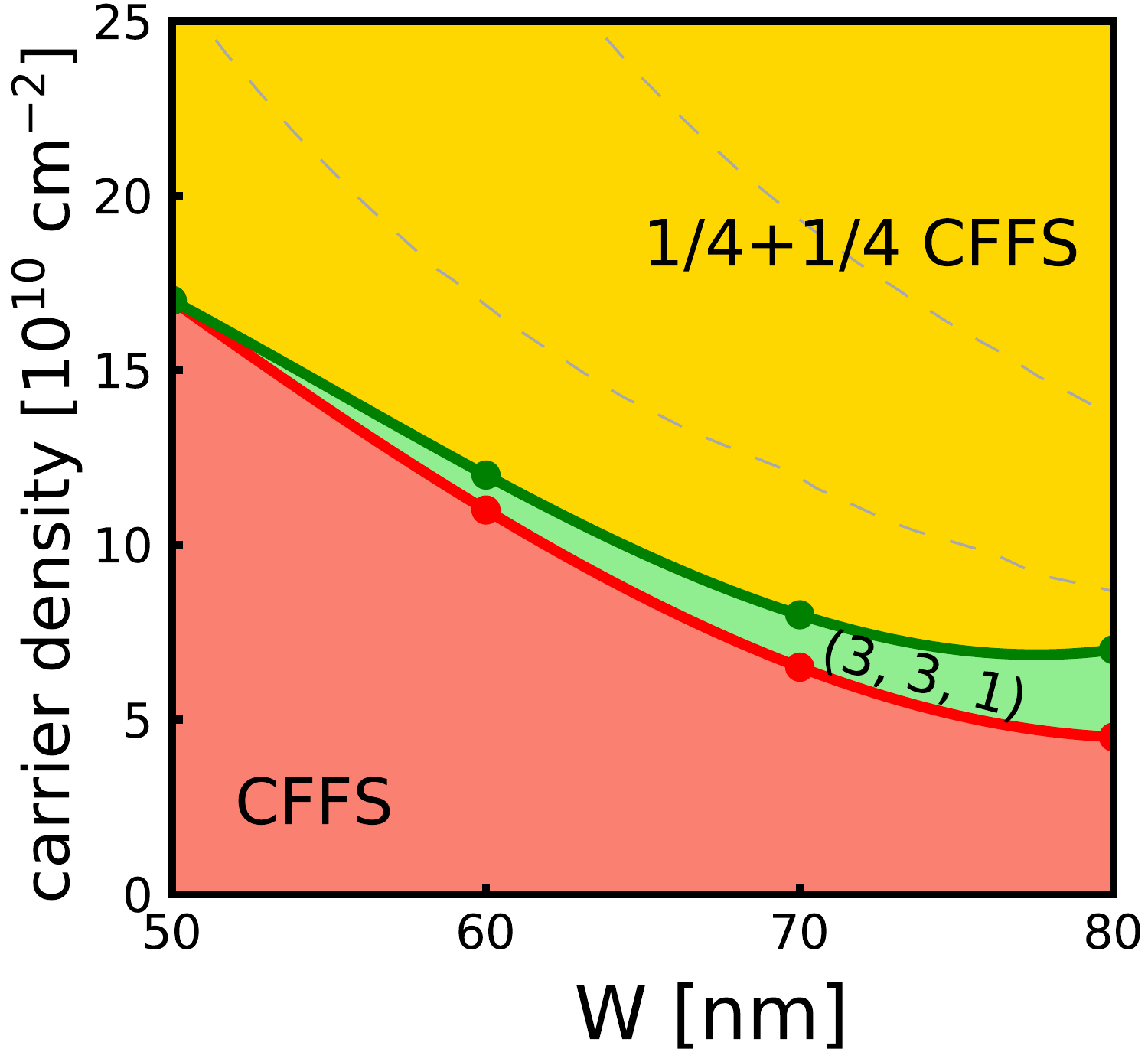}
			\caption{The full phase diagram of states at half filling, calculated by VMC method on the sphere. The uncertainty in the density at the transition point is approximately $1\times 10^{10} \text{cm}^{-2}$. The region where experiments find an incompressible state~\cite{Shabani09b} is indicated by light dashed grey lines.}
			\label{sphere_VMC_PT_2}
		\end{figure}

		All the above calculations have been performed in the torus geometry. In this section, we present results from our VMC calculations in the spherical geometry. The energy extrapolations are shown in Fig.~\ref{VMC_extrap_CFFS_sphere} for the one-component CFFS, Fig.~\ref{VMC_extrap_Pfaffian_sphere} for the Pfaffian state, Fig.~\ref{VMC_extrap_331_sphere} for the $(3,3,1)$ state, Fig.~\ref{VMC_extrap_2CFFS_sphere} for the $1/4+1/4$ CFFS and \ref{VMC_extrap_singlet_sphere} for the single component CFFS. Figs.~\ref{sphere_VMC_FW_1} and \ref{sphere_VMC_FW_2} depict the energies as a function of density for several quantum well widths. The resulting phase diagrams within the one-component and the two-component regimes are shown in Figs.~\ref{sphere_VMC_PT_1}. While the phase diagram of two-component states is almost identical to that on the torus, the phase diagram of one-component states is different: in particular, a phase transition occurs from the one-component CFFS to the Pfaffian state at sufficiently large densities. The final phase diagram shown in Fig.~\ref{sphere_VMC_PT_2} is similar to but slightly different from, the VMC phase diagram obtained from the torus geometry, shown in the main text. 
		
		We believe that the results from the torus geometry are more reliable for the following reasons. (i) As one can see, the thermodynamic extrapolations in the spherical geometry are not as linear as in the torus geometry, and thus entail greater uncertainty in the thermodynamic limit. This is because the finite width effect is only considered in the calculation of the electron-electron repulsion, whereas the electron-background and background-background interactions are chosen to be the same as those for zero-width well, for the simplicity of the calculation. (The form of the background-background interactions in the spherical geometry can be found in the appendix of Ref.~[\onlinecite{Jain07}], while in the torus geometry, the electron-background and background-background interactions are included through Ewald summation which assumes the same form for all interactions.) (ii) The torus geometry is better for the CFFS states. While one can directly construct the CFFS on the torus by attaching flux quanta to electron Fermi sea for any particle number, one must work with the Jain states of filling factor $\nu=\frac{n}{2n+1}$ and take the limit $n\to \infty$ to obtain the energy of the CFFS. Alternatively, one can consider systems with zero effective flux and take the thermodynamic limit~\cite{Rezayi94,Balram15b,Balram17}. The filled shell systems occur at particle numbers $N_e=4,9,16,25,36, ...$. However, due to the complexity of the wave functions, we cannot go beyond $N_e=36$ in VMC. This size limitation makes the energy comparisons less reliable. (iv) Finally, when it comes to DMC, very few CFFS systems are accessible in the spherical geometry, making thermodynamic extrapolations even more unreliable. For that reason, we have not performed DMC calculations in the spherical geometry.

		
		\section{Jacobi $\theta$ function and its periodicity}\label{theta_function_definition}
		
		Here we list the definition and properties of the Jacobi $\theta$ function, following the conventions in the text book by David Mumford\cite{Mumford07}. In general, the $\theta$ function is defined as 
		\begin{equation}
		\begin{aligned}
		&\theta_{a,b}(z|\tau)\\
		&=\sum_{n=-\infty}^{+\infty}\exp\left[\pi i(n+a)^2 \tau+2\pi i(n+a)(z+b)\right],
		\end{aligned}
		\end{equation}
		which satisfies the periodicity properties: 
		\begin{equation}
		\begin{aligned}
		&\theta_{a,b}(z+1|\tau)=e^{2\pi a i}\theta_{a,b}(z|\tau)\\
		\end{aligned}
		\end{equation}
		and
		\begin{equation}
		\begin{aligned}
		&\theta_{a,b}(z+\tau|\tau)\\
		&=\exp\left[-\pi i \tau-2\pi i(z+b)\right]\theta_{a,b}(z|\tau)
		\end{aligned}
		\end{equation}
		For simplicity of notation, we have dropped the subscripts and defined $\theta_{1/2, 1/2}\left(z|\tau\right)=\theta\left(z|\tau\right)$ in the main text. 
		The other three Jacobi theta functions for the Pfaffian states on a torus are defined as follows:
		\begin{equation}
		\begin{aligned}
		\theta_2\left(z|\tau\right)&=\theta\left(z+1/2|\tau\right)\\
		\theta_3\left(z|\tau\right)&=e^{i\pi \tau/4}e^{i\pi z}\theta\left(z+1/2+i/2|\tau\right)\\
		\theta_4\left(z|\tau\right)&=e^{i\pi \tau/4}e^{i\pi z}\theta\left(z+i/2|\tau\right)\\
		\end{aligned}
		\end{equation}
		
		\section{Quasi-degeneracy of the Pfaffian state on the torus}
		\label{PF_DEGENERACY}
		
\begin{figure}[H]
			\includegraphics[width=\columnwidth]{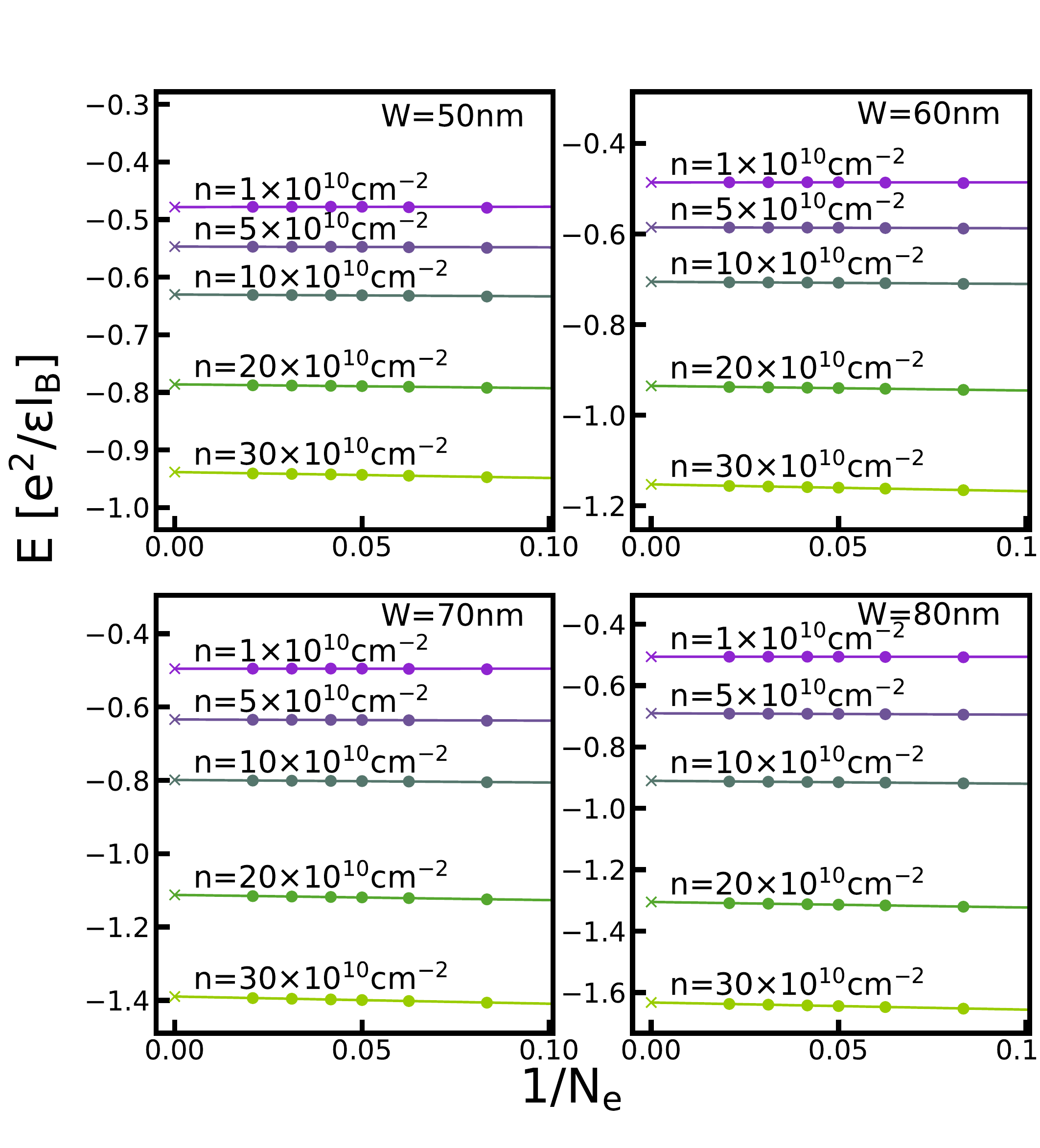}
			\caption{The energy of the Pfaffian (2) state [in which $\theta_a(z)$ is chosen to be $\theta_3(z)$] as a function of $1/N_e$. Each energy is obtained by VMC method with the effective interaction defined in Eq.\,\ref{V_eff_explicit}.}
			\label{Pf_degeneracy_1}
		\end{figure}

\begin{figure}[H]
			\includegraphics[width=\columnwidth]{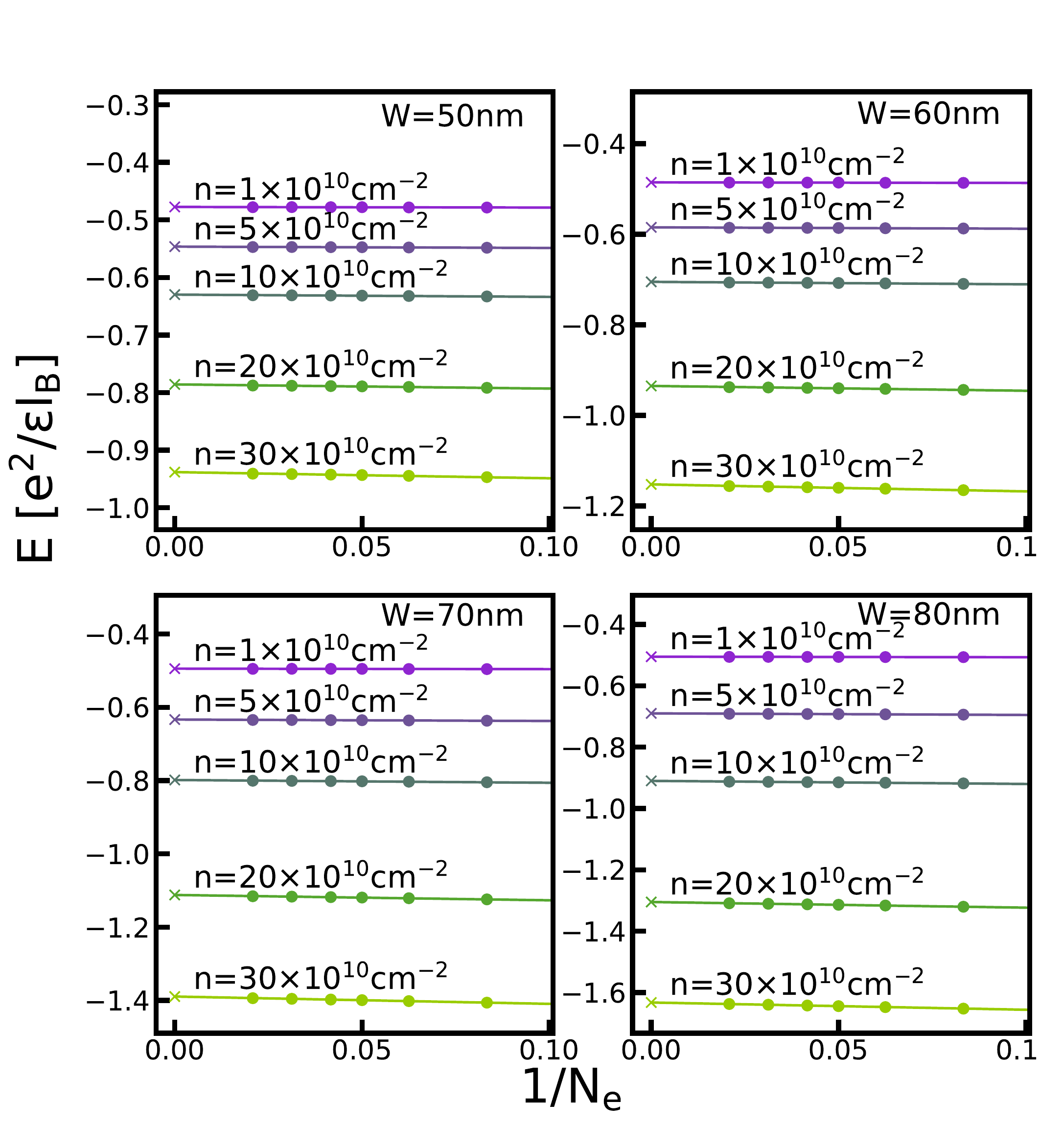}
			\caption{The energy of the Pfaffian (3) state [in which $\theta_a(z)$ is chosen to be $\theta_4(z)$] as a function of $1/N_e$. Each energy is obtained by VMC method with the effective interaction defined in Eq.\,\ref{V_eff_explicit}.}
			\label{Pf_degeneracy_2}
		\end{figure}

\begin{figure}[H]
			\includegraphics[width=\columnwidth]{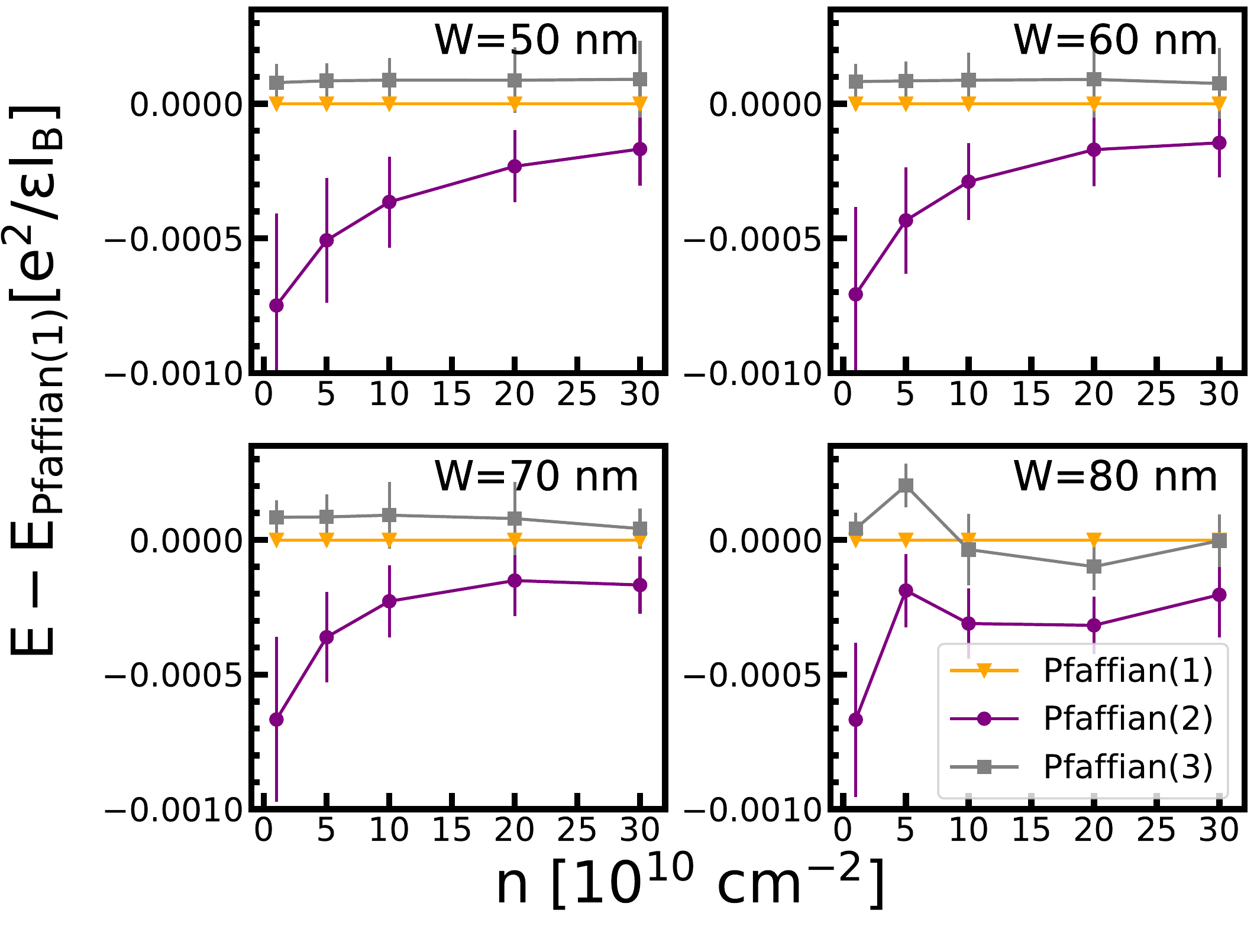}
			\caption{Comparison of the VMC thermodynamic energies of the three Pfaffian states.}
			\label{Pf_degeneracy_3}			
		\end{figure}

		We have given in Eq.\,\ref{Pfaffian_wfn} the explicit form for three Pfaffian wave functions, called Pfaffian (1), Pfaffian (2), and Pfaffian (3), which correspond to the choices $a=2$, 3 and 4, respectively. These are not related by CM translation, and as a result, have different Coulomb energy expectation values for finite systems. We have calculated the thermodynamic limits for the energies of these three states by the VMC method. We present the extrapolations of the VMC energies of the Pfaffian (2) and Pfaffian (3) in Figs.~\ref{Pf_degeneracy_1} and \ref{Pf_degeneracy_2} for various quantum well widths and densities. We compare these energies with the energy of the Pfaffian (1) (Fig.~\ref{VMC_EXTRAP_PFAF}) in Fig.~\ref{Pf_degeneracy_3}. At the lowest density of $n=10^{10} \text{cm}^{-2}$, the energy differences can be on the order of $\sim 0.008\pm0.003 e^2/\epsilon l_B$, which is approximately $1/4$ of the energy difference between the one-component CFFS and Pfaffian (1).  As the carrier density increases, the differences between the various Pfaffian wave functions quickly drop to $\sim 0.0001 e^2/\epsilon l_B$.  (Peterson{\it et al.}\cite{Peterson08} have also found similar behavior as a function of the well-width in their ED studies.)  Around transition densities in experiments, the difference is smaller than the uncertainty of either 2D-DMC or 3D-DMC, which is generally of the order of $0.001 e^2/\epsilon l_B$. Due to this fact, we conclude that at least in this work the choice on the $\theta_a(z)$ in the Pfaffian does not affect our result, and we have used Pfaffian (1) with $a=2$ in our calculations.

		\section{Exact diagonalization studies for the LDA interaction}
		\label{ED_Ajit}
		
		In the main article, we found that, within the single component states, VMC with the LDA interaction (without LLM) supports the CFFS state in the entire parameter range considered. In this section, we present results obtained from exact diagonalization for the LDA interaction.  Before we go on to the states at $\nu=1/2$, we first show that the $1/3$ Laughlin, and the $2/5$ and $3/7$ Jain states are robust to the effects of finite-width and density changes in the LLL. Using the pseudopotentials of the interaction obtained from the finite-width LDA discussed above [with parameters $W=18-70 \text{nm}$ and $n=1\times 10^{10}-30\times 10^{10}$ cm$^{-2}$], we obtain the exact ground states at $1/3$, $2/5$, and $3/7$ in the LLL at the $1/3$ Laughlin, $2/5$ Jain, and $3/7$ Jain fluxes, respectively.  All our calculations are carried out for a system of $N_e=12$ electrons which is the largest system for which the $2/5$ and $3/7$ Jain states (obtained by a brute-force projection to the LLL) have been constructed in the Fock space~\cite{Yang19a}.
		
		We also evaluate the charge and neutral gaps for the same system of $N_e=12$ electrons using exact diagonalization.  The neutral gap is defined as the difference in energies of the two lowest-lying states of the system of $N$ electrons at the flux $2Q_{gs}$ corresponding to the incompressible ground state.  The charge gap is defined as $\Delta_{c} = [E(2Q_{gs}+1)+E(2Q_{gs}-1)-2E(2Q_{gs})]/n_{q}$, where $E(2Q)$ is the background-subtracted~\cite{Balram20} ground-state energy of $N_e$ electrons at flux $2Q$, and $n_{q}$ is the number of quasiparticles (quasiholes) created by the removal (insertion) of a single flux quantum in the ground state.  The charge gap measures the energy required to create a far-separated quasiparticle-quasihole pair in the ground-state. The value of $n_{q}$ is one, two, and three, for the $1/3$ Laughlin, $2/5$, and $3/7$ Jain states respectively.  
		
\begin{figure*}[htpb]
			\begin{center}			
				\includegraphics[width=0.32\textwidth]{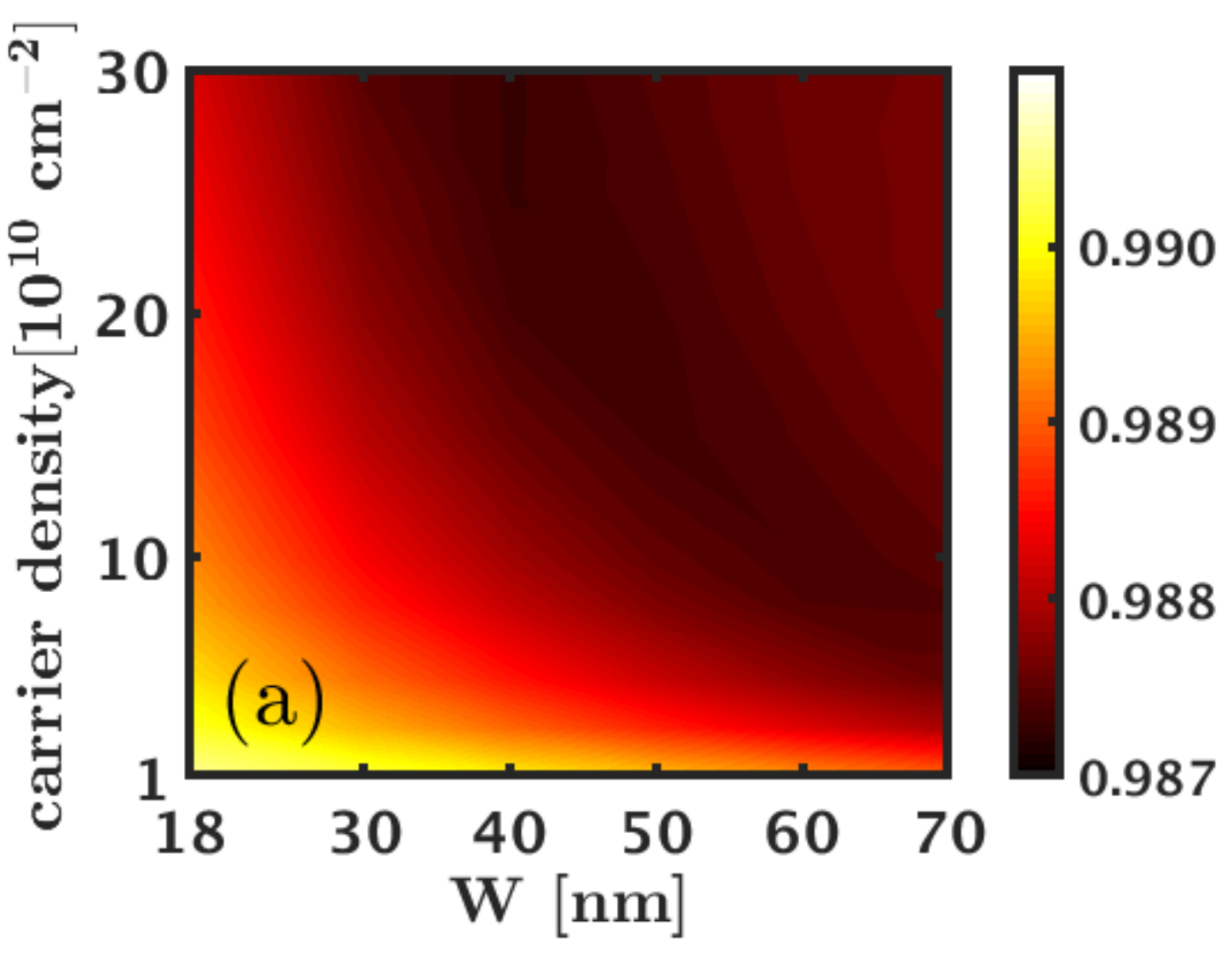}
				\includegraphics[width=0.32\textwidth]{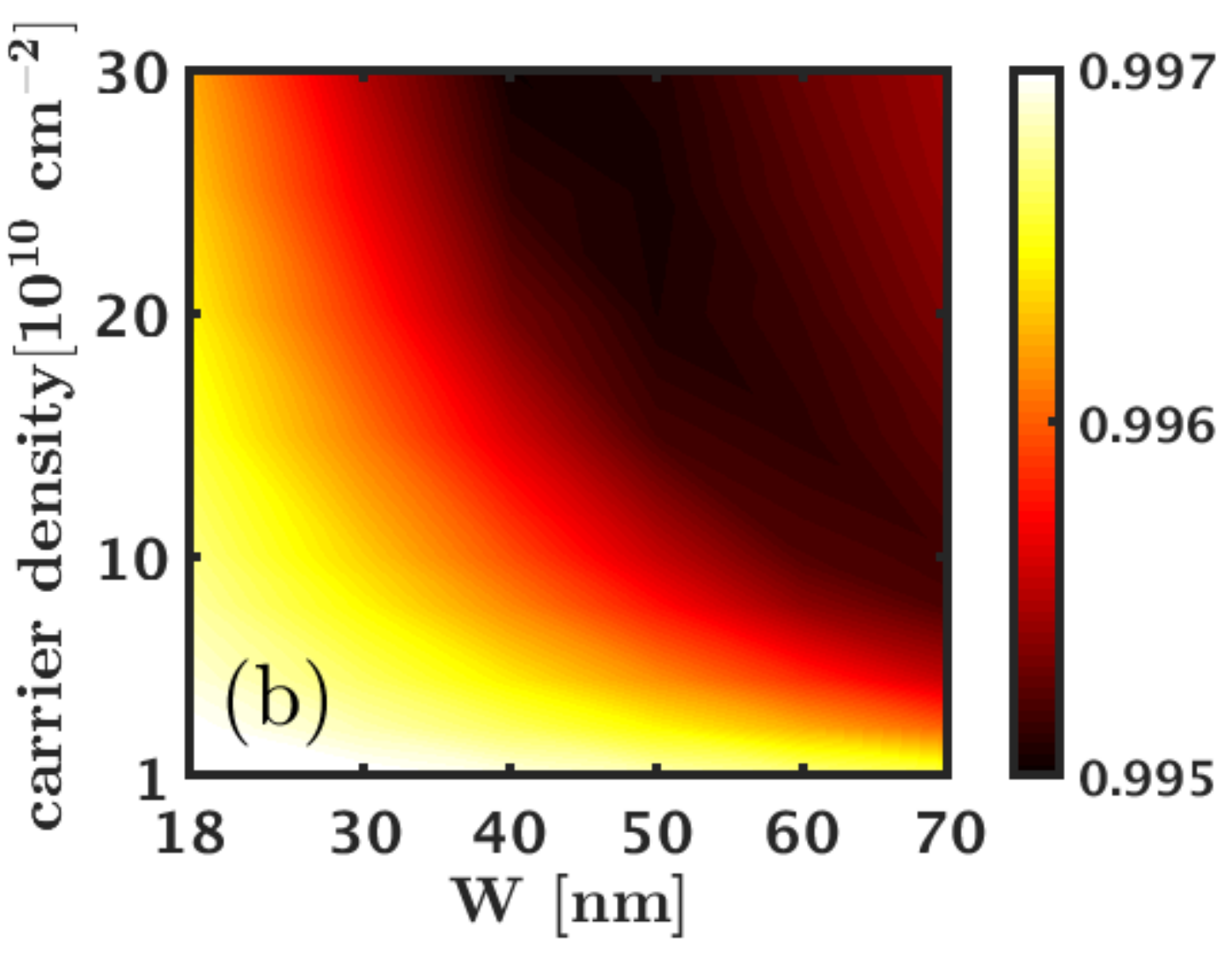}
				\includegraphics[width=0.32\textwidth]{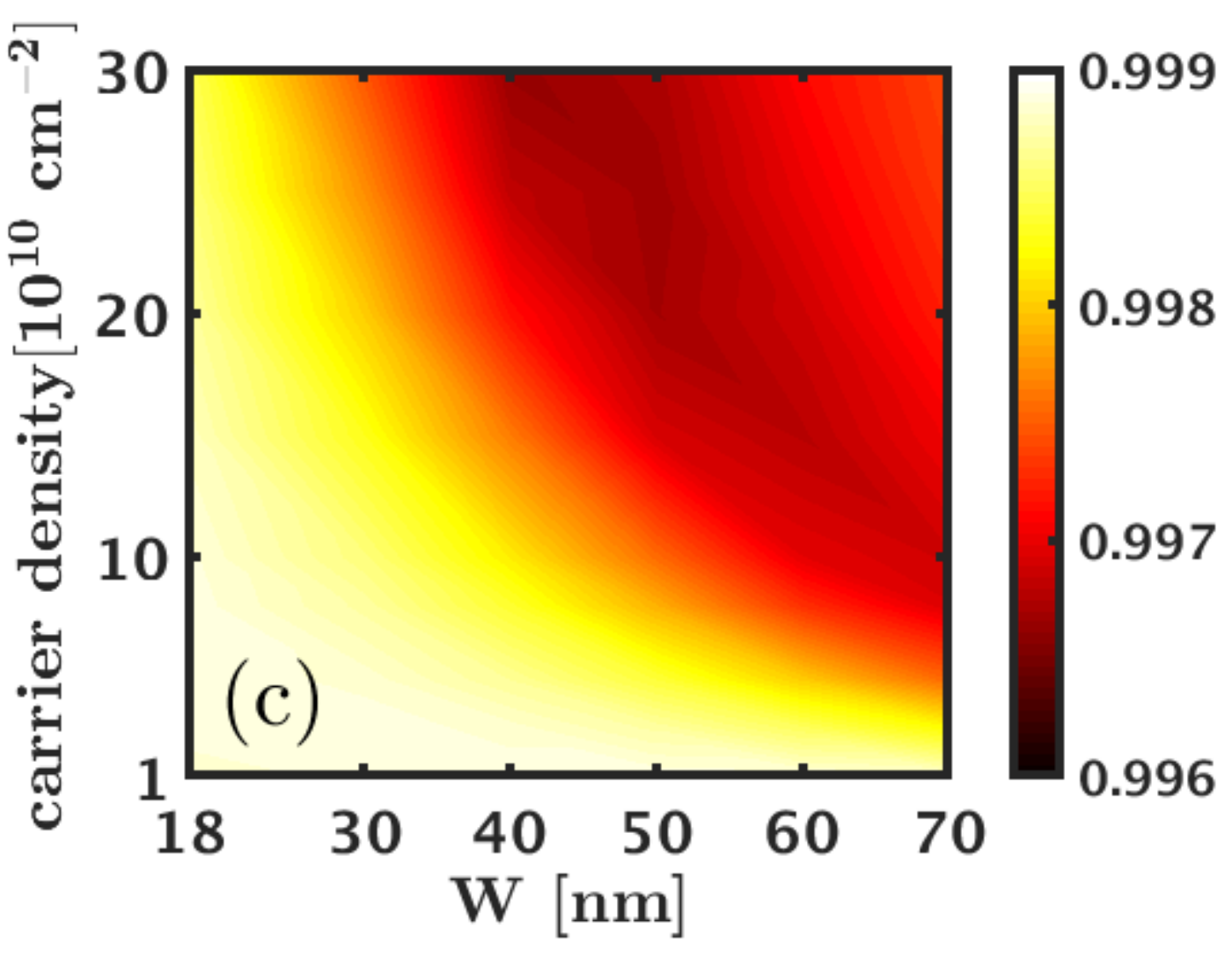}\\
				\vspace{0.3cm}
				\includegraphics[width=0.32\textwidth]{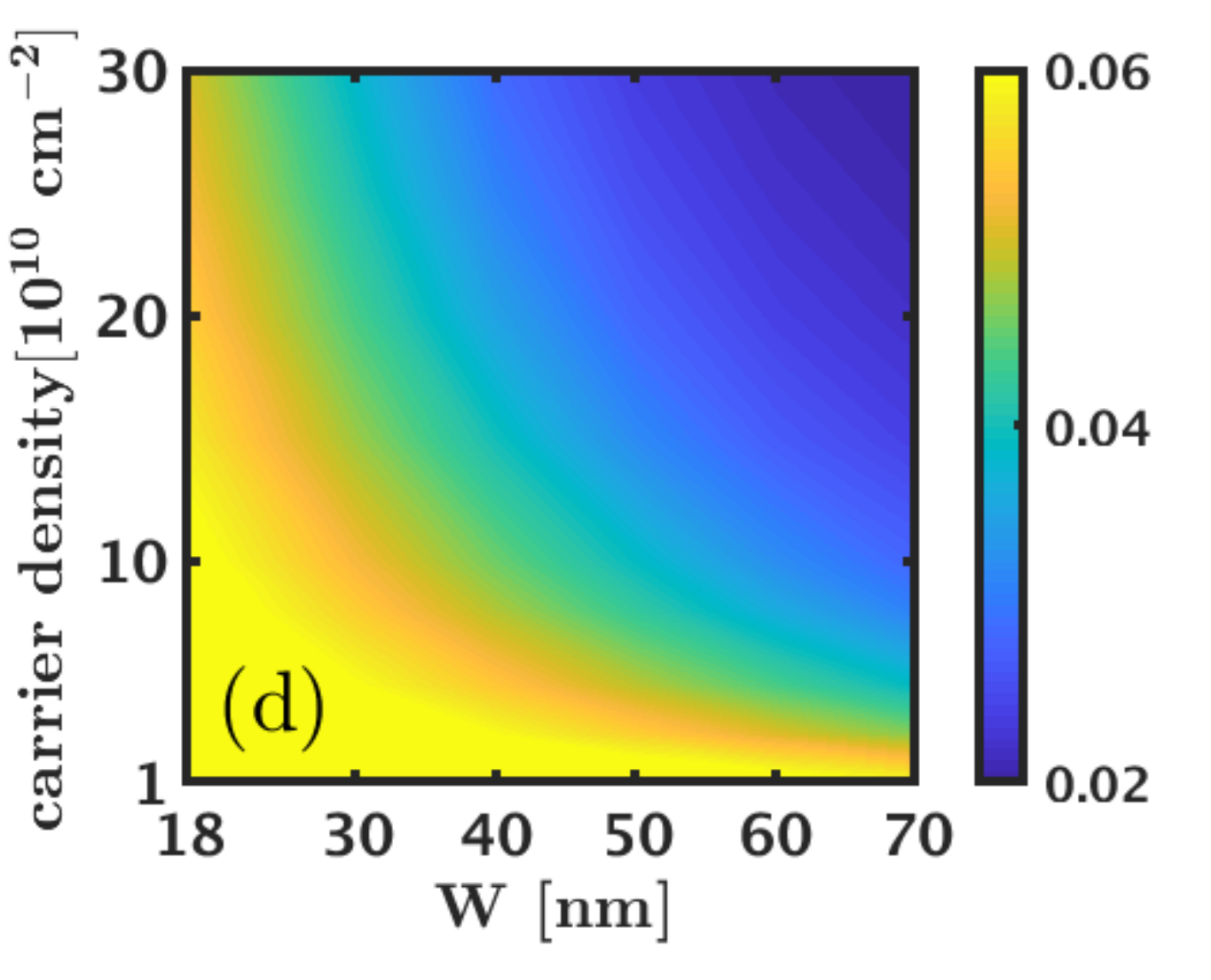}
				\includegraphics[width=0.32\textwidth]{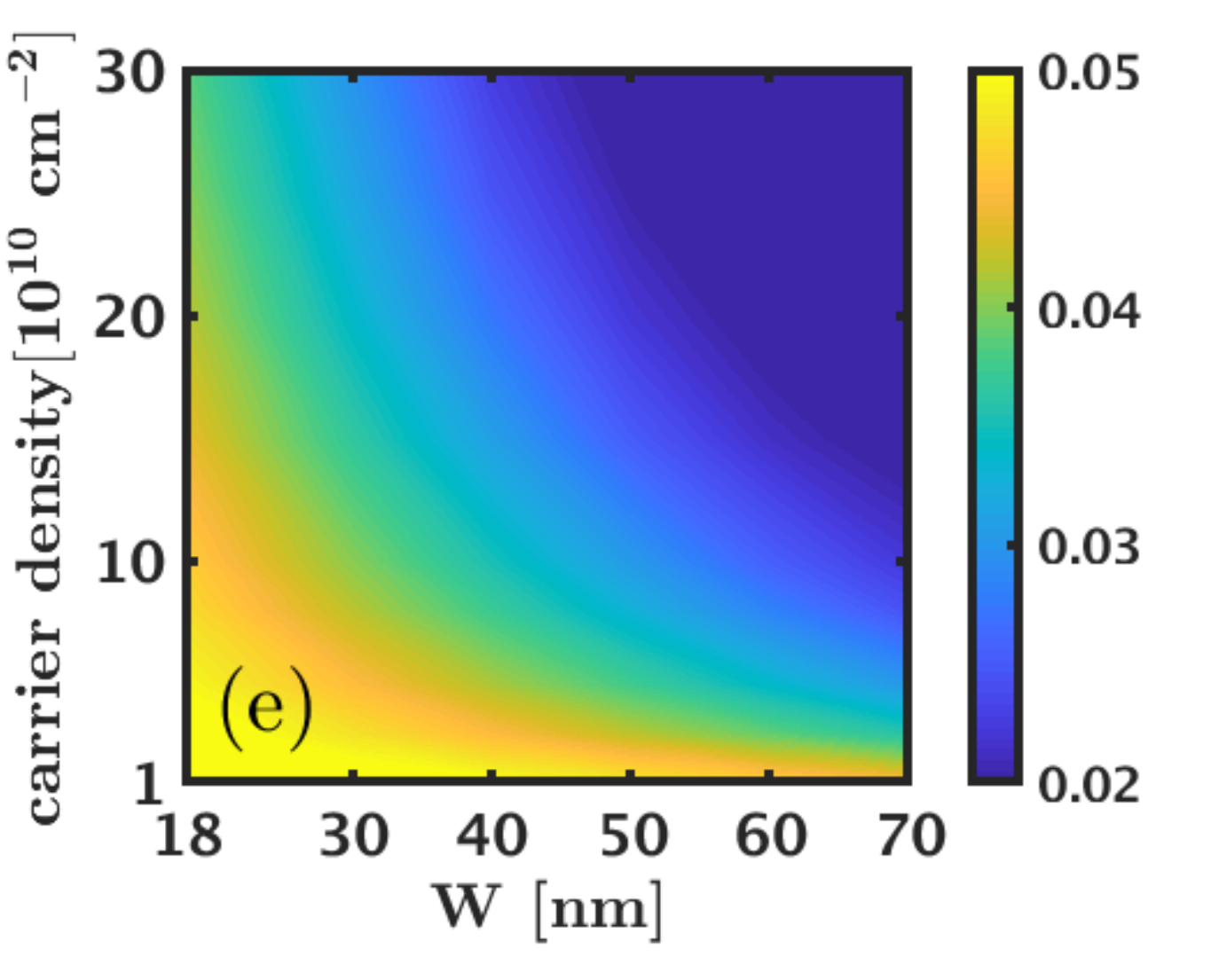}
				\includegraphics[width=0.32\textwidth]{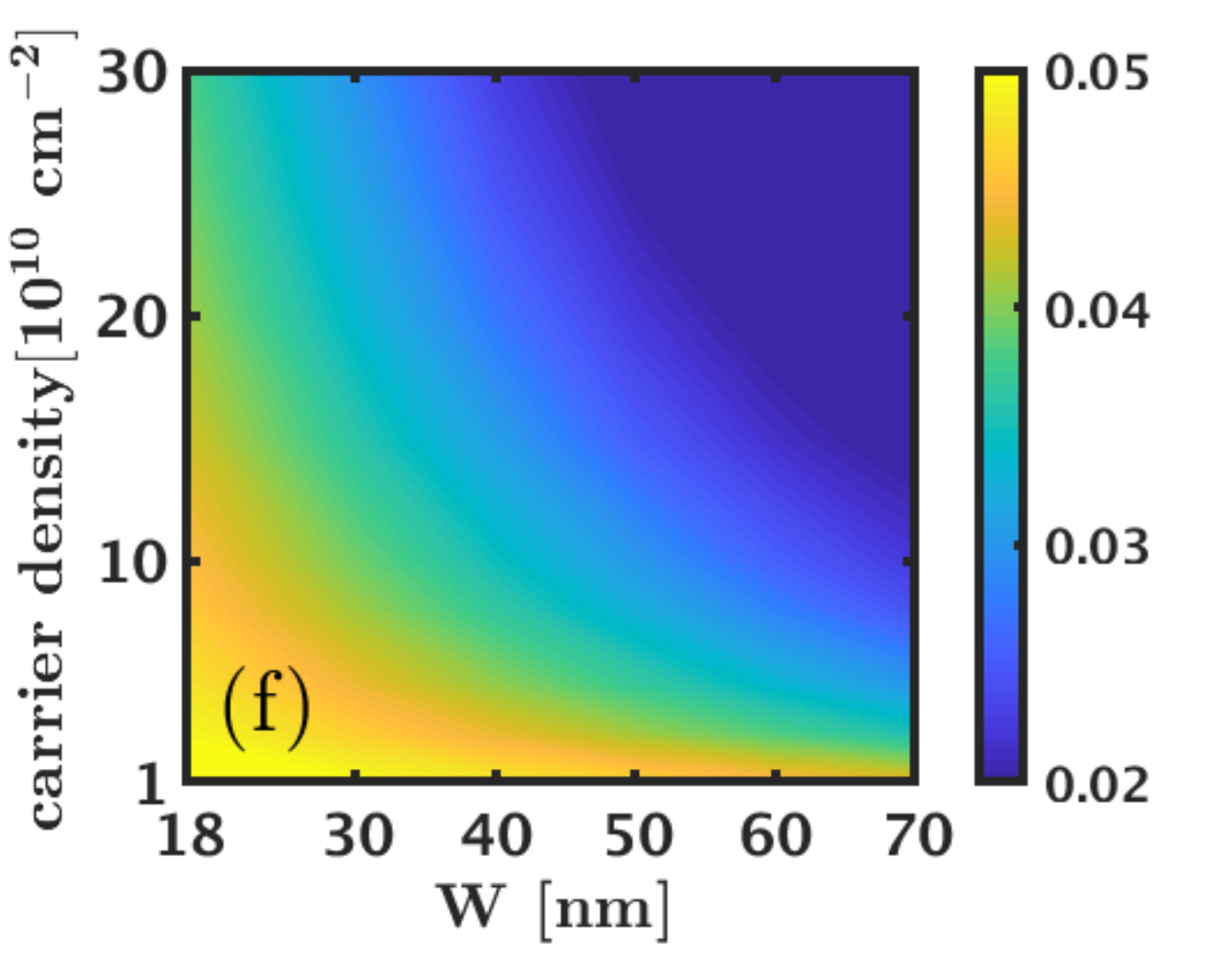} \\
				\vspace{0.3cm}
				\includegraphics[width=0.32\textwidth]{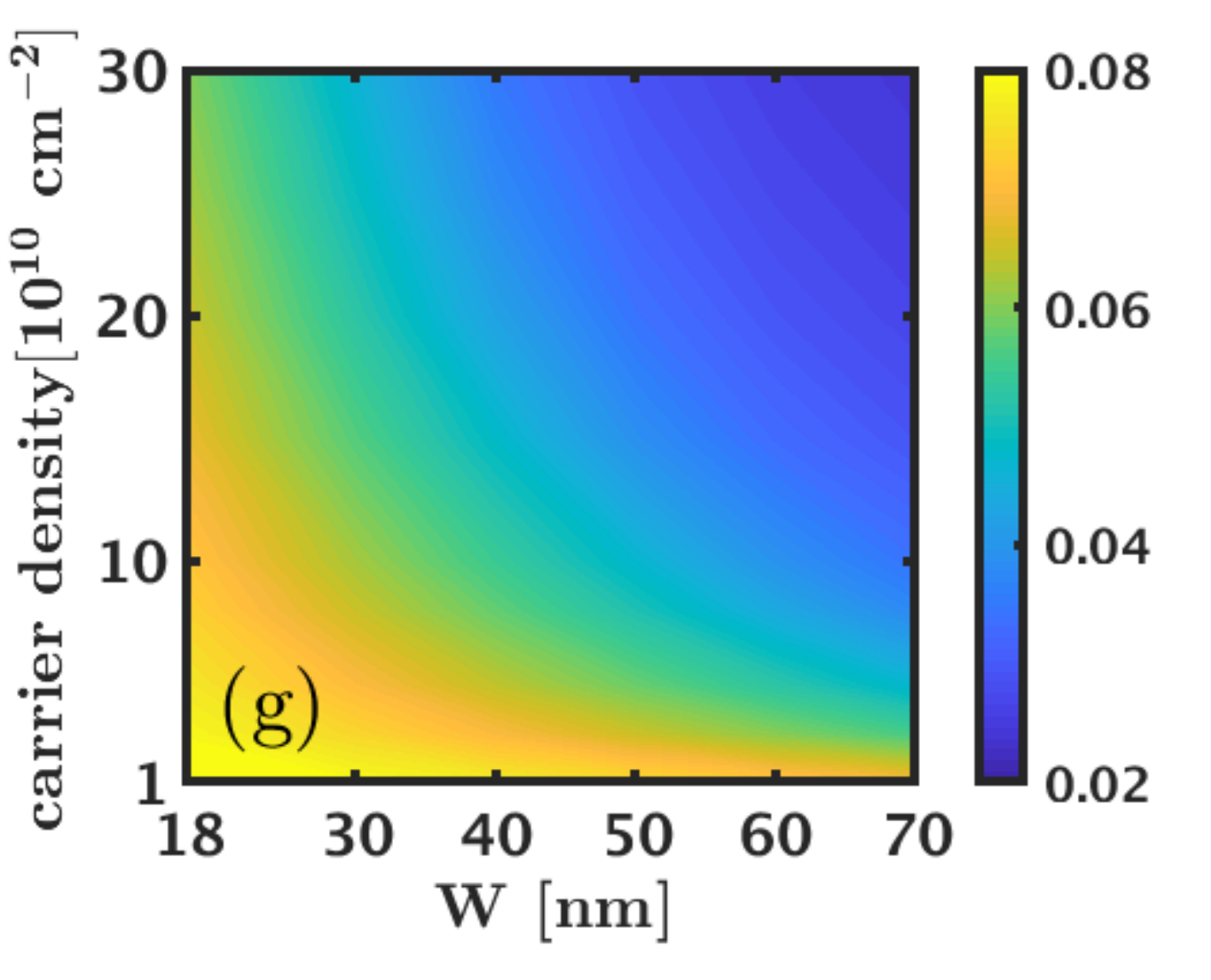}
				\includegraphics[width=0.32\textwidth]{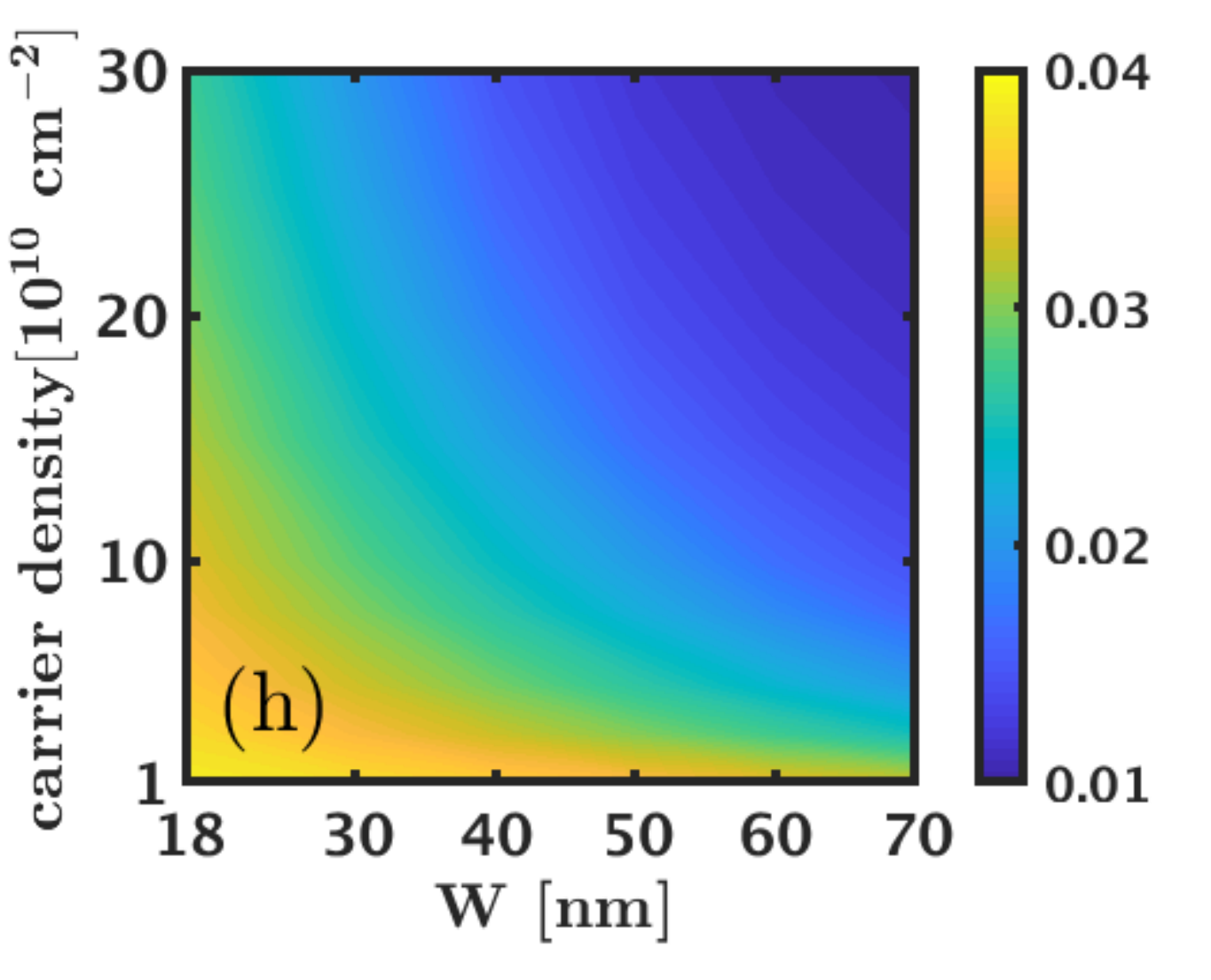}
				\includegraphics[width=0.32\textwidth]{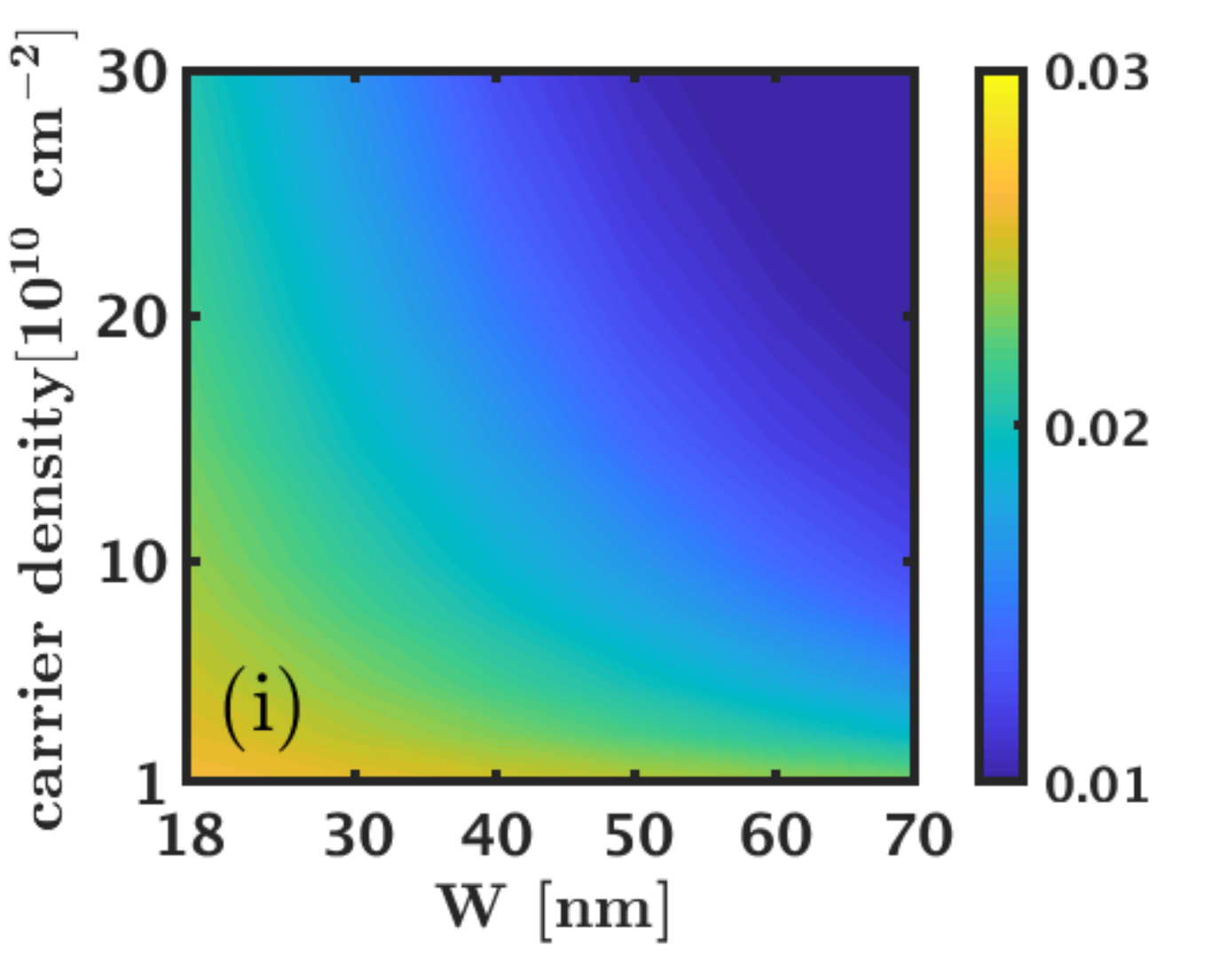}
			\end{center}
			\caption{
			Overlaps with the exact lowest Landau level ground state [top panels (a), (b), and (c)], neutral gaps [middle panels (d), (e), and (f)] and charge gaps [bottom panels (g), (h), and (i)] in the spherical geometry for the $\nu=1/3$ Laughlin [left panels (a), (d), and (g)] $\nu=2/5$ Jain [center panels (b), (e), and (i)] and $\nu=3/7$ Jain [right panels (c), (f), and (i)] state evaluated using the pseudopotentials of the finite-width interaction obtained using a local density approximation (LDA). All the panels are for $N_{e}=12$ electrons.}
			\label{fig: Laughlin_Jain_overlaps_finite_width_LDA}
		\end{figure*}

		The results for the overlaps and gaps obtained from exact diagonalization using the LDA pseudopotentials at $\nu=1/3$, $2/5$ and $3/7$ are shown in Fig. ~\ref{fig: Laughlin_Jain_overlaps_finite_width_LDA} (note that the scales on different plots are different). We find that the $1/3$ Laughlin, $2/5$ and $3/7$ Jain states provide a near-perfect representation of the exact ground state at all widths and densities considered. Furthermore, these states support robust charge and neutral gaps, which indicates that they are stable to perturbations in the interaction arising from finite-width corrections and density variations. These results are consistent with the experimental observation of incompressible states at $1/3$, $2/5$, and $3/7$ in wide quantum wells~\cite{Shabani09b}.

\begin{figure*}[htpb]
			\begin{center}
				\includegraphics[width=0.32\textwidth]{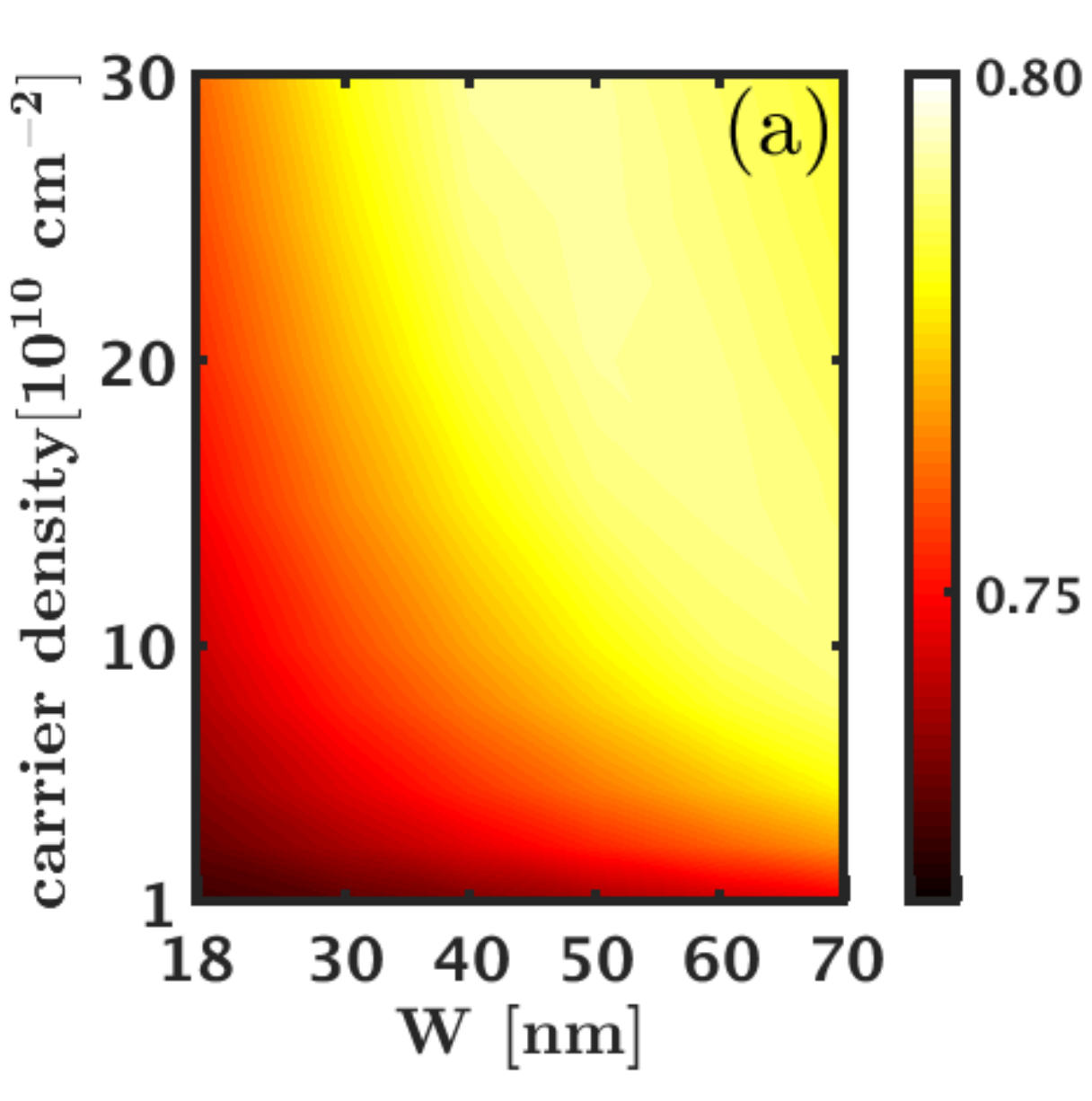}
				\includegraphics[width=0.32\textwidth]{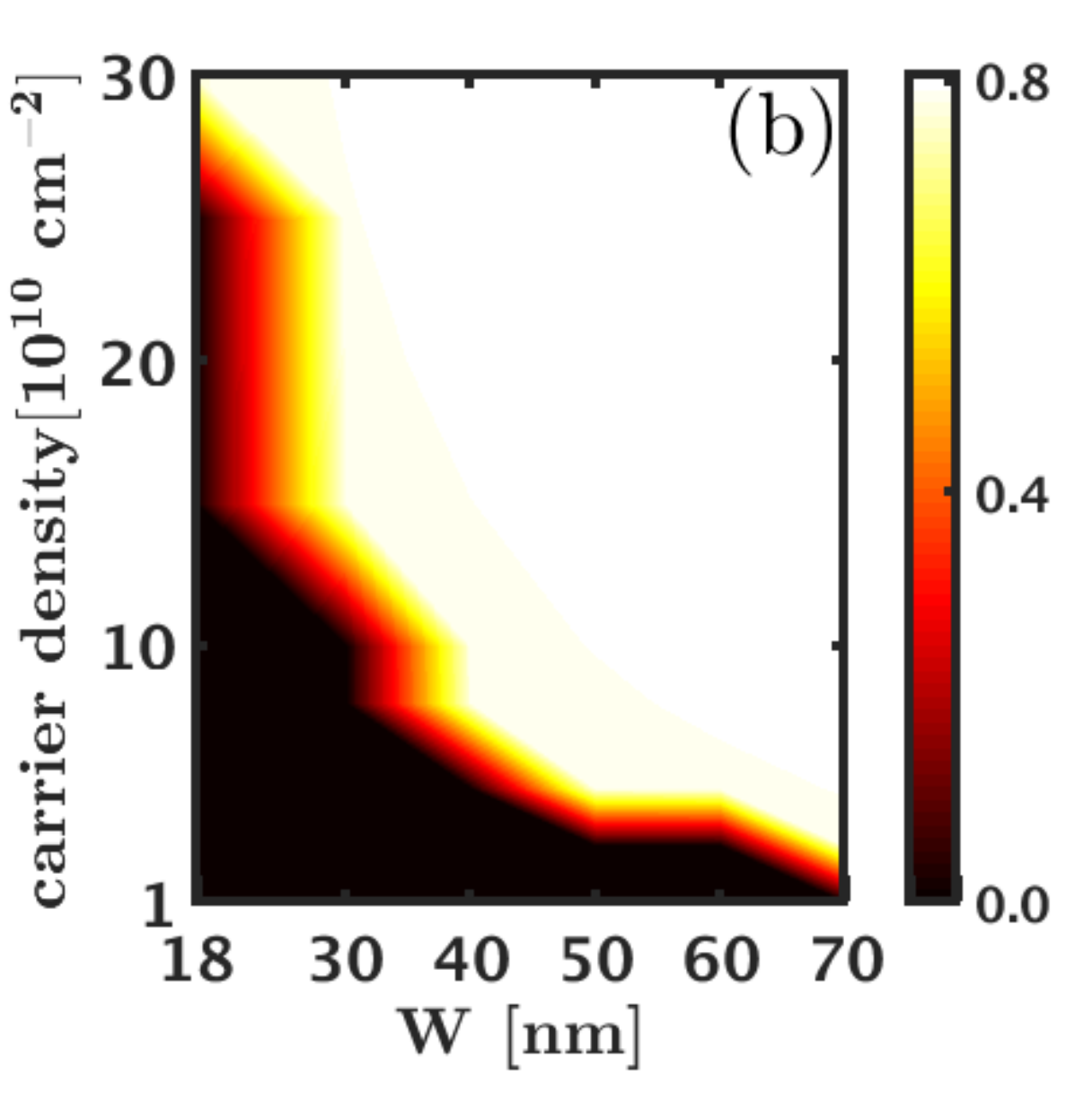}
				\includegraphics[width=0.32\textwidth]{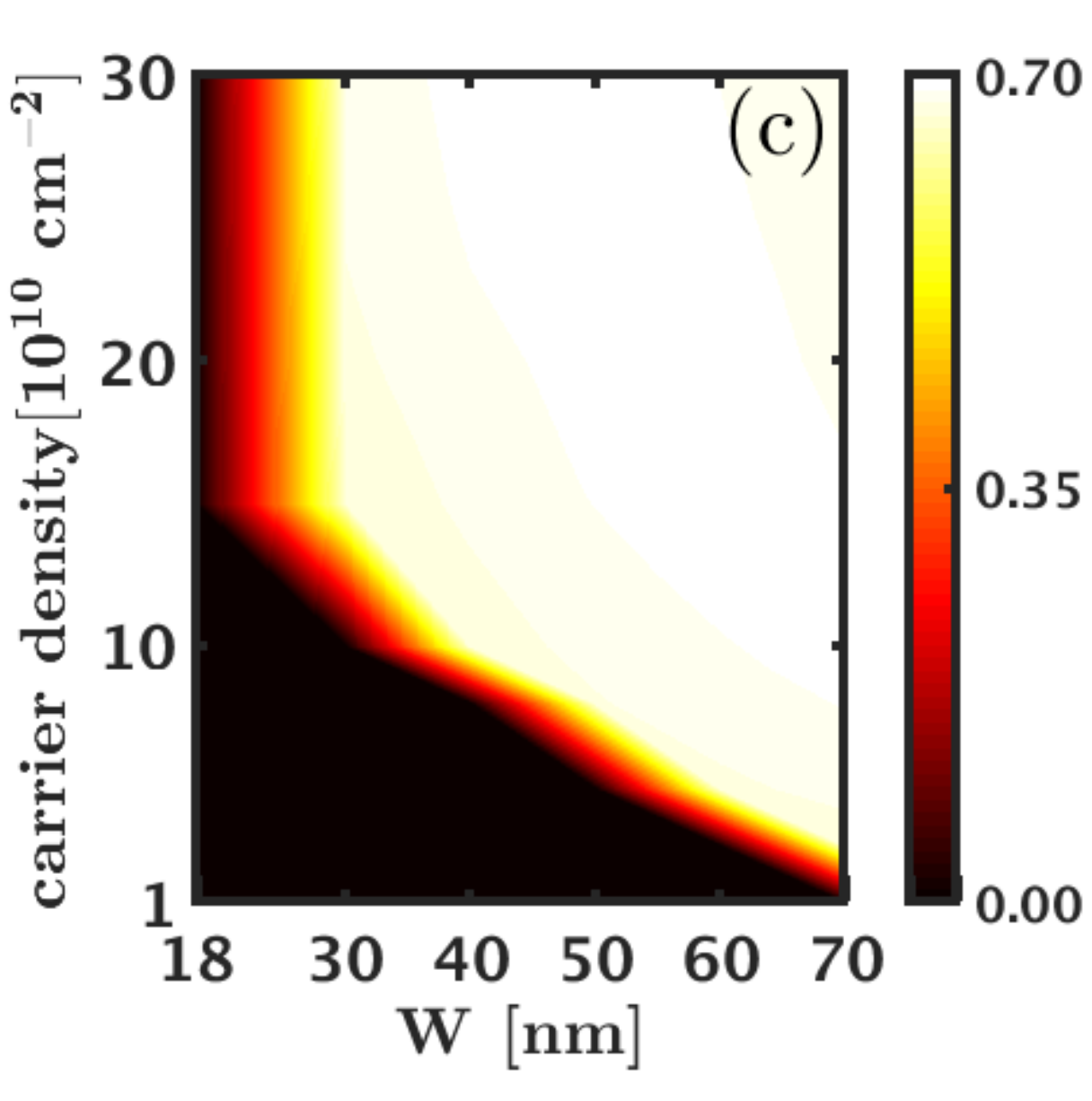} \\
				\vspace{0.3cm}
				\includegraphics[width=0.32\textwidth]{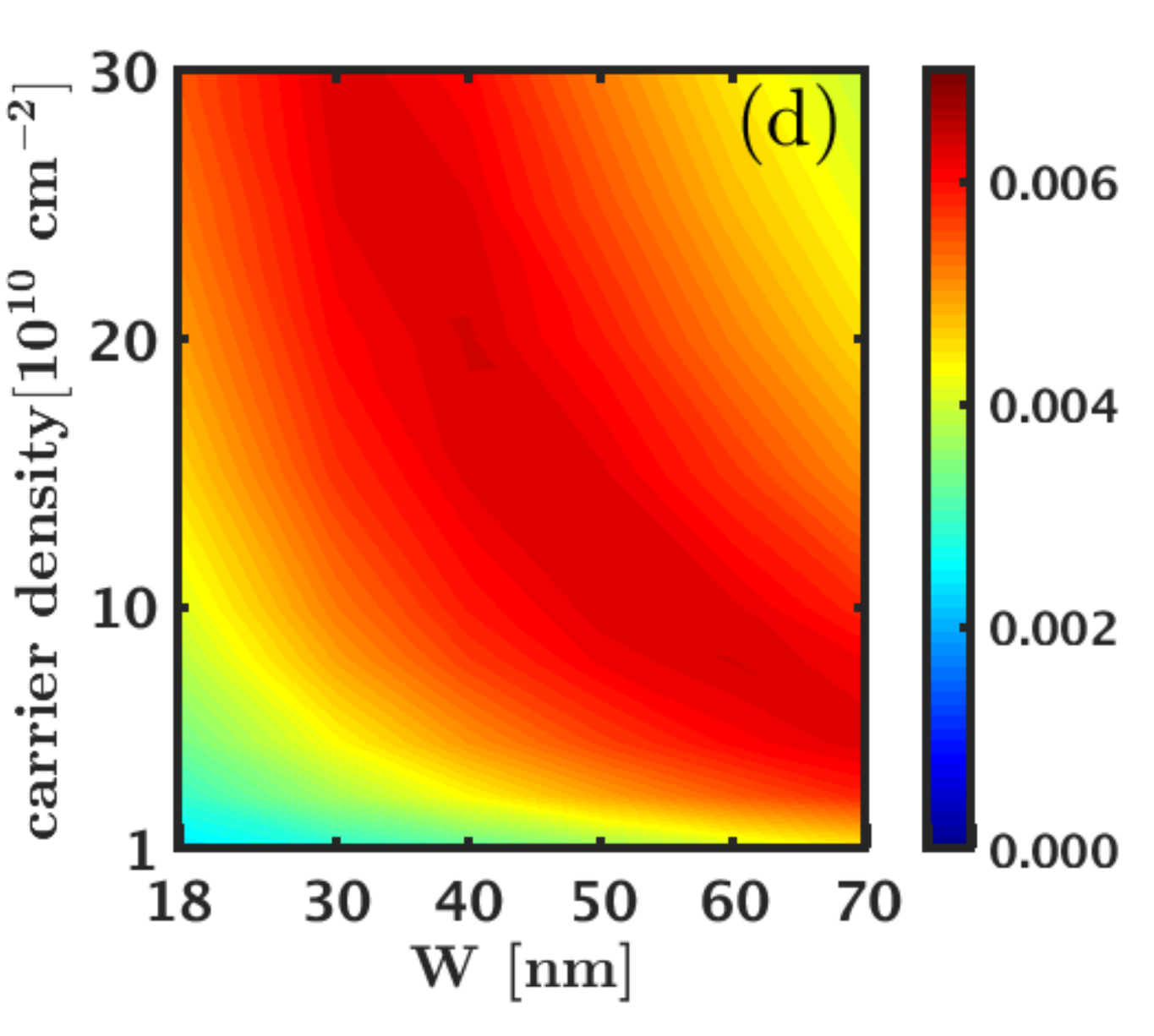}
				\includegraphics[width=0.32\textwidth]{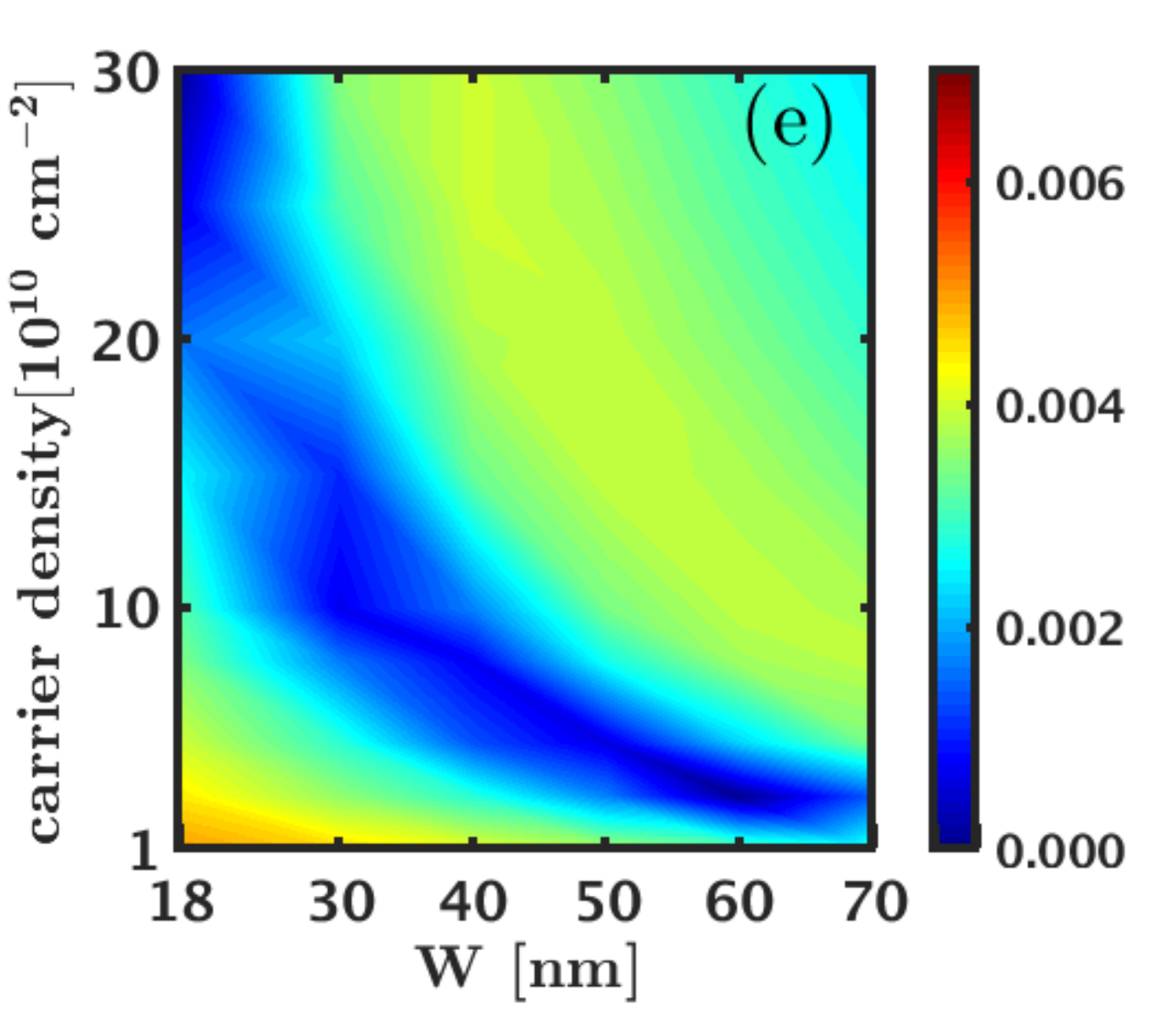}
				\includegraphics[width=0.32\textwidth]{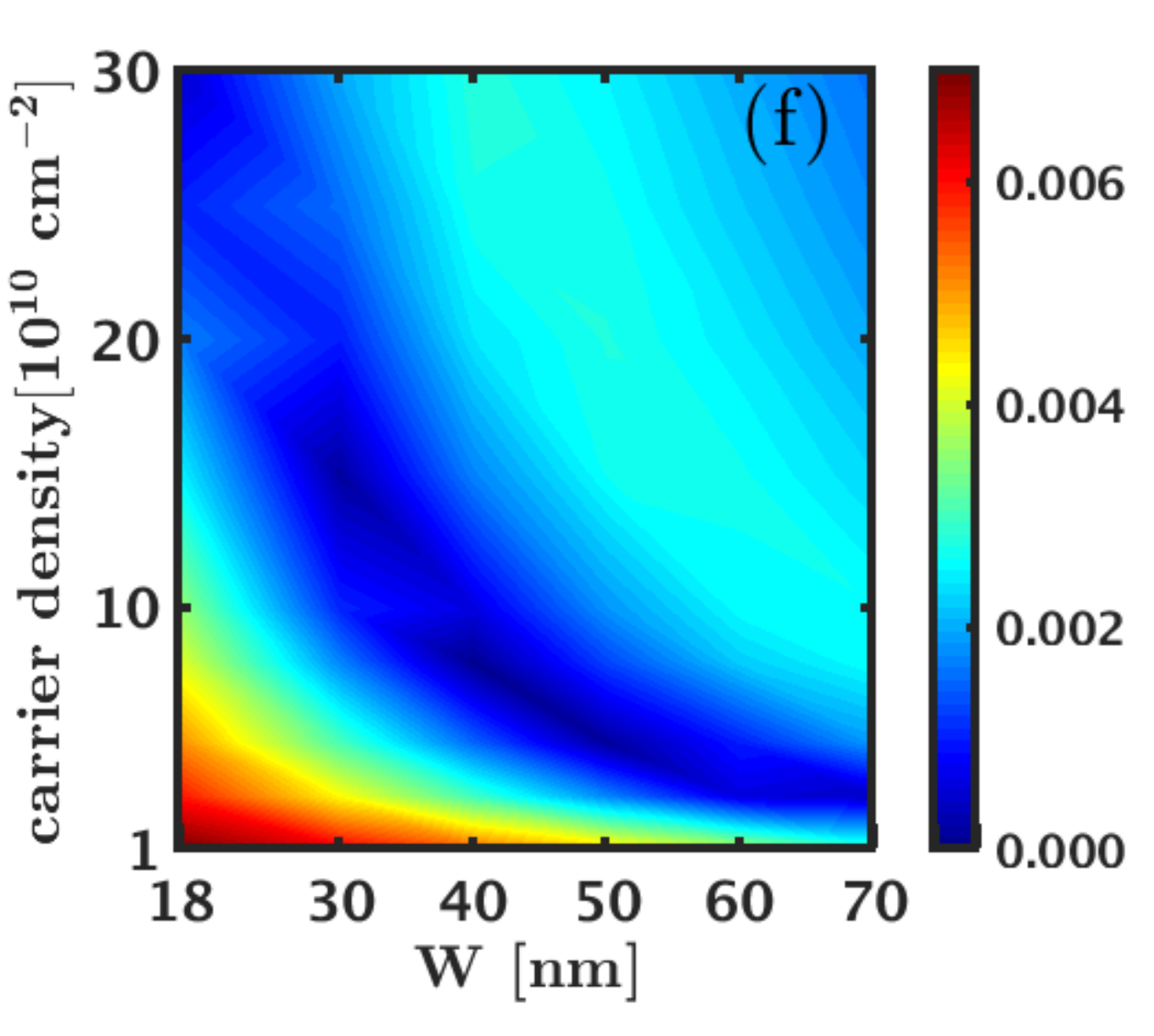}\\
				\vspace{0.3cm}
				\includegraphics[width=0.32\textwidth]{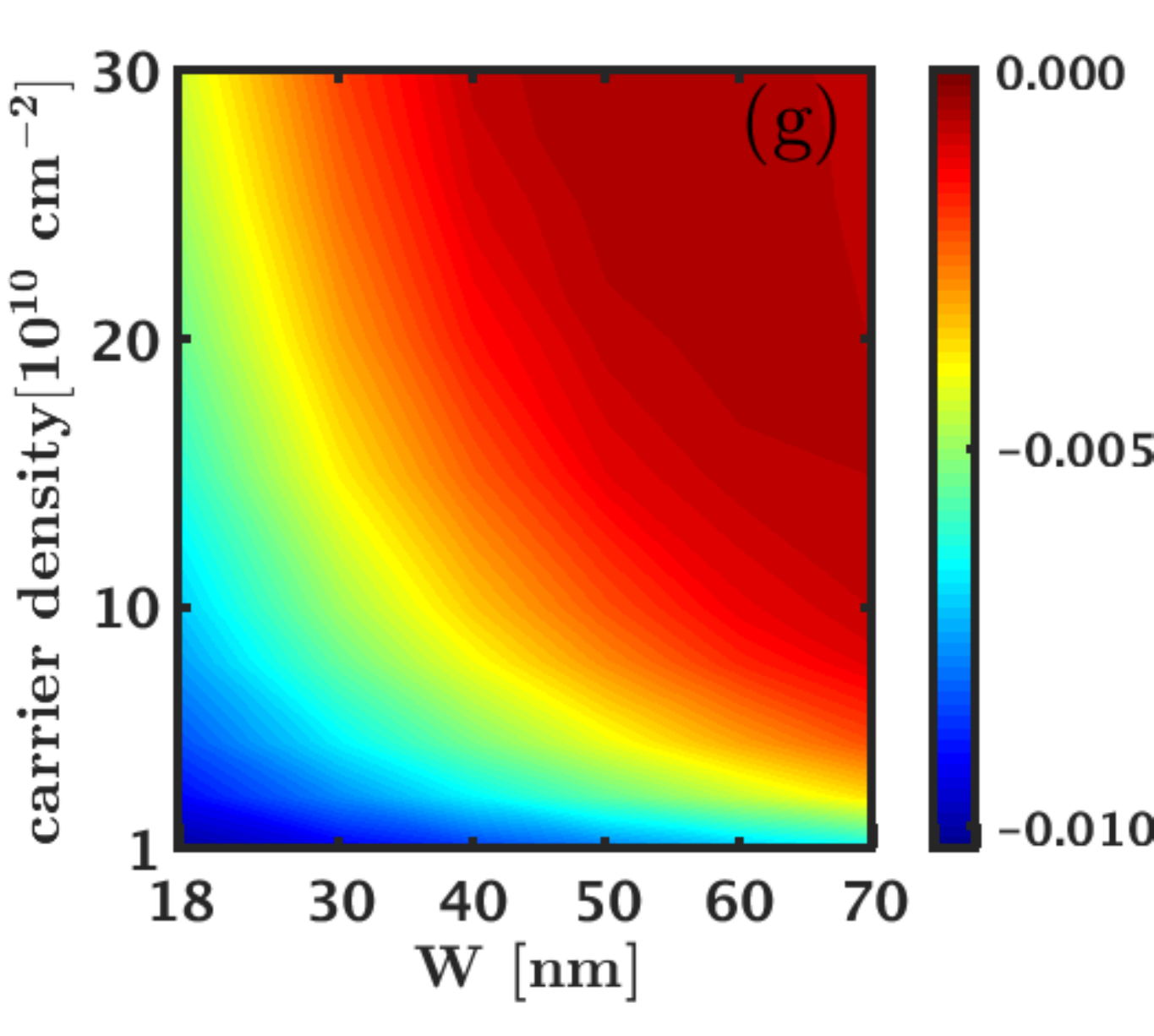}
				\includegraphics[width=0.32\textwidth]{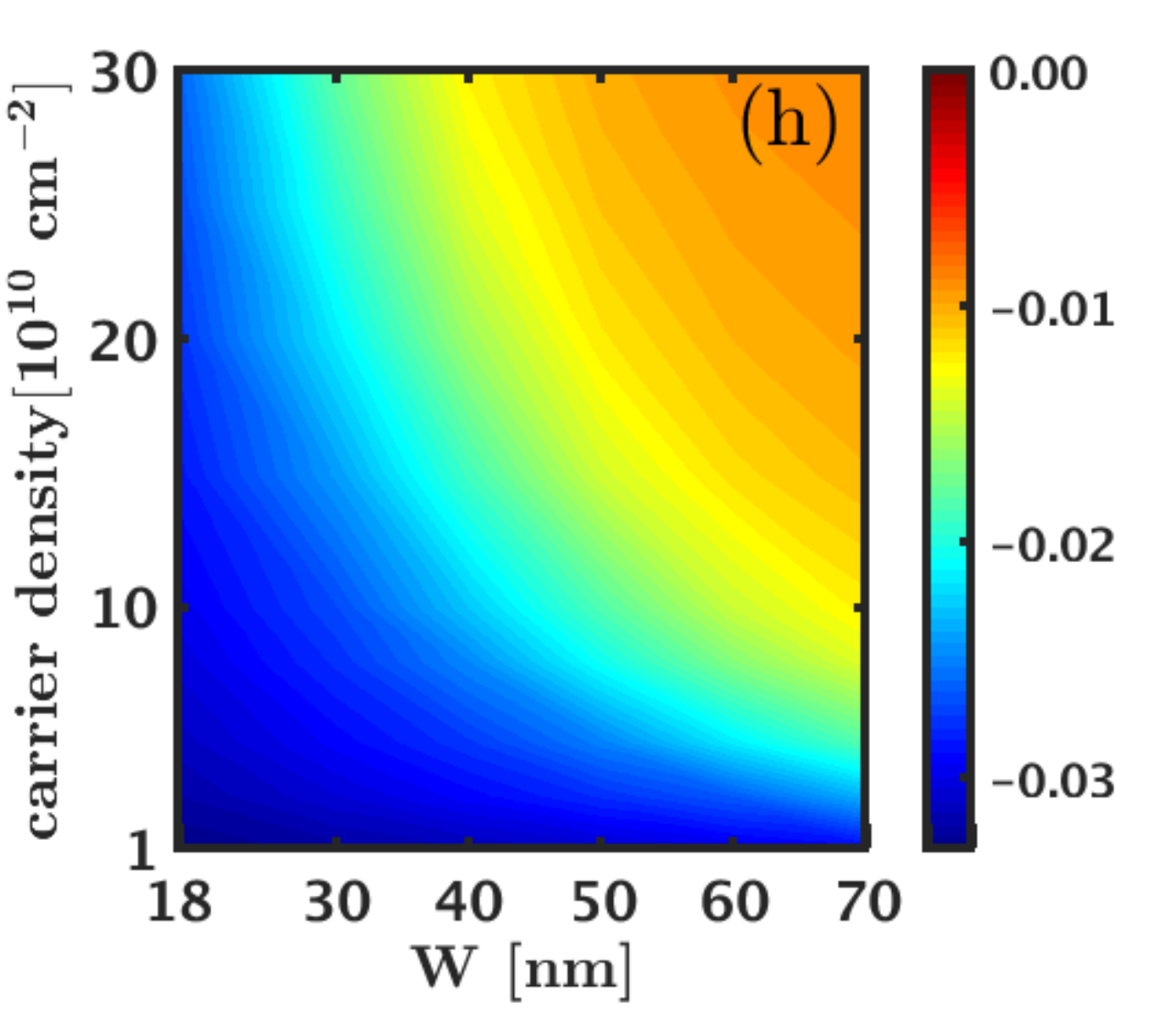}
				\includegraphics[width=0.32\textwidth]{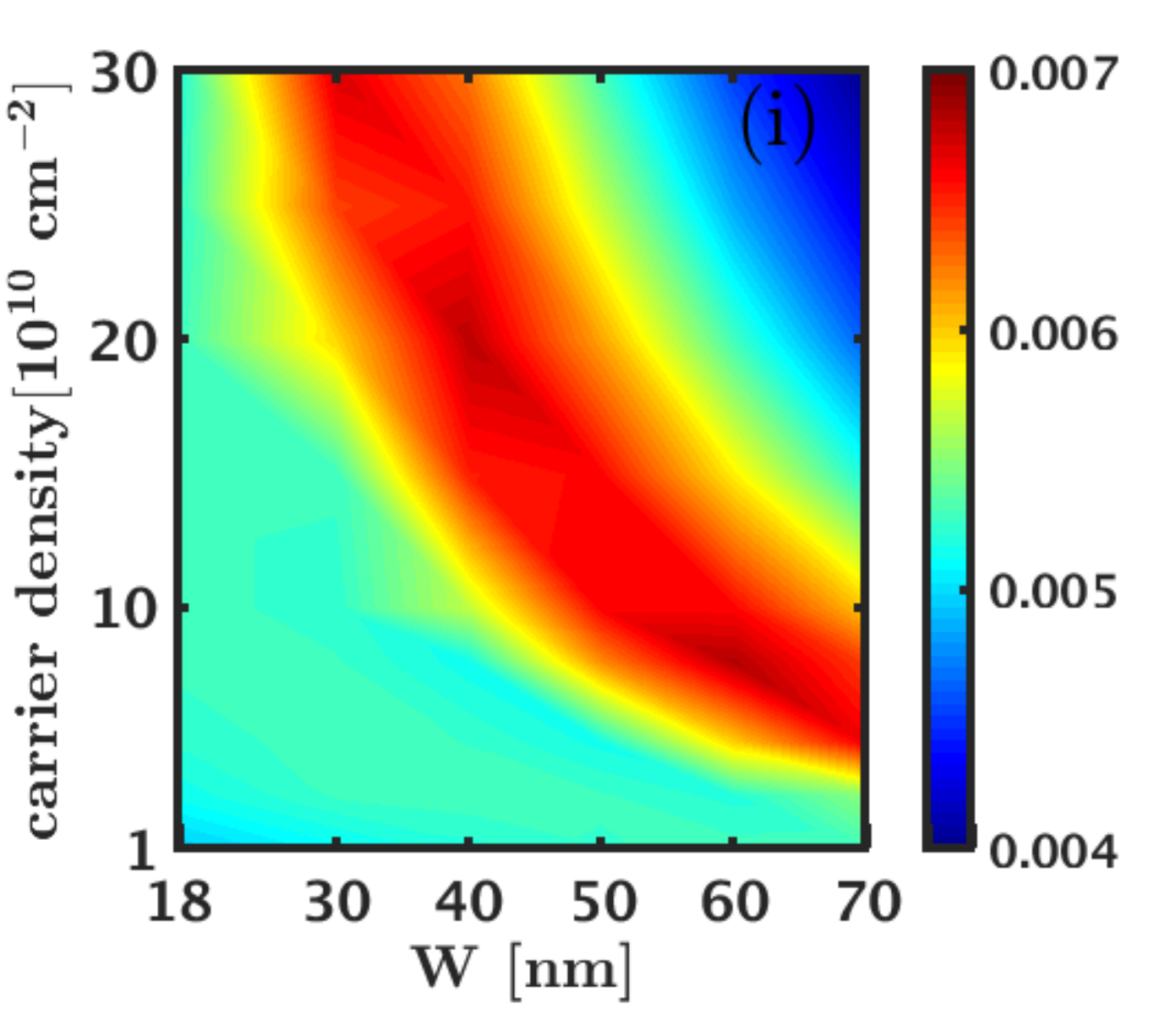}
			\end{center}
			\caption{Overlaps of the $\nu=1/2$ Moore-Read Moore-Read Pfaffian state with the exact lowest Landau level ground state [top panels (a), (b), and (c)], neutral gaps [middle panels (d), (e), and (f)] and charge gaps [bottom panels (g), (h), and (i)] in the spherical geometry evaluated using the pseudopotentials of the finite-width interaction obtained using a local density approximation (LDA). The left, center and right panels correspond to $N_{e}=14$ [panels (a), (d), and (g)],~$16$ [panels (b), (e), and (h)] and $18$ [panels (c), (f), and (i)] respectively.}
			\label{fig: MR_Pfaffian_overlaps_finite_width_LDA}
		\end{figure*}

		We next consider the 1/2 state and evaluate its charge and neutral gaps as well as its overlaps with the Moore-Read Pfaffian wave function as a function of the width and density.  Here we consider the three systems of $N_e=14,~16$ and $18$ electrons that do not alias with any of the Jain states~\cite{Scarola02b}. The overlap maps shown in Fig. ~\ref{fig: MR_Pfaffian_overlaps_finite_width_LDA} indicate that the overlap of the Pfaffian state with the exact ground state increases with increasing width and density and reaches a value comparable to the overlap of the Pfaffian wave function with the 5/2 Coulomb ground state~\cite{Balram20}.  We next look at the neutral and charge gaps ($n_{q}=2$) of the 1/2 Moore-Read Pfaffian state.  These results, shown in Fig. ~\ref{fig: MR_Pfaffian_overlaps_finite_width_LDA}, suggests that the 1/2 Moore-Read Pfaffian state does not consistently, i.e. for all values of $N_e$, support a robust charge / neutral gap.  For the system of $N_e=14$ and $N_e=16$ electrons, we find that the charge gap is negative for most widths and densities, which indicates that the 1/2 Moore-Read Pfaffian state is not stabilized for these interactions. Even for the system of $N_e=18$ electrons,  where the charge gaps are positive, the 1/2 Moore-Read Pfaffian state has a gap that is an order of magnitude lower than that of the Laughlin and Jain states. Thus, we conclude that the LDA interaction does not stabilize the 1/2 Moore-Read Pfaffian state in the LLL for the LDA interaction (without LLM).
		
\begin{figure}[htpb]
			\begin{center}
				\includegraphics[width=0.47\textwidth]{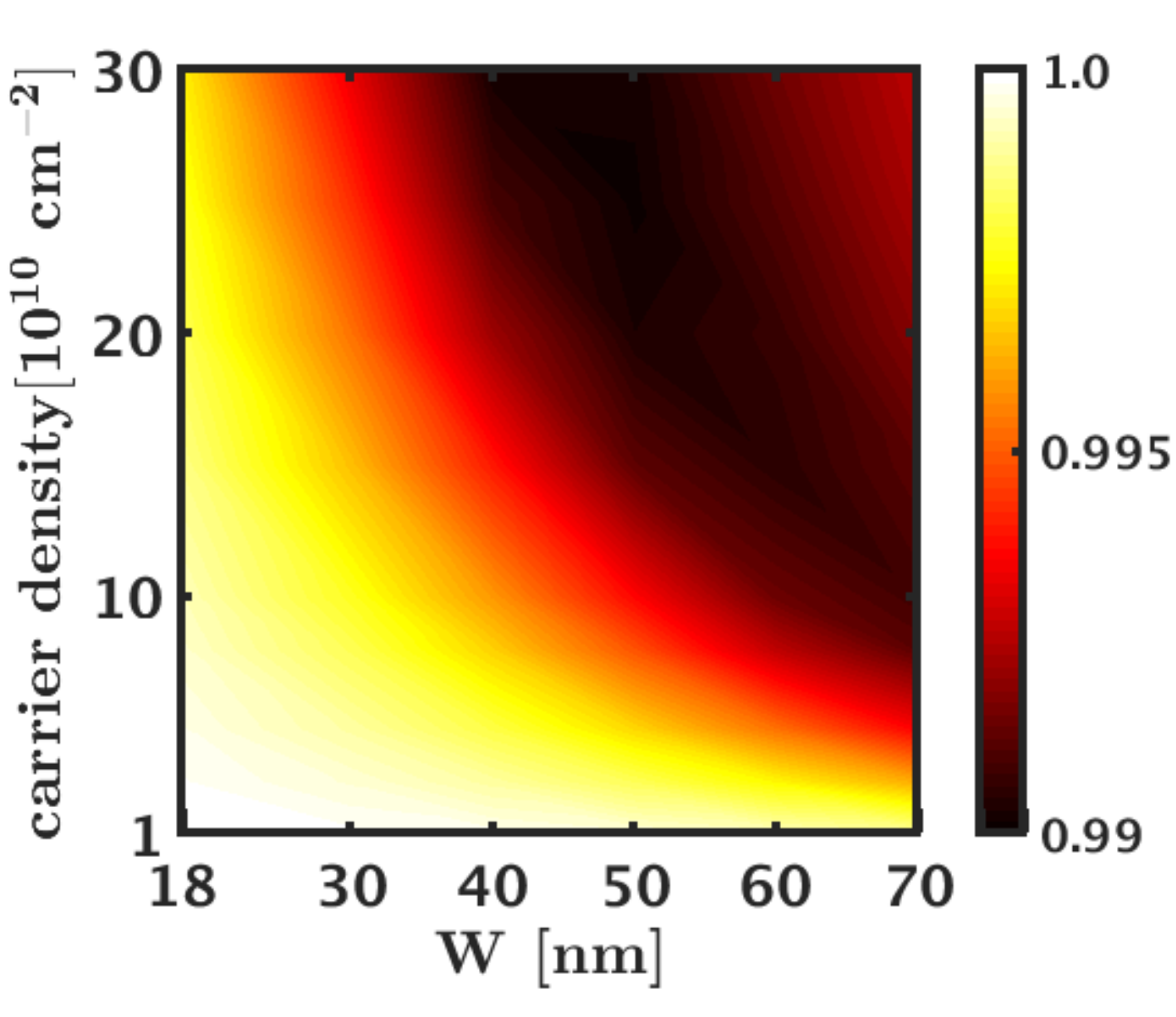}
			\end{center}
			\caption{Overlaps of the composite fermion Fermi sea (zero-width Coulomb ground state in the lowest Landau level [see text]) with the ground state of the finite-width LDA interaction for $N_{e}=14$ electrons at flux $2Q=25$. The overlap is essentially unity in the entire range of widths and densities considered.}
			\label{fig: CFFS_overlaps_finite_width_LDA}
		\end{figure}

		Finally, we turn to the CFFS state at $\nu=1/2$ and consider its overlap with the exact ground state.  For this purpose, we consider the exact zero-width LLL Coulomb ground state of $N_e=14$ electrons at $2Q=2N_e-3$, since this system has a uniform ($L=0$) ground state.  We take this ground state to represent the CFFS state and calculate its overlap with the exact LDA ground state as a function of width and density. These overlaps are shown in Fig. ~\ref{fig: CFFS_overlaps_finite_width_LDA} and are essentially unity in the entire parameter space we have considered. (For comparison, the overlap of the Moore-Read Pfaffian state with the exact zero-width LLL Coulomb ground state for this system size is $0.72$~\cite{Balram20}.)
		
		To summarize, our exact diagonalization results are consistent with the VMC results given in the main article. In the entire parameter range that we explored, the CFFS has almost unit overlap with the exact ground state. Thus the CFFS state is favored over the Moore-Read Pfaffian state for all the LDA interactions that we have looked at in the absence of LLM.
		\section{Additional details on the diffusion Monte Carlo}
		\label{DMC_algorithm}
		The fixed-phase DMC, which is a generalization of the standard DMC method~[\onlinecite{Mitas98, Foulkes01}], was developed in Ref.~[\onlinecite{Ortiz93}] and also described in Refs.~[\onlinecite{Zhang16, Zhao18}]. The method we use in this paper is based on these articles. Here we give some details that are specific to our work.
		
		We use parameters appropriate for Gallium Arsenide. We express lengths in units of $l_B$ and energies in units of $\frac{e^2}{\epsilon l_B}$. The local energy for a 2D system is simply $E_L(\vec{R})=\frac{N_e}{2\kappa}+V_\text{Ewald}(\vec{R})$ and for a 3D system an extra term $\sum_i{E}_\text{trans}\left(  w_i \right)$ is introduced due to the transverse degree of freedom:	
		\begin{equation}
		E_L\left(\vec R \right)=\frac{N_e}{2 \kappa}+V_\text{Ewald}(\vec{R})+\sum_i{E}_\text{trans}\left(  w_i \right)
		\end{equation}
		where $N_e/2\kappa$ is the cyclotron energy for $N_e$ particles in the initial trial state. $V_\text{Ewald}(\vec{R})$ is the Coulomb interaction extended periodically in the x-y plane; it satisfies open boundary conditions in the transverse dimension as appropriate for our 3D quantum wells (for  2D systems we simply set all $w_i$'s to be $0$). Its explicit form is given below in Appendix~\ref{Ewald_V}.		
		The transverse local energy of a one-component state is given by:
		\begin{equation}
		\begin{aligned}
		&E_\text{trans}\left( w \right)\\
		=&\begin{cases}\frac{1}{\kappa} \frac{\pi^2}{2 W^2} \left(9-\frac{8}{1+\alpha-2\alpha \cos(2\pi  w/W)}\right), &| w|<W/2\\
		\infty, &| w|\geq W/2
		\end{cases}
		\end{aligned}
		\end{equation}
		For two-component states, the energies for the left-layer and right-layer are as follows:
		\begin{equation}
		\begin{aligned}
		&E_\text{trans}^L\left( w\right)=\begin{cases}\frac{1}{\kappa} \frac{\beta (2 W-\beta w)}{2 W^2 w},&-W/2< w<0\\
		\infty,  & w\geqslant0
		\end{cases}\\
		&E_\text{trans}^R\left( w\right)=\begin{cases}\frac{1}{\kappa} \frac{\beta \left[2 W-\beta (w0-w)\right]}{2 W^2 (W-w)},&0< w<W/2\\
		\infty,  & w\leqslant 0
		\end{cases}
		\end{aligned}
		\end{equation}
		We use the mixed estimator method\cite{Foulkes01} to calculate the ground state energy.
		
		\section{Transverse distribution of fully antisymmetrized two-component states}
		In the main text, we make the approximation that the two transverse basis wave functions of two-component states do not overlap, i.e. they are located entirely either in the left or the right half of the quantum well. The approximation becomes quantiatively valid when the well-width or the density is very high, in which case which both the lowest symmetric and asymmetric subbands have vanishing density at the center, and the linear combinations of them form the left- and right-layer bases. This approximation simplifies the calculation because
the system's energy can be evaluated without doing an antisymmetrization over all particles. 

\begin{figure}
			\includegraphics[width=\columnwidth]{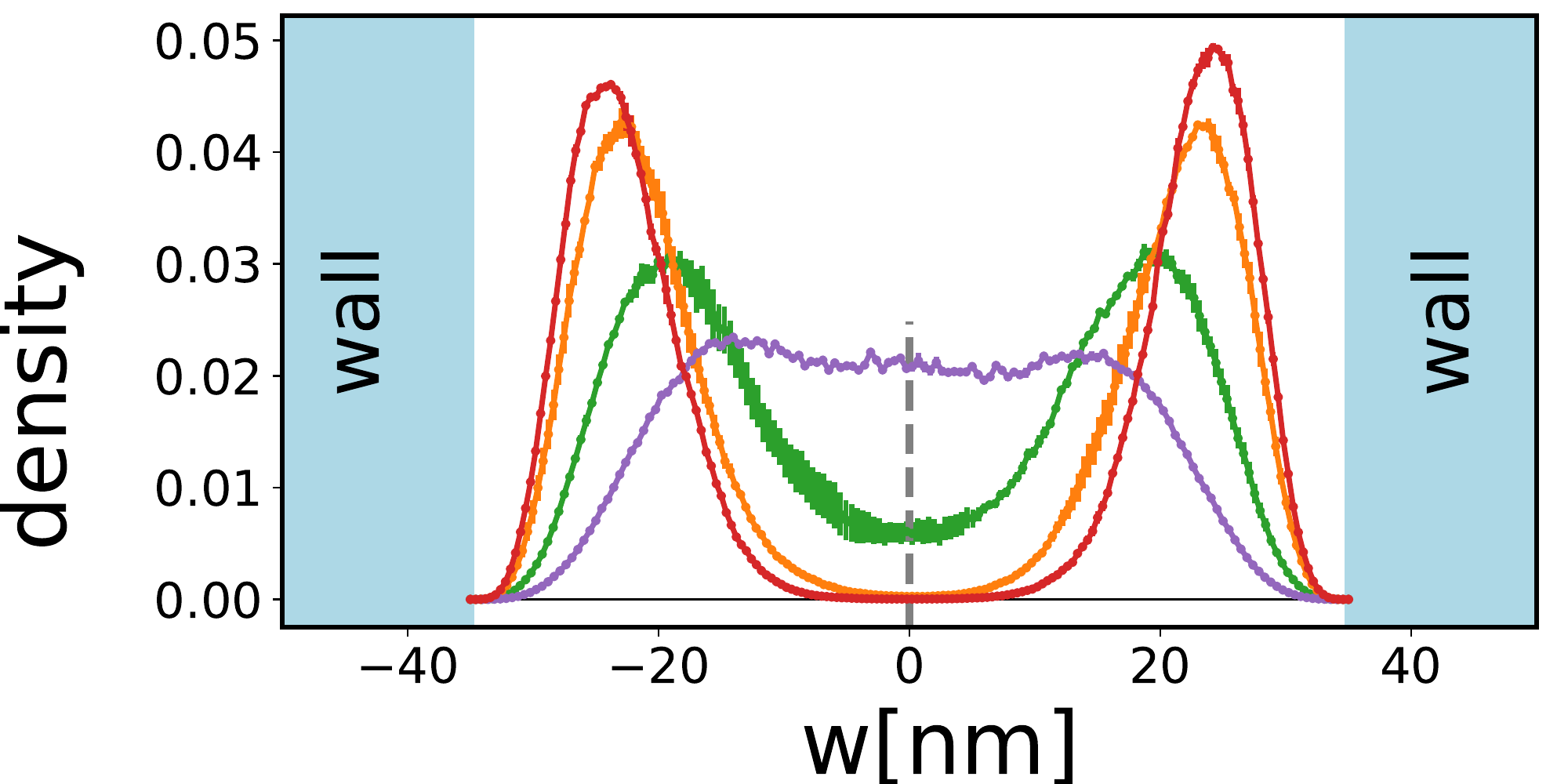}
			\includegraphics[width=\columnwidth]{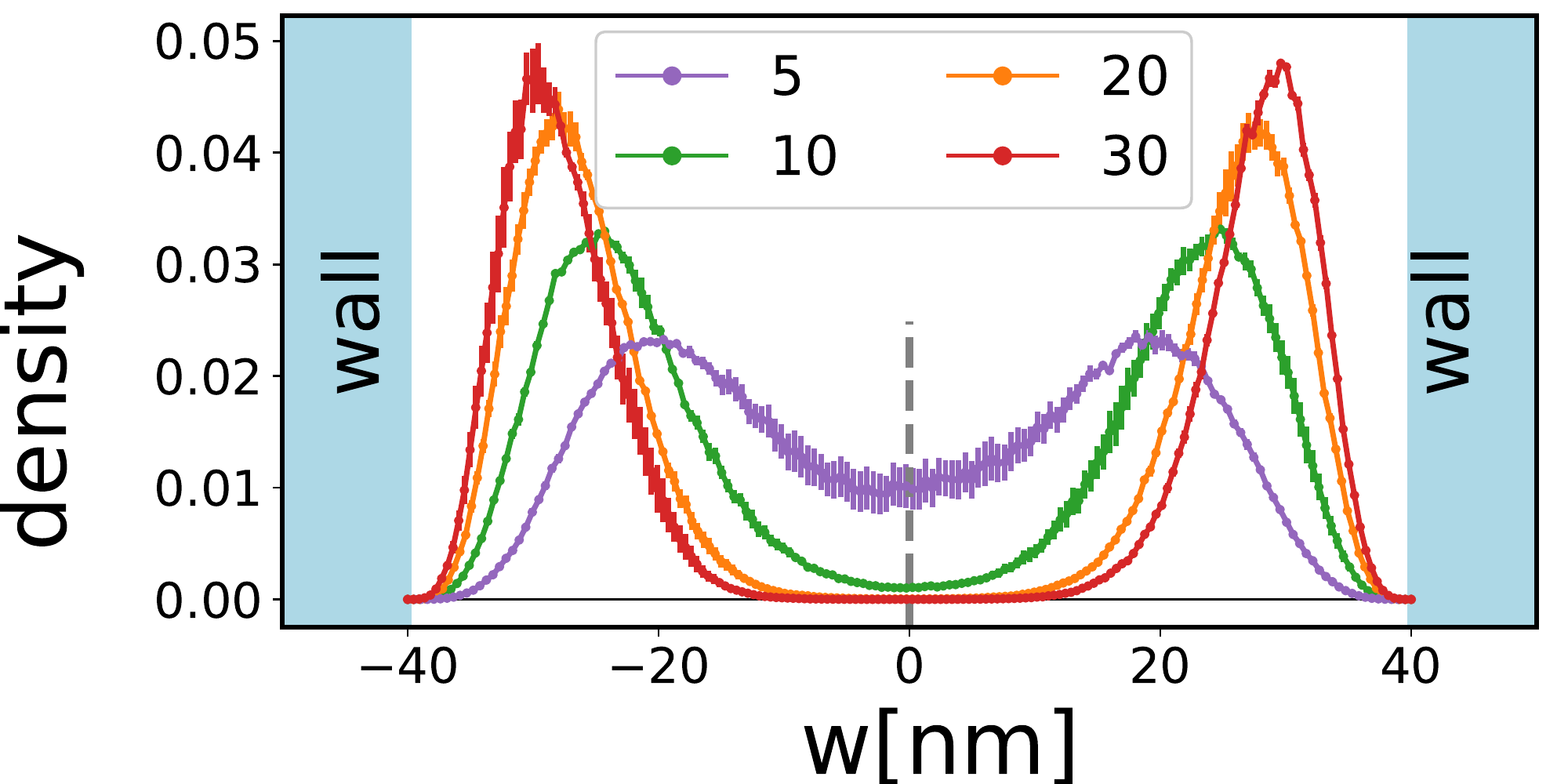}
			\caption{The transverse density profiles of the $(3,3,1)$-like state for 4-particle system of the widths $W=70 \text{nm}$ (top) and $W=80 \text{nm}$ (bottom). The legend shows the carrier density in units of $10^{10} \text{cm}^{-2}$.}
			\label{Full_Antisymmtrized}
		\end{figure}

In this section, we test the dependence of the transverse density on the well-width and the carrier density numerically with fully-antisymmetrized wave functions in 3D space and ascertain to what extent the system can be approximated with two non-overlapping bases. Because the number of permutations increases rapidly with the system size, and because one does not have analytical ways to simplify the calculation of the drift velocity in the 3D-DMC, we estimate that the study of a system with more than 8-10 particles is out of our reach. 
Fortunately, we have found that the system's transverse distribution is largely insensitive to the size of the system and the type of the in-plane wave function. Therefore we study a 4-particle system with its in-plane wave function given by the $(3, 3, 1)$ state. We choose transverse wave functions that are not strictly orthogonal, i.e. incorporate a small tunneling between the two layers. Specifically, we choose
		\begin{equation}
		\begin{aligned}
			\psi_L\left( w_i \right)&=\frac{w_i}{W} \exp{\left[-8 \frac{w_i}{W}\right]}\\
			\psi_R\left( w_i \right)&=(1-\frac{w_i}{W}) \exp{\left[-8\left(1-\frac{w_i}{W}\right)\right]}\\
		\end{aligned}
		\end{equation}
		Here we have shifted the quantum well's location to the range $[0, W]$ for simplicity. 
		We do not enforce the central density to be zero; as a result, whether the system is a well-defined bilayer is determined by the diffusion process itself. 
		The bases chosen here are not strictly orthogonal but they are still linearly independent. If the final distribution breaks into two well-separated density lobes, then it indicates that the system can be treated as a two-component state. On the contrary, if the final distribution is not well-separated, then one should not treat the system as a two-component state. [This is the reason why we call the state $(3, 3, 1)$-like state rather than $(3, 3, 1)$ state in the caption of Fig.~\ref{Full_Antisymmtrized}.] This also offers an estimation of the width and density beyond which the system can be treated as a two-component state. Our 3D-DMC results for the density are shown in Fig.~\ref{Full_Antisymmtrized}. As one can see, the system is only well-separated and has negligible density in the center when $n\gtrsim 2\times 10^{11} \text{cm}^{-2}$ for $W=70 \text{nm}$ and $n\gtrsim 1\times 10^{11} \text{cm}^{-2}$ for $W=80 \text{nm}$. Recalling that in the main text we show a phase transition from a one-component state to a two-component state occurring around $n=2.2\times10^{11} \text{cm}^{-2}$ for $W=70 \text{nm}$ $n=1.5\times10^{11} \text{cm}^{-2}$ for $W=80 \text{nm}$, this calculation of the fully-antisymmetrized state justifies our approximation in the main text.
		
		\section{Thermodynamic extrapolations of energy}
		
\begin{figure}[H]
			\includegraphics[width=\columnwidth]{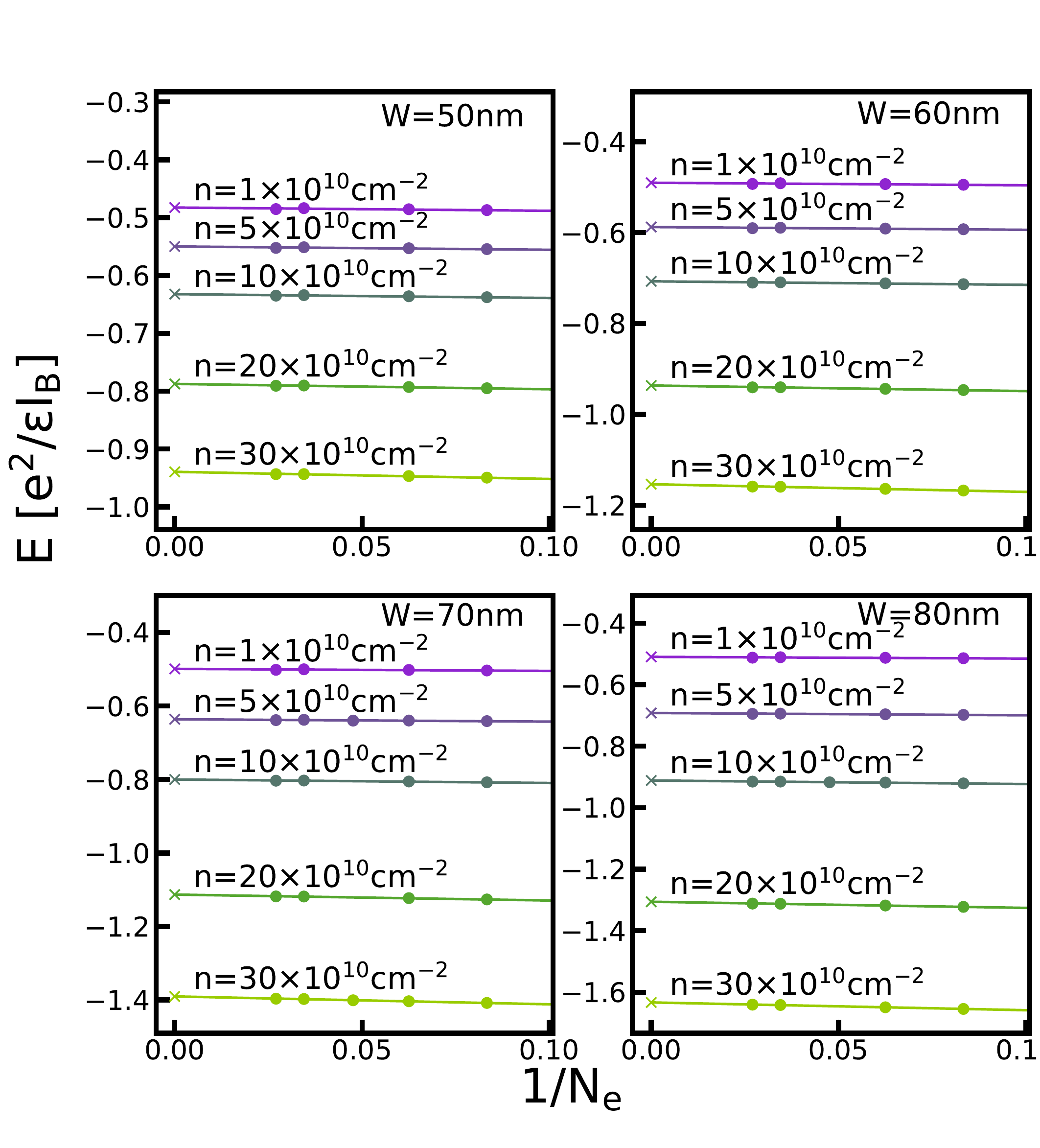}
			\caption{The VMC energy of the one-component CFFS as a function of $1/N_e$.}
			\label{VMC_EXTRAP_CFFS}
		\end{figure}

\begin{figure}[H]
			\includegraphics[width=\columnwidth]{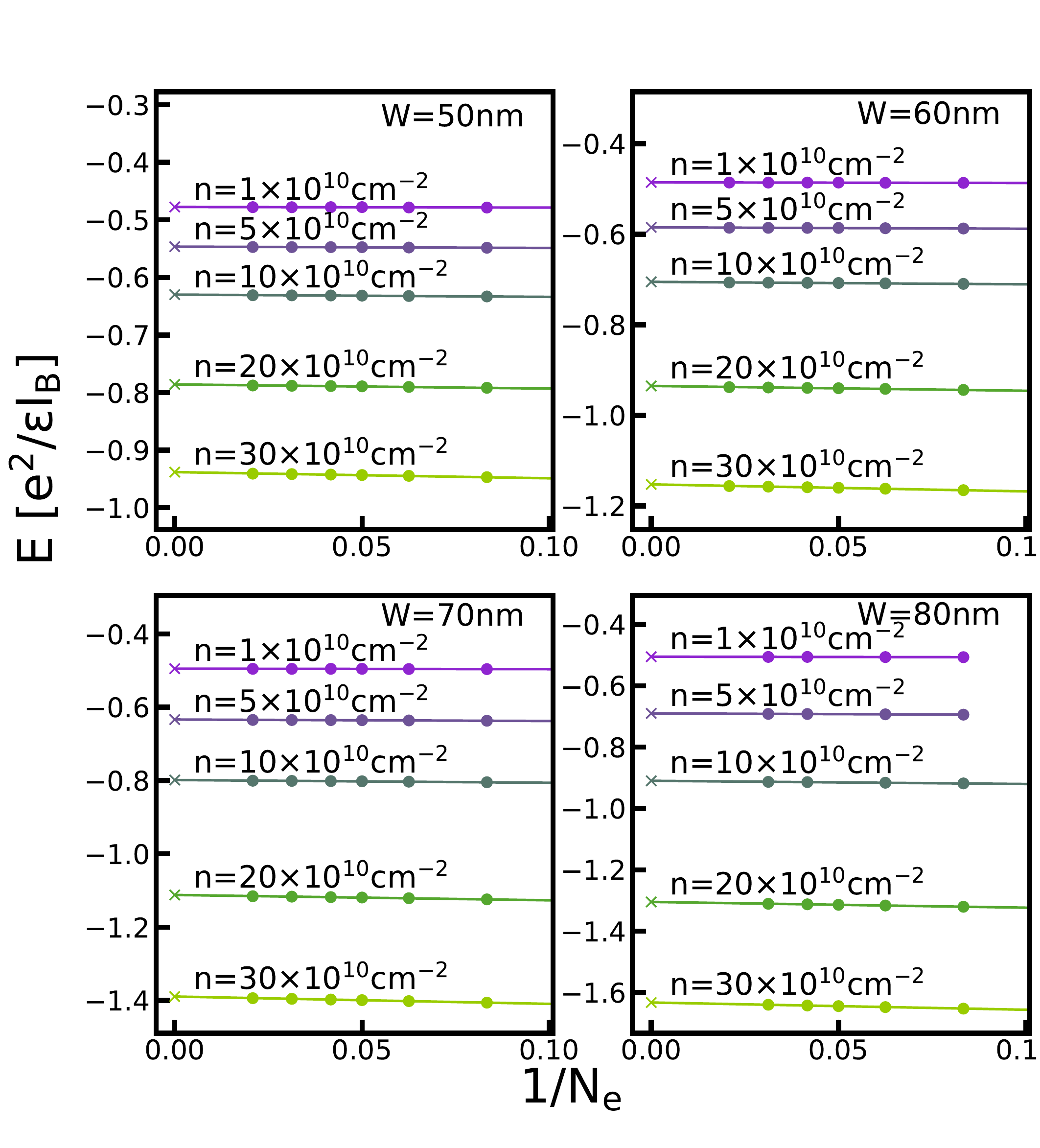}
			\caption{The VMC energy of the one-component Pfaffian as a function of $1/N_e$.}
			\label{VMC_EXTRAP_PFAF}
		\end{figure}

\begin{figure}[H]
			\includegraphics[width=\columnwidth]{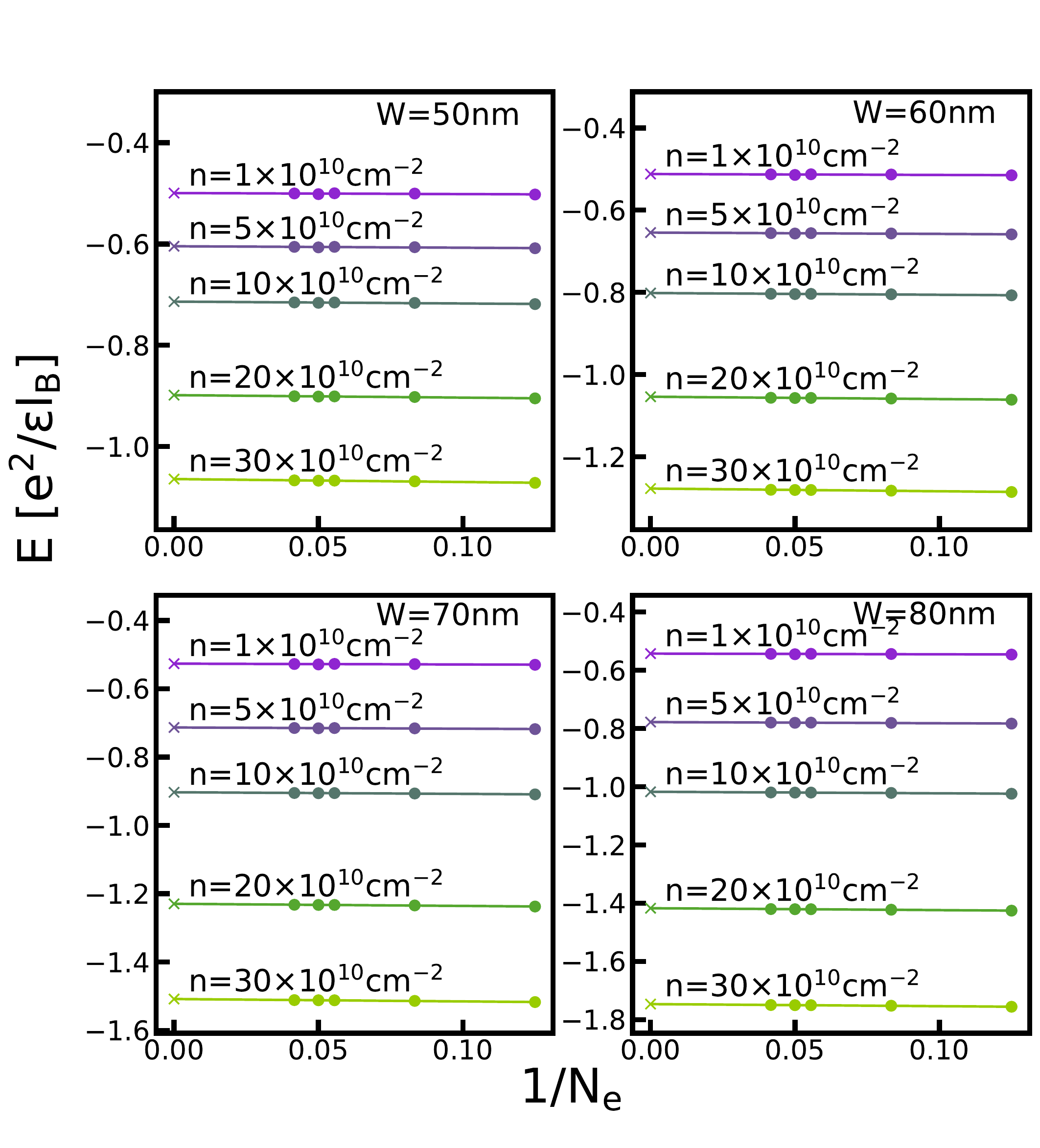}
			\caption{The VMC energy of the two-component $(3, 3, 1)$ as a function of $1/N_e$.}
			\label{VMC_EXTRAP_331}
		\end{figure}

\begin{figure}[H]
			\includegraphics[width=\columnwidth]{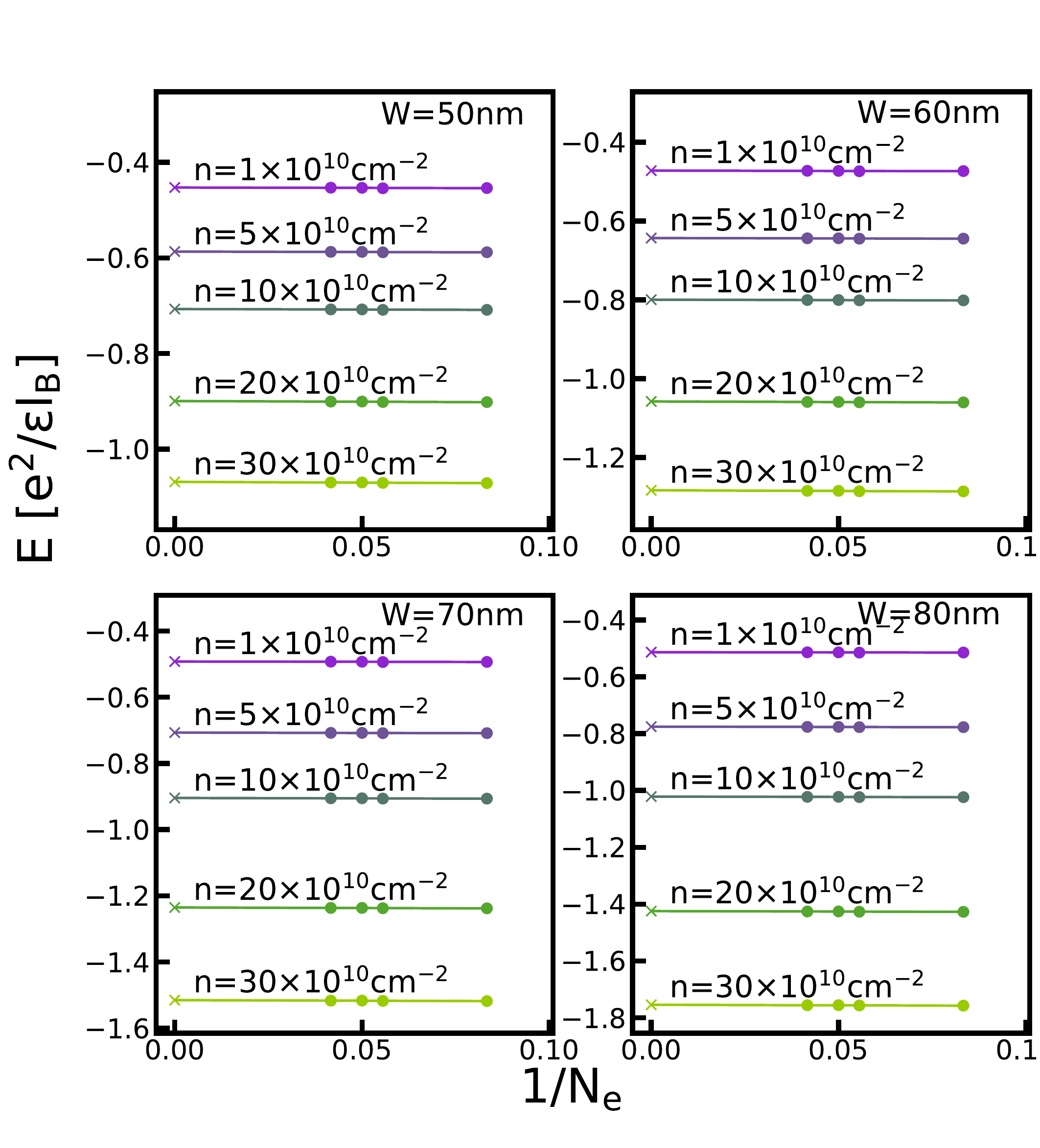}
			\caption{The VMC energy of the two-component $1/4+1/4$ CFFS as a function of $1/N_e$.}
			\label{VMC_EXTRAP_BICFFS}
		\end{figure}

\begin{figure}[H]
			\includegraphics[width=\columnwidth]{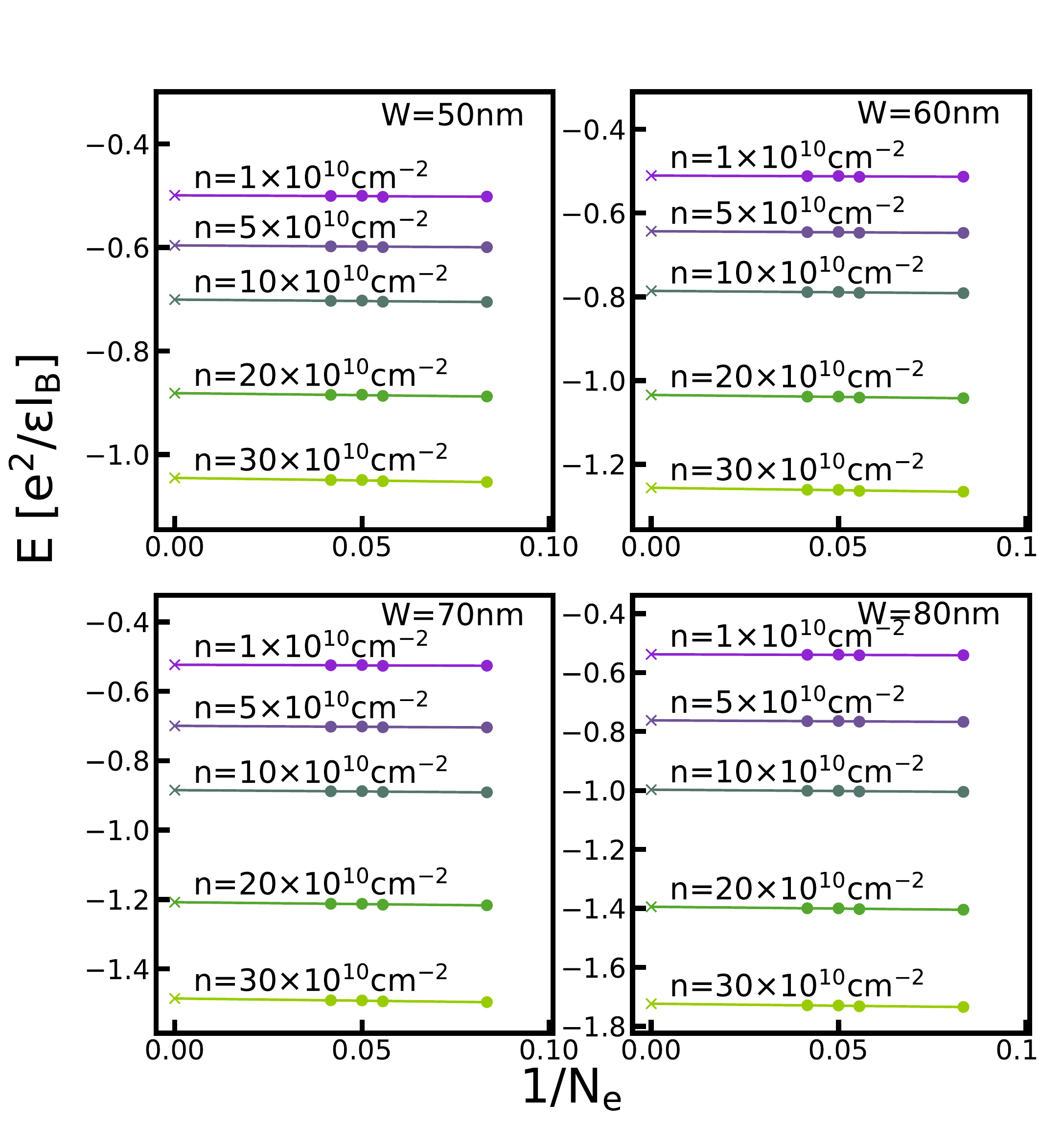}
			\caption{The VMC energy of the two-component pseudo-spin singlet CFFS as a function of $1/N_e$.}
			\label{VMC_EXTRAP_SINGLET}
		\end{figure}

\begin{figure}[H]
			\includegraphics[width=\columnwidth]{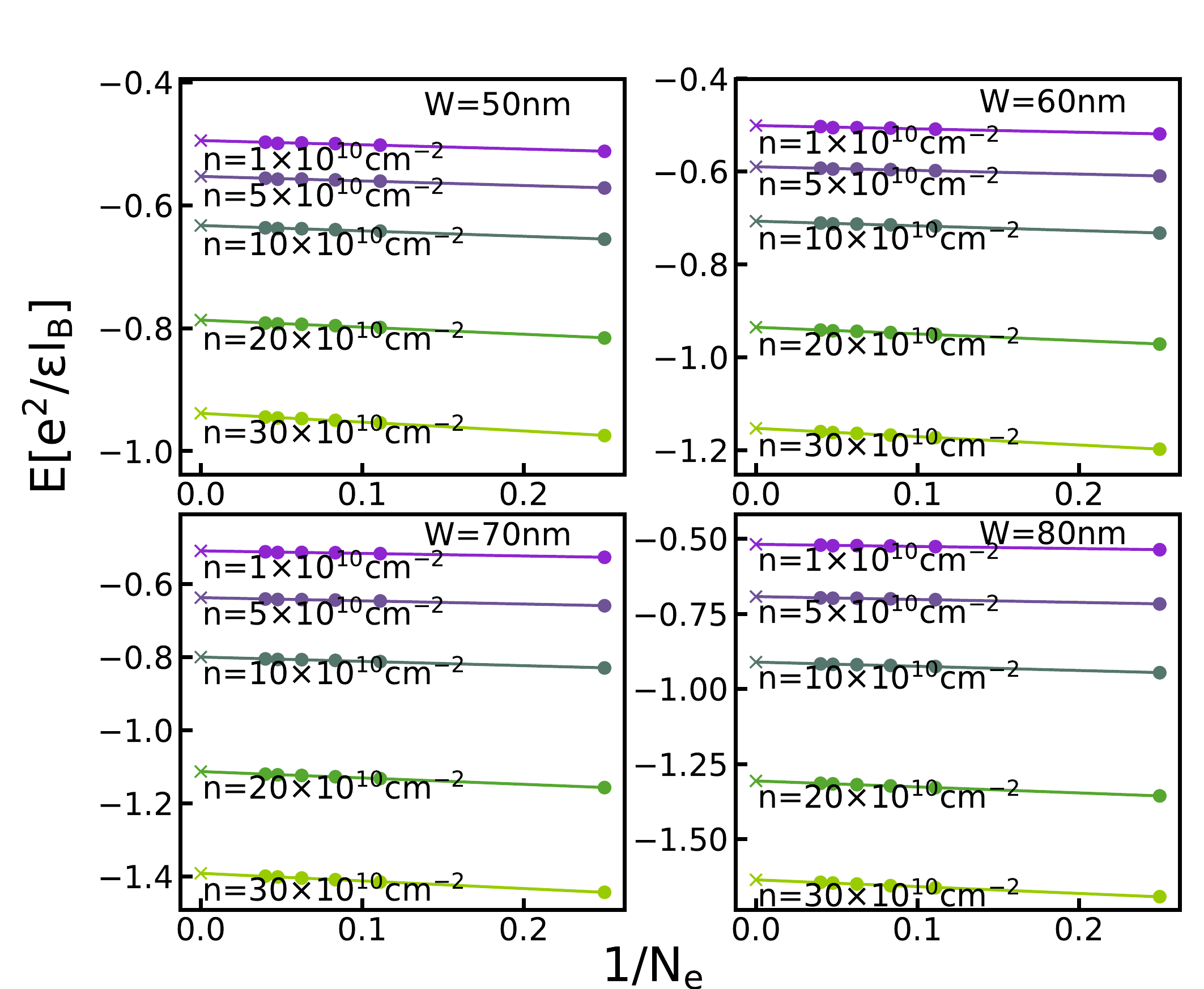}
			\caption{2D-DMC energy of the one-component CFFS as a function of $1/N_e$.}
			\label{2D_DMC_EXTRAP_CFFS}
		\end{figure}

\begin{figure}[H]
			\includegraphics[width=\columnwidth]{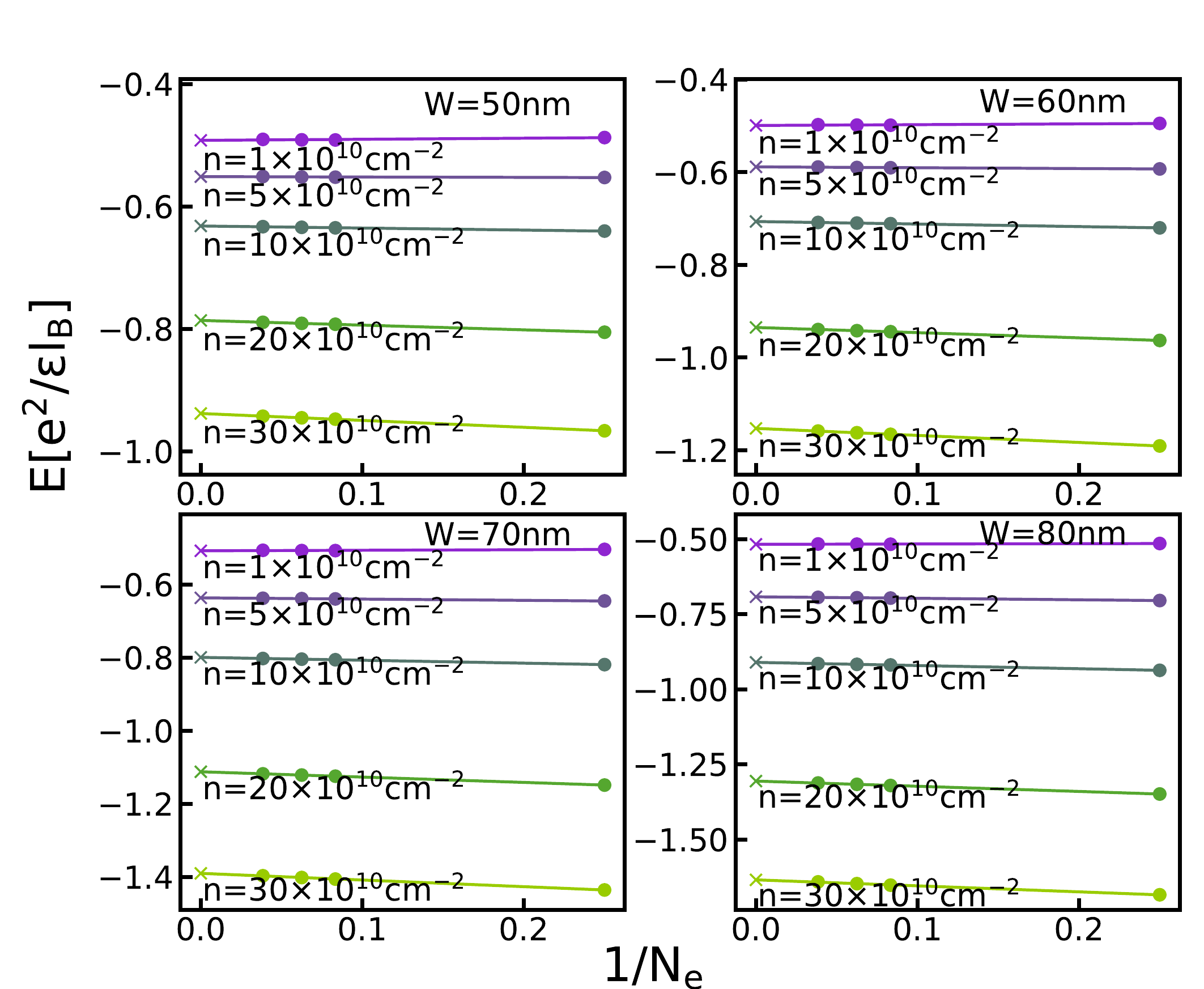}
			\caption{2D-DMC energy of the one-component Pfaffian as a function of $1/N_e$.}
			\label{2D_DMC_EXTRAP_PFAF}
		\end{figure}

\begin{figure}[H]
			\includegraphics[width=\columnwidth]{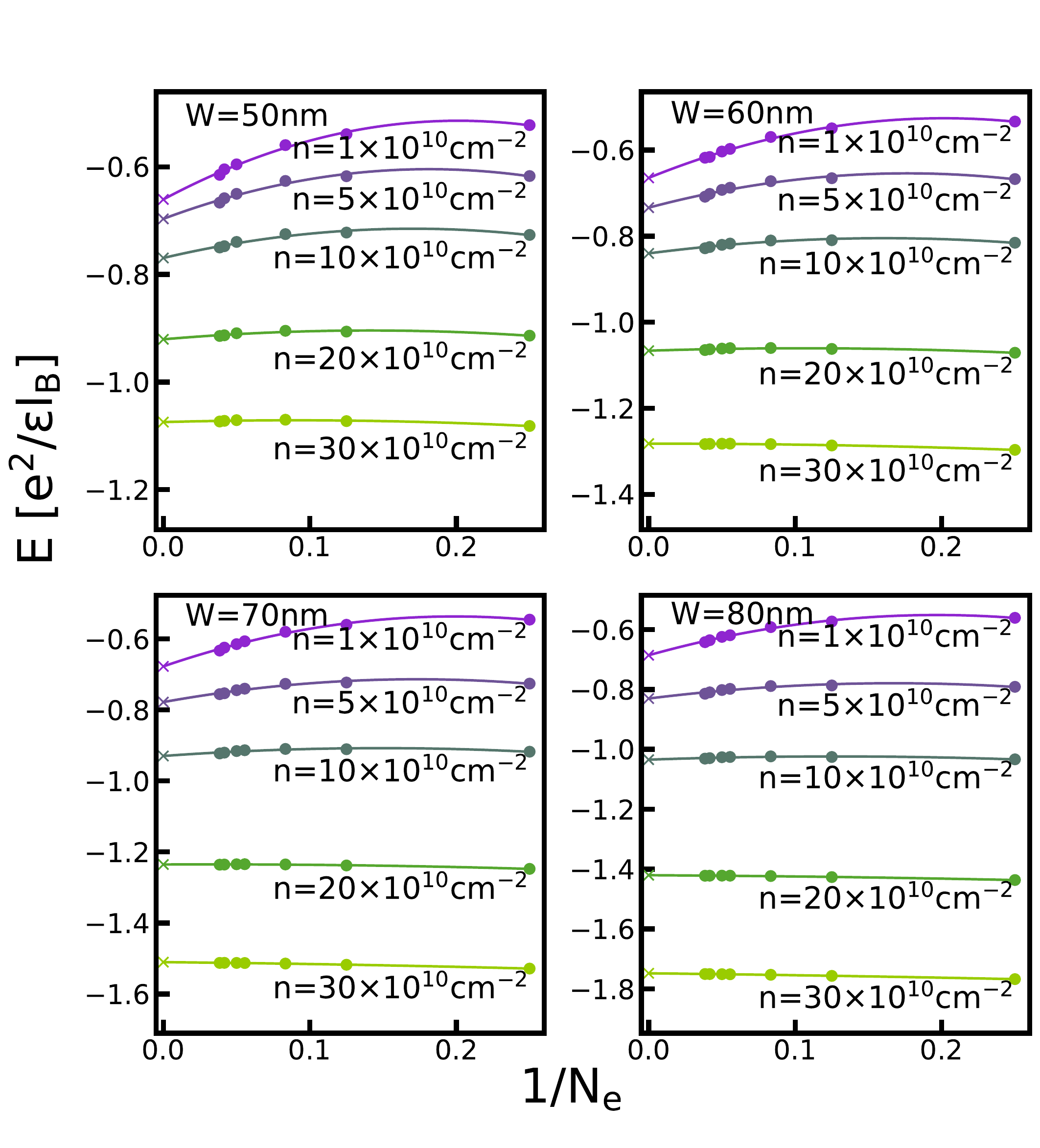}
			\caption{2D-DMC energy of the two-component $(3, 3, 1)$ as a function of $1/N_e$.}
			\label{2D_DMC_EXTRAP_331}
		\end{figure}

\begin{figure}[H]
			\includegraphics[width=\columnwidth]{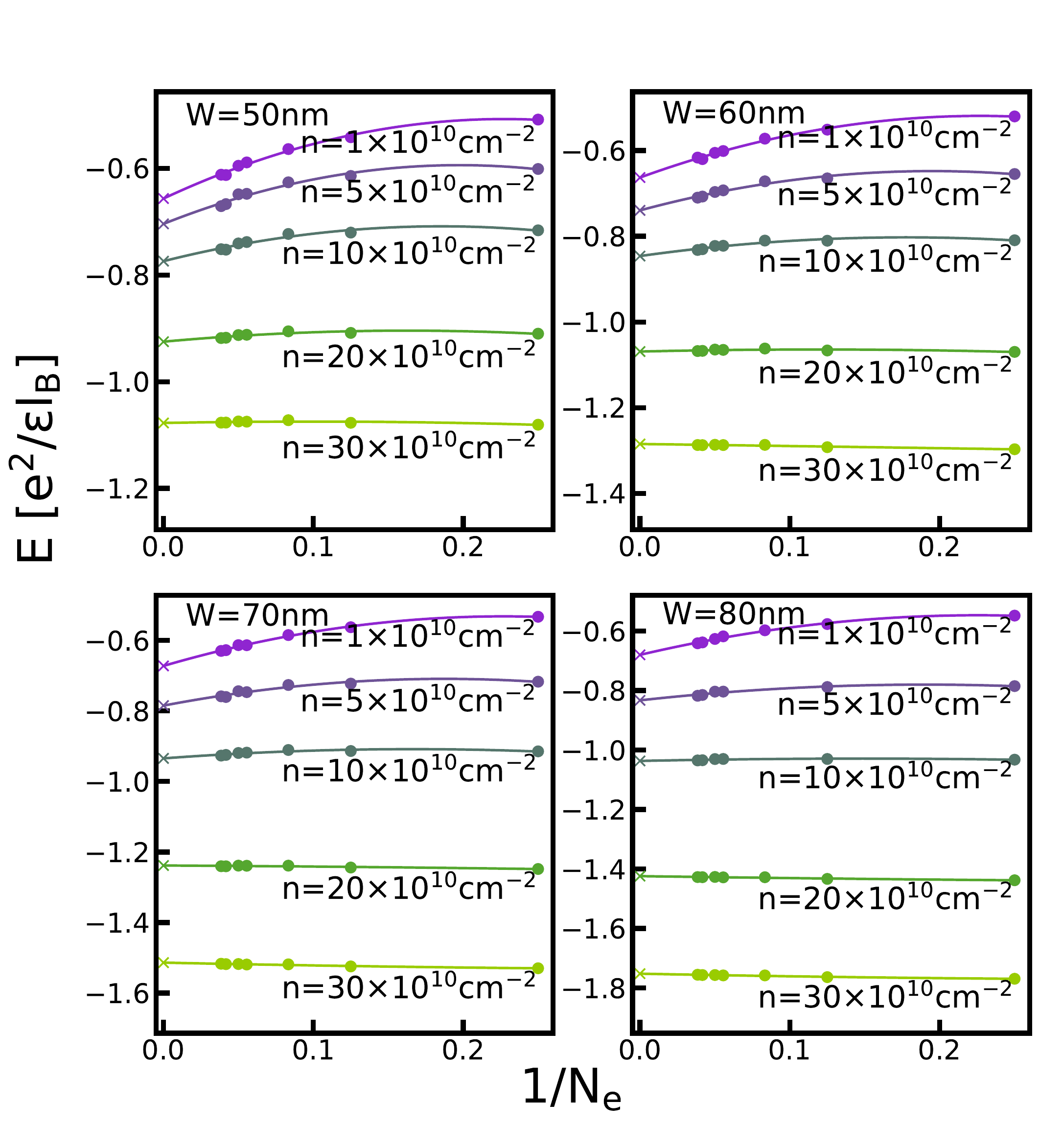}
			\caption{2D-DMC energy of the two-component $1/4+1/4$ CFFS as a function of $1/N_e$.}
			\label{2D_DMC_EXTRAP_BICFFS}
		\end{figure}

\begin{figure}[H]
			\includegraphics[width=\columnwidth]{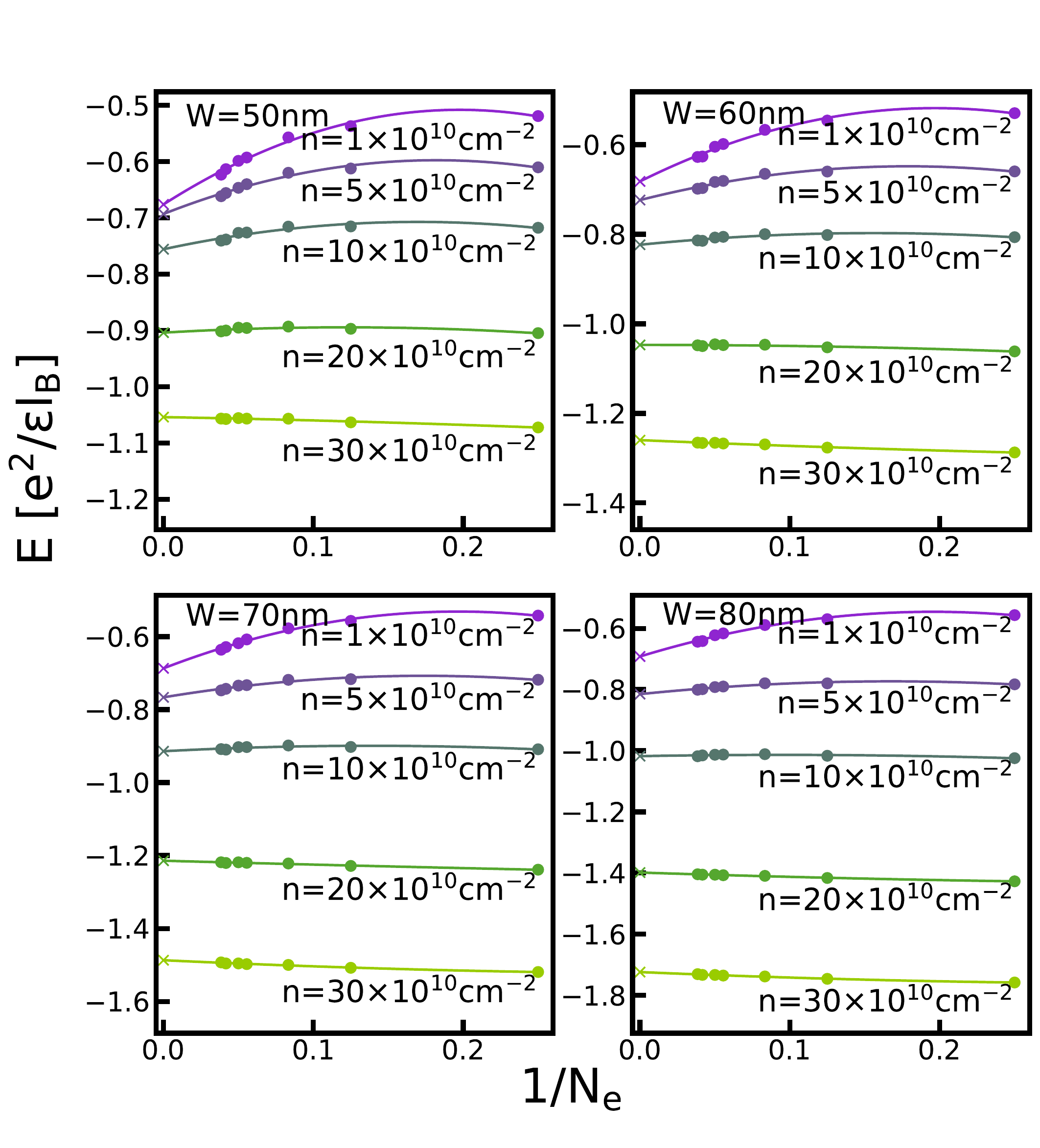}
			\caption{2D-DMC energy of the two-component pseudo-spin singlet CFFS as a function of $1/N_e$.}
			\label{2D_DMC_EXTRAP_SINGLET}
		\end{figure}

\begin{figure}[H]
			\includegraphics[width=\columnwidth]{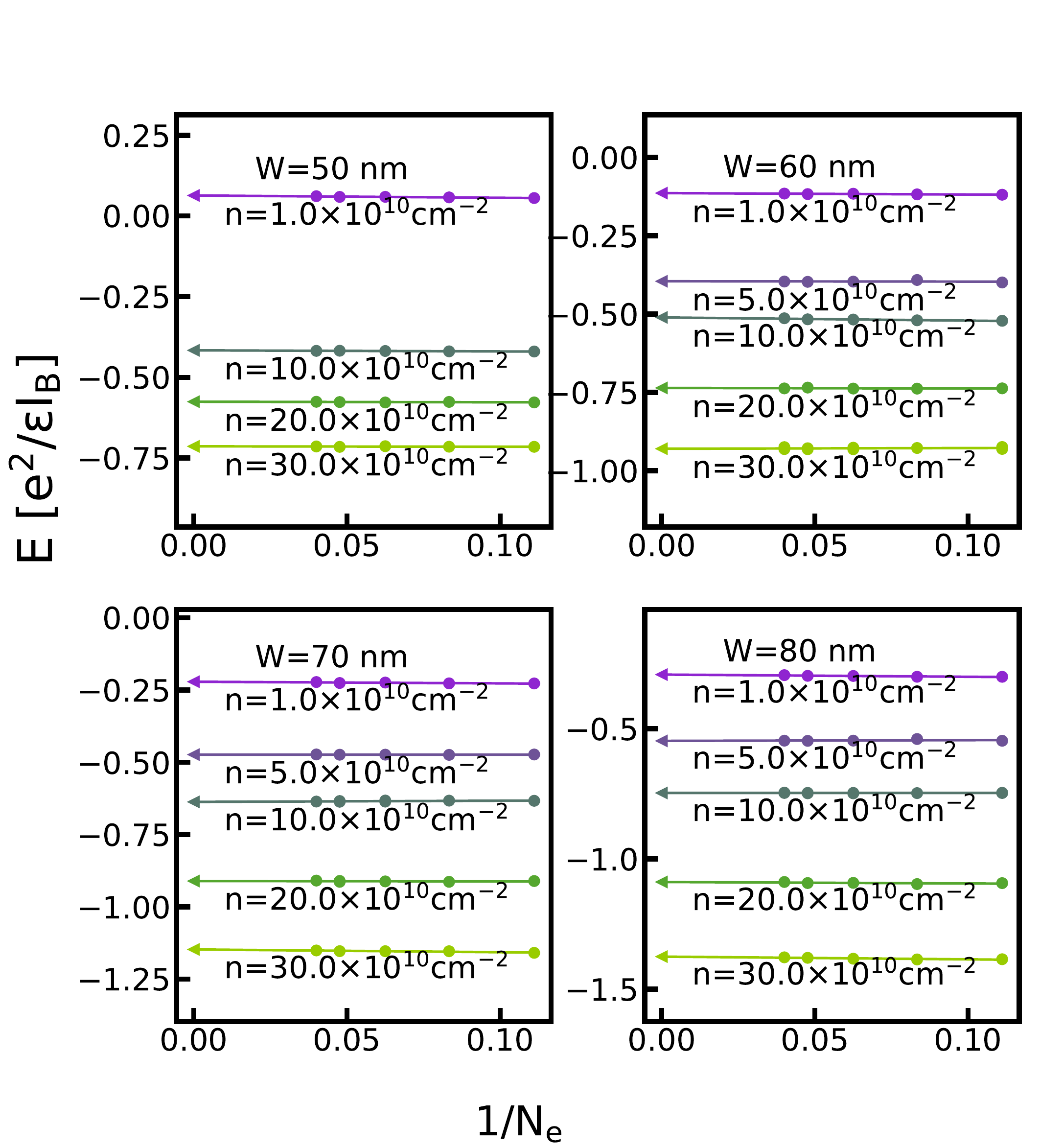}
			\caption{3D-DMC energy of the one-component CFFS as a function of $1/N_e$.}
			\label{3D_DMC_EXTRAP_CFFS}
		\end{figure}

\begin{figure}[H]
			\includegraphics[width=\columnwidth]{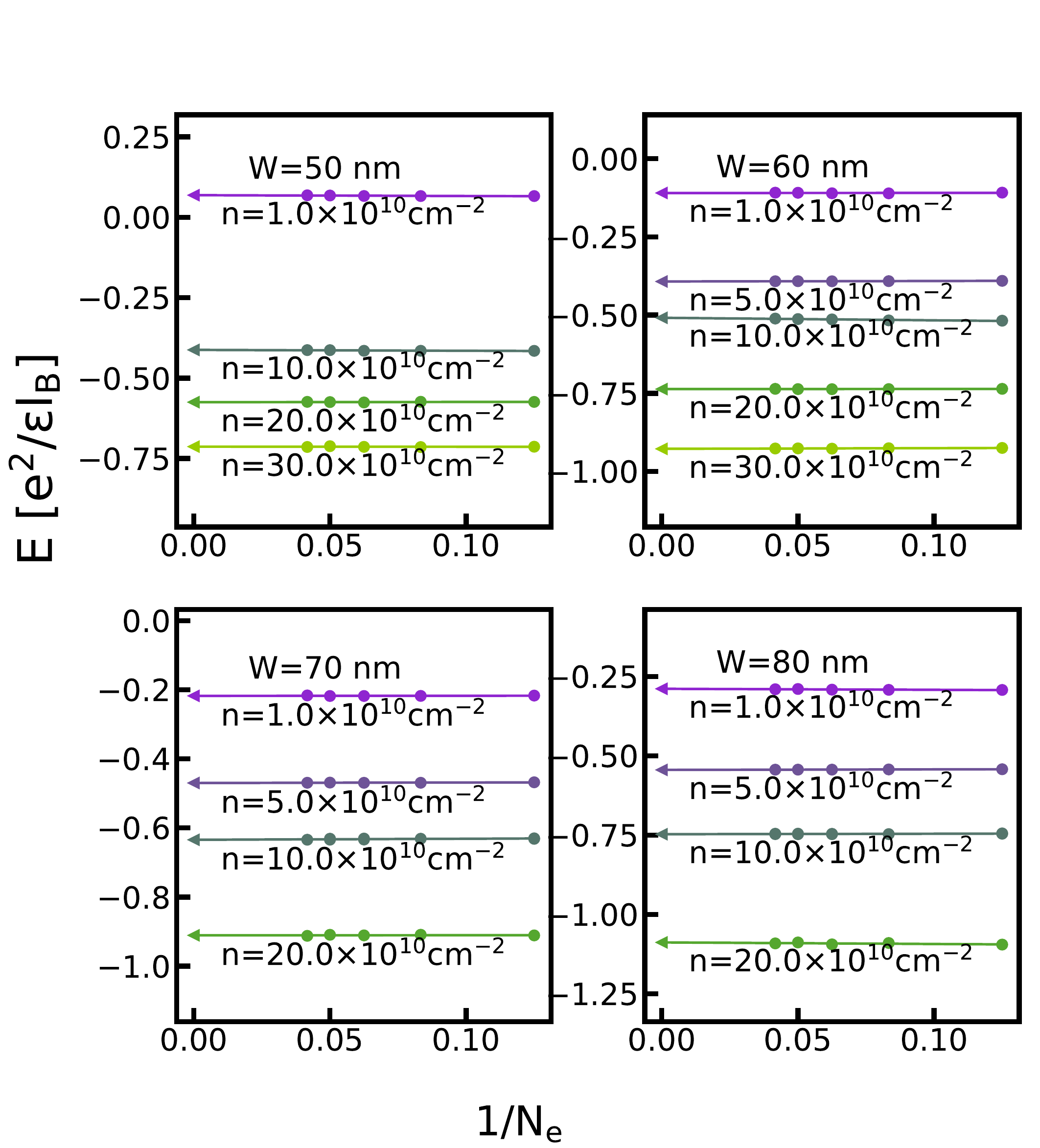}
			\caption{3D-DMC energy of the one-component Pfaffian as a function of $1/N_e$.}
			\label{3D_DMC_EXTRAP_PFAF}
		\end{figure}

\begin{figure}[H]
			\includegraphics[width=\columnwidth]{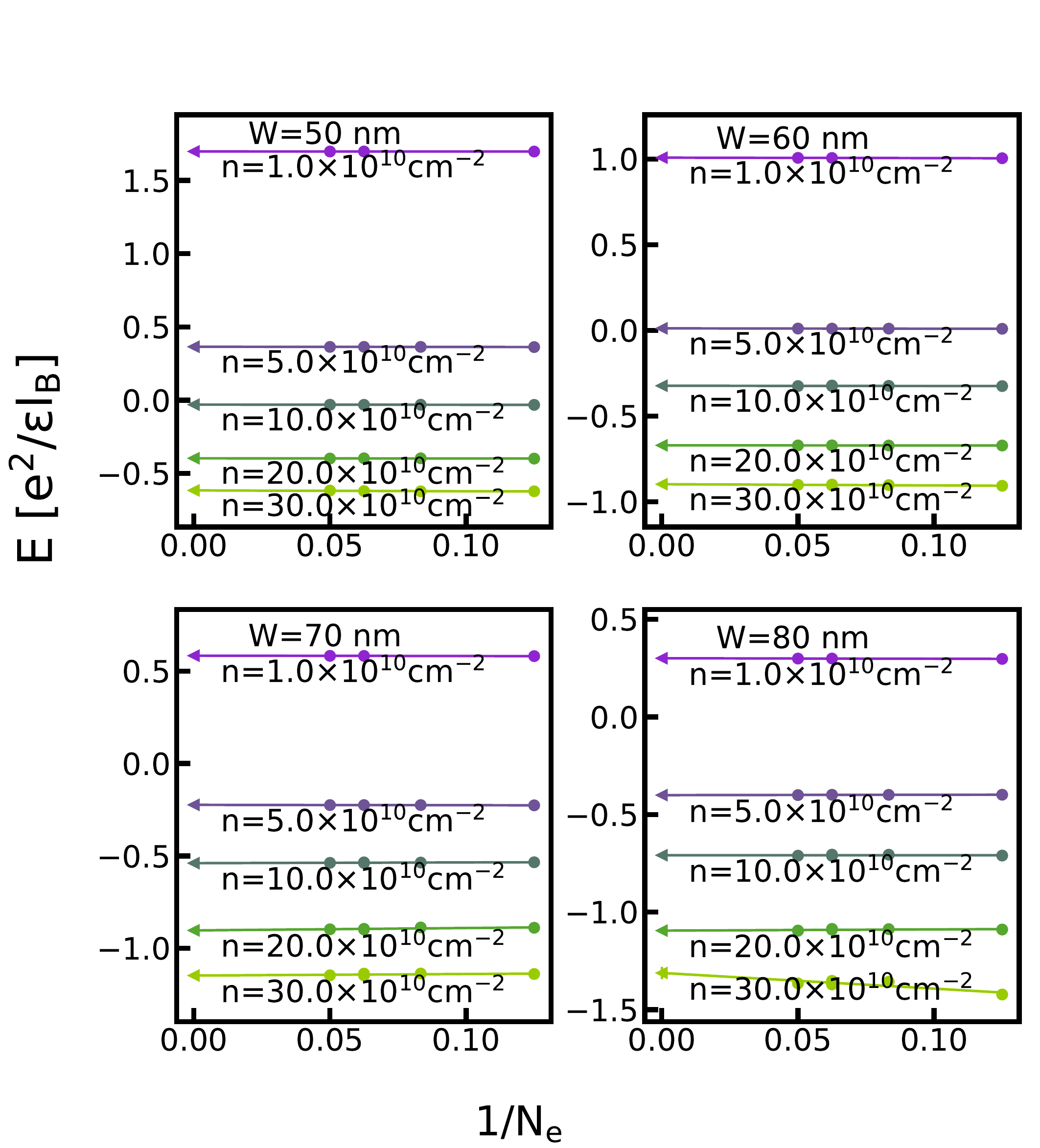}
			\caption{3D-DMC energy of the two-component $(3, 3, 1)$ as a function of $1/N_e$.}
			\label{3D_DMC_EXTRAP_331}
		\end{figure}

\begin{figure}[H]
			\includegraphics[width=\columnwidth]{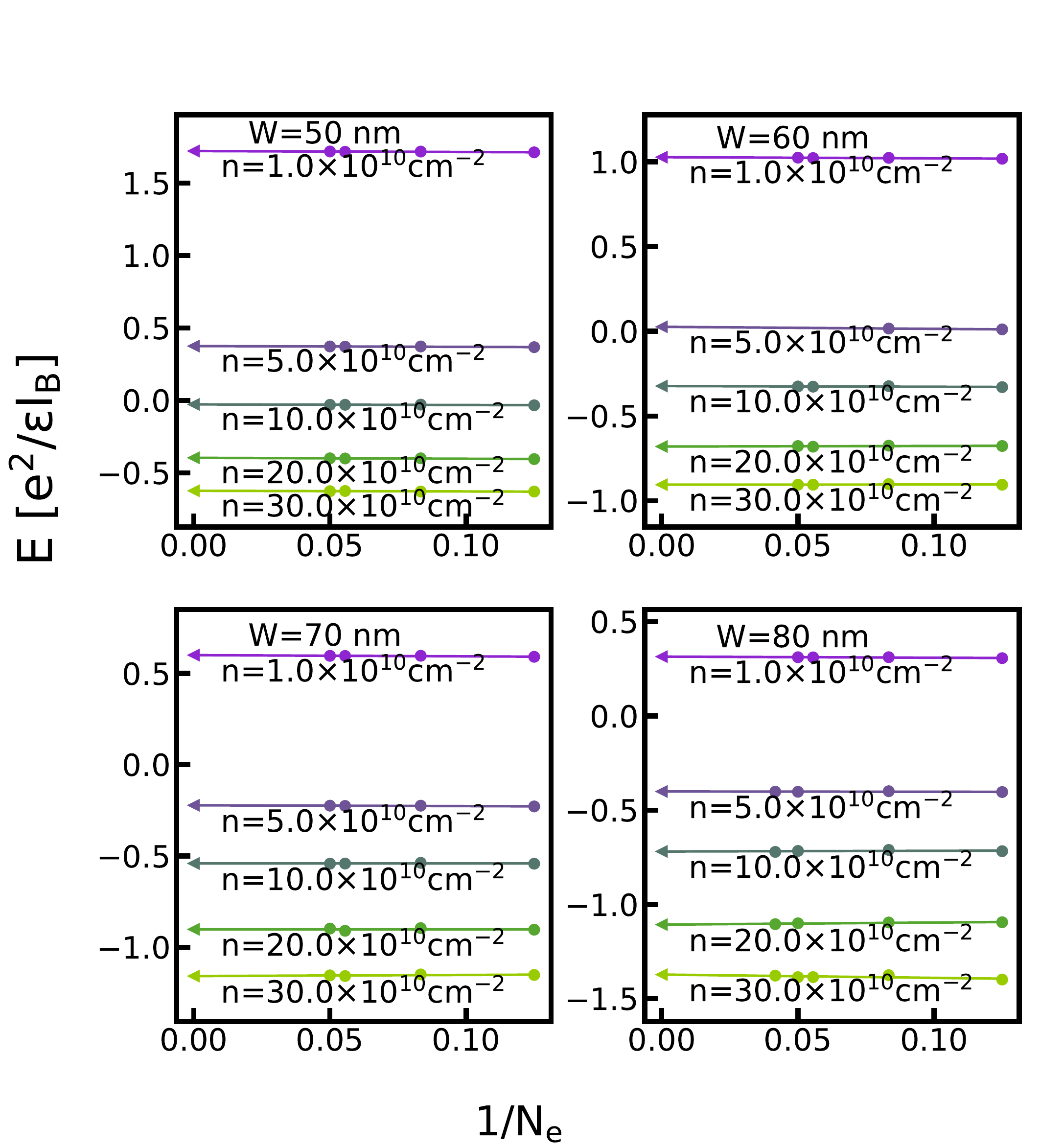}
			\caption{3D-DMC energy of the two-component $1/4+1/4$ CFFS as a function of $1/N_e$.}
			\label{3D_DMC_EXTRAP_BICFFS}
		\end{figure}

\begin{figure}[H]
			\includegraphics[width=\columnwidth]{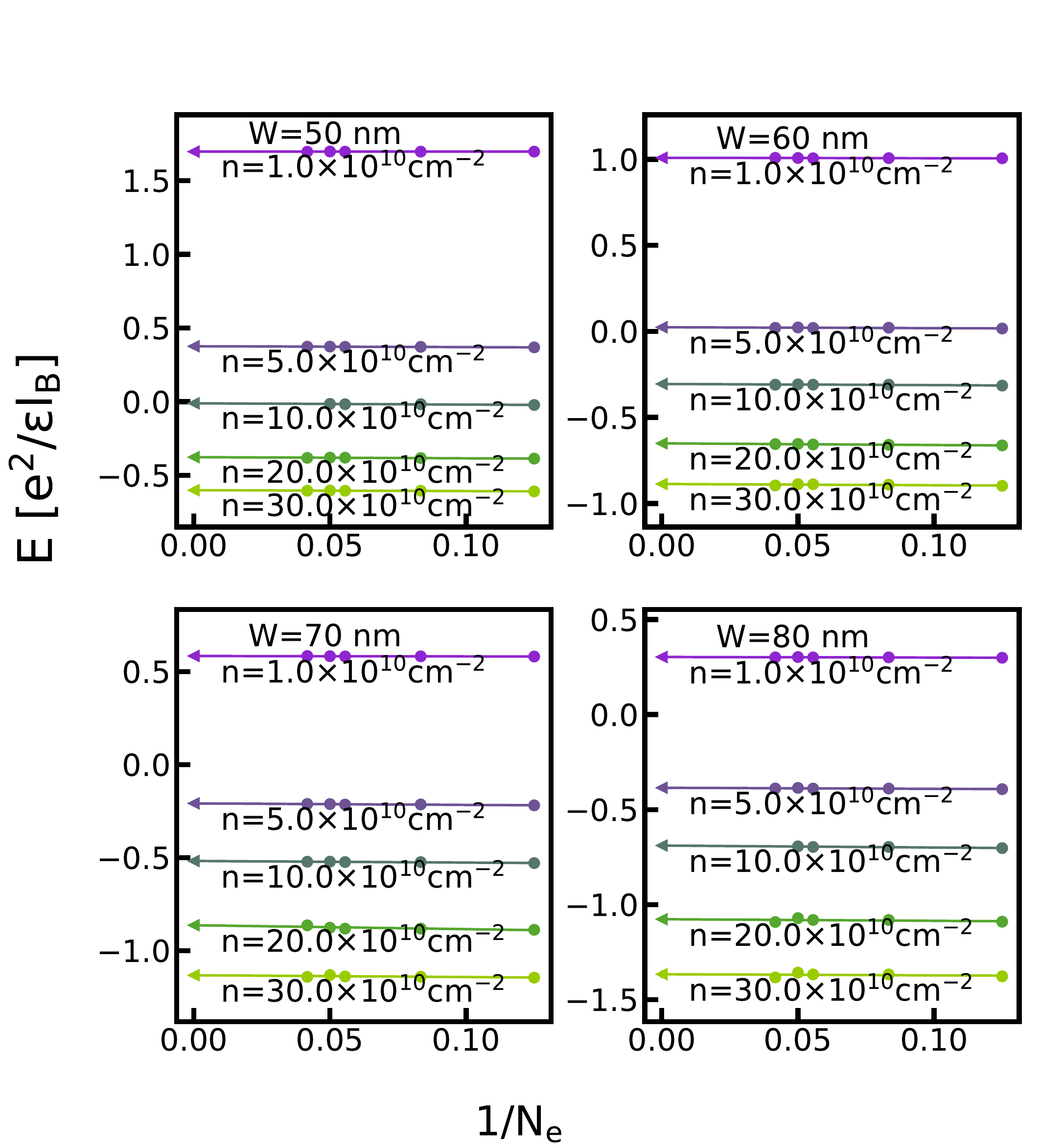}
			\caption{3D-DMC energy of the two-component pseudo-spin singlet CFFS as a function of $1/N_e$.}
			\label{3D_DMC_EXTRAP_SINGLET}
		\end{figure}

		The phase diagrams in the main text are obtained by comparing the energies of different states in the thermodynamic limit. For completeness, we show the extrapolations of the energies of various states calculated by either VMC, 2D-DMC, or 3D-DMC in this section.
		Figs.~\ref{VMC_EXTRAP_CFFS}-\ref{VMC_EXTRAP_SINGLET} show the energy extrapolation for the VMC calculation; 
		Figs.~\ref{2D_DMC_EXTRAP_CFFS}-\ref{2D_DMC_EXTRAP_SINGLET} show the energy extrapolation for the 2D-DMC calculation; and 
		Figs.~\ref{3D_DMC_EXTRAP_CFFS}-\ref{3D_DMC_EXTRAP_SINGLET} show the energy extrapolation for the 3D-DMC calculation.
		

		

		\section{The system size dependence of transverse density}
		\label{transverse density}

\begin{figure}[H]
			\includegraphics[width=\columnwidth]{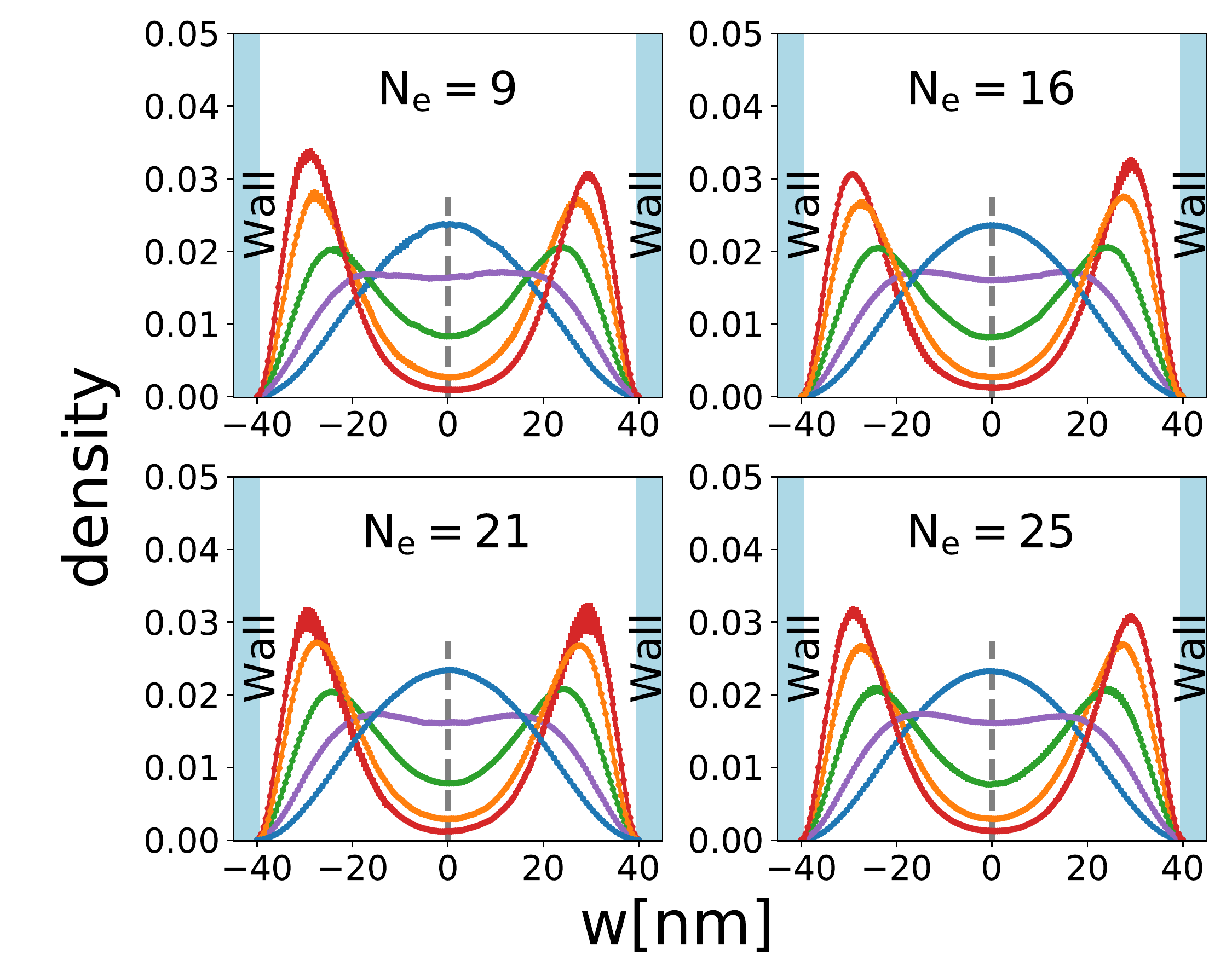}
			\caption{The transverse density profiles of the CFFS state for several particle numbers. They are identical within the statistical uncertainty. The different colors are for different densities, following the same color scheme as in Fig.~\ref{Density_1L}.}
			\label{FINITE_SIZE_DEN}
		\end{figure}

		The profiles of the transverse density for the CFFS, obtained from our 3D-DMC calculation, are shown in Fig.~\ref{FINITE_SIZE_DEN} for several system sizes. These show that the 3D-DMC transverse density has negligible dependence on the system size. This conclusion also applies to other states considered in this paper. We, therefore, believe that the various transverse density profiles shown in this article represent the thermodynamic limit.

		\section{Periodic Coulomb interaction and Ewald Summation}
		\label{Ewald_V}
		
		On the torus, we must work with a periodic version of the Coulomb interaction. A naive strategy is to extend the normal Coulomb potential periodically. Although this approach is theoretically possible, it is impractical because of slow convergence.
		The Ewald-summation method overcomes this difficulty.
		The idea is to split the Coulomb interaction into a short-ranged part and a long-ranged part. The short-ranged part can be summed in real space quickly; the long-ranged part in the real space becomes short-ranged in the momentum space, hence can be summed conveniently in the momentum space.
		 We follow Yeh's approach\cite{Yeh99a} in which a generalized summation is explicitly formalized including the transverse dimension with an open boundary:
		\begin{widetext}
			\begin{equation}
			\begin{aligned}
			V_\text{Ewald}=&\frac{1}{2}\sideset{}{'}\sum_{i,j=1}^{N_e} \sum_{|\vec{m}=0|}^{\infty}q_i q_j \frac{\text{erfc}(\alpha|\vec r_{ij}+\vec m|)}{|\vec r_{ij}+\vec m|}+\frac{\pi}{2A}\sum_{i,j=1}^{N_e}\sum_{\vec h\neq 0}q_i q_j\frac{\cos(\vec h\cdot \vec r_{ij})}{h}\\
			&\times \left\{ \exp{(h z_{ij})}\text{erfc}(\alpha z_{ij} +\frac{h}{2\alpha})+\exp{(-h z_{ij})}\text{erfc}(-\alpha z_{ij}+\frac{h}{2\alpha}) \right\}\\&-\frac{\pi}{A} \sum_{i=1}^{N_e}\sum_{i=1}^{N_e}q_i q_j\left\{ z_{ij} \text{erf}(\alpha z_{ij})+\frac{1}{\alpha\sqrt{\pi}}\exp(-\alpha^2z_{ij}^2)\right\}-\frac{\alpha}{\sqrt{\pi}}\sum_{i=1}^{N_e}q_i^2
			\end{aligned} \label{Ewaldsum}
			\end{equation}
		\end{widetext}
		The prime on the summation $\sideset{}{'}\sum_{i,j=1}^{N_e}$ is to remind us that terms with $i=j$ are included only for $\vec{m}\neq 0$. 
		It is worth noting that this definition of the interaction properly includes the charge-neutrality condition, i.e. it contains the electron-electron, background-background repulsion, and the electron-background attraction. To be more explicit, the omission of the term with $\vec{h}=0$ in the summation and the last of Eq.~\ref{Ewaldsum} term are due to the electron-background and background-background interaction. We refer the reader to Refs.~[\onlinecite{Heyes97}] and [\onlinecite{Parry75}], 
		for a thorough discussion of the technical aspects of this method.

	\end{appendix}
	
	%
	

\end{document}